\documentclass[]{book}

\usepackage[final]{neel}

\usepackage{makeidx}         
\RequirePackage{afterpage}

\usepackage{subfigure}
\usepackage{layouts}
\RequirePackage{array} 
\RequirePackage{tabularx}
\RequirePackage{ltablex} 
\RequirePackage{longtable}  

\hypersetup{
  pdftitle={Magnetic nanowires and nanotubes},    	
  pdfauthor={Michal Stano and Olivier Fruchart},   	
  pdfsubject={Book chapter}, 						  	        
  pdfkeywords={magnetic nanowire, magnetic nanotube, synthesis, magnetic properties}   
}

\RequirePackage[authoryear,round]{natbib}
\bibpunct[; ]{~(}{)}{,}{a}{}{;} 

\RequirePackage[fit]{truncate}
\RequirePackage{fancyhdr}

\graphicspath{{fig/}}
\def\figurename{Fig.}

\newcommand{\nocontentsline}[3]{}
\newcommand{\tocless}[2]{\bgroup\let\addcontentsline=\nocontentsline#1{#2}\egroup}

\renewcommand\thesection{\arabic{section}.}

\RequirePackage{suffix}

\begin{document}

\fancyhead{}




\setcounter{tocdepth}{3}
\renewcommand\thechapter{}
\setlength{\headheight}{15.2pt}
\renewcommand{\chaptername}{}%
\makeatletter
\renewcommand{\thefigure}{\@arabic\c@figure}
\renewcommand{\thetable}{\@arabic\c@table}
\renewcommand{\theequation}{\@arabic\c@equation}
\@addtoreset{figure}{chapter}%
\@addtoreset{table}{chapter}%
\@addtoreset{equation}{chapter}\makeatother

\newcommand\chapterauthor[1]{\printchapterauthor{#1}}
\WithSuffix\newcommand\chapterauthor*[1]{\printchapterauthor{#1}}

\makeatletter
\newcommand{\printchapterauthor}[1]{%
  {\parindent0pt\vspace*{-25pt}%
  \linespread{1.1}\large\scshape#1%
  \par\nobreak\vspace*{35pt}}
  \@afterheading%
}
\renewcommand{\chaptermark}[1]{\markboth{\protect\textsc{\large Magnetic nanowires and nanotubes}}{}}
\renewcommand{\sectionmark}[1]{\markboth{\protect\textsc{Magnetic nanowires and nanotubes}}{}\markright{\thesection\ #1}}

\tocless\chapter{Magnetic nanowires and nanotubes}

\chapterauthor{Michal Sta\v{n}o\footnote{CEITEC - Central European Institute of Technology, Brno University of Technology, 612 00 Brno, Czech Republic} and Olivier Fruchart\footnote{Univ. Grenoble Alpes, CNRS, CEA, Grenoble INP, INAC-Spintec, F-38000 Grenoble, France}}
\makeatother




We propose a review of the current knowledge about the synthesis, magnetic properties and applications of magnetic cylindrical nanowires and nanotubes. By \textsl{nano} we consider diameters reasonably smaller than a micrometer. At this scale, comparable to micromagnetic and transport length scales, novel properties appear. At the same time, this makes the underlying physics easier to understand due to the limiter number of degrees of freedom involved. The three-dimensional nature and the curvature of these objects contribute also to their specific properties, compared to patterns flat elements. While the topic of nanowires and later nanotubes started now decades ago, it is nevertheless flourishing, thanks to the progress of synthesis, theory and characterization tools. These give access to ever more complex and thus functional structures, and also shifting the focus from material-type measurements of large assemblies, to single-object investigations. We first provide an overview of common fabrication methods yielding nanowires, nanotubes and structures engineered in geometry~(change in diameter, shape) or material~(segments, core-shell structures), shape or core-shell. We then review their magnetic properties: global measurements, magnetization states and switching, single domain wall statics and dynamics, and spin waves. For each aspect, both theory and experiments are surveyed. We also mention standard characterization techniques useful for these.  We finally mention emerging applications of magnetic nanowires and nanotubes, along with the foreseen perspectives in the topic.

\medskip
\noindent Keywords: magnetic nanotube, magnetic nanowire, core-shell, domain-wall

\tableofcontents

\mainmatter

\makeatletter
\pagestyle{fancy}
\fancyhead{}
\fancyfoot{}
\renewcommand{\chaptermark}[1]{\markboth{Magnetic nanowires and nanotubes}{}}
\renewcommand{\sectionmark}[1]{\markboth{Magnetic nanowires and nanotubes}{}\markright{\thesection\ #1}}
\fancyhead[RE]{\truncate{.90\headwidth}{\sc \nouppercase{\leftmark}}}  
\fancyhead[RO,LE]{\bfseries \thepage}
\fancyhead[LO]{\truncate{.90\headwidth}{\sc \nouppercase{\rightmark}}} 
\makeatother

\section{Preamble}
\label{sec_intro}

There exists a number of reviews and key references covering magnetic nanow\-ires~(NWs) and nanotubes~(NTs). \citet{bib-FER1999a} proposed the earliest review on the topic, considering only magnetic nanowires. The templates of choices were then track-etched polymer membranes, combined with electroplating for the synthesis. Investigations concerned the magnetic anisotropy and reversal, and a deep focus on magneto-transport (anisotropic magnetoresistance -- AMR, giant magnetoresistance -- GMR), at a time when GMR was a hot topic. At that time most measurements were global, yet with a few pioneering single-object reports.  In the \textsl{handbook of nanophysics: nanotubes and nanowires}, \citet{bib-SAT2010} provided among many other chapters, a few ones on magnetic tubes and wires (\eg, chapters 14, 22, and 31) covering mainly synthesis, modelling of magnetic properties and spin waves, and global magnetometry measurements. \citet{bib-IVA2013b} made a review of magnetization reversal in nanowires, with extensive reference to literature results. \citet{bib-SOU2014} made a review on alumina templates, and electroplating to fabricate wires and tubes. Although these aspects are generic, they included also a review of magnetic investigations: anisotropy~(shape and interactions in arrays, magnetocrystalline, magneto-elastic), magnetization reversal, and finally covered applications. The importance of magnetic microscopy is already larger at that time. A key contribution is the book edited by \citet{bib-VAZ2015}, with 25 focused chapters written by different authors. Chapters concern synthesis, microwires and magneto-impedance, simulations of magnetization states and domain wall~(DW) motion, applications.

\begin{figure}[htbp]
\centering
    \includegraphics[width=1.0\linewidth]{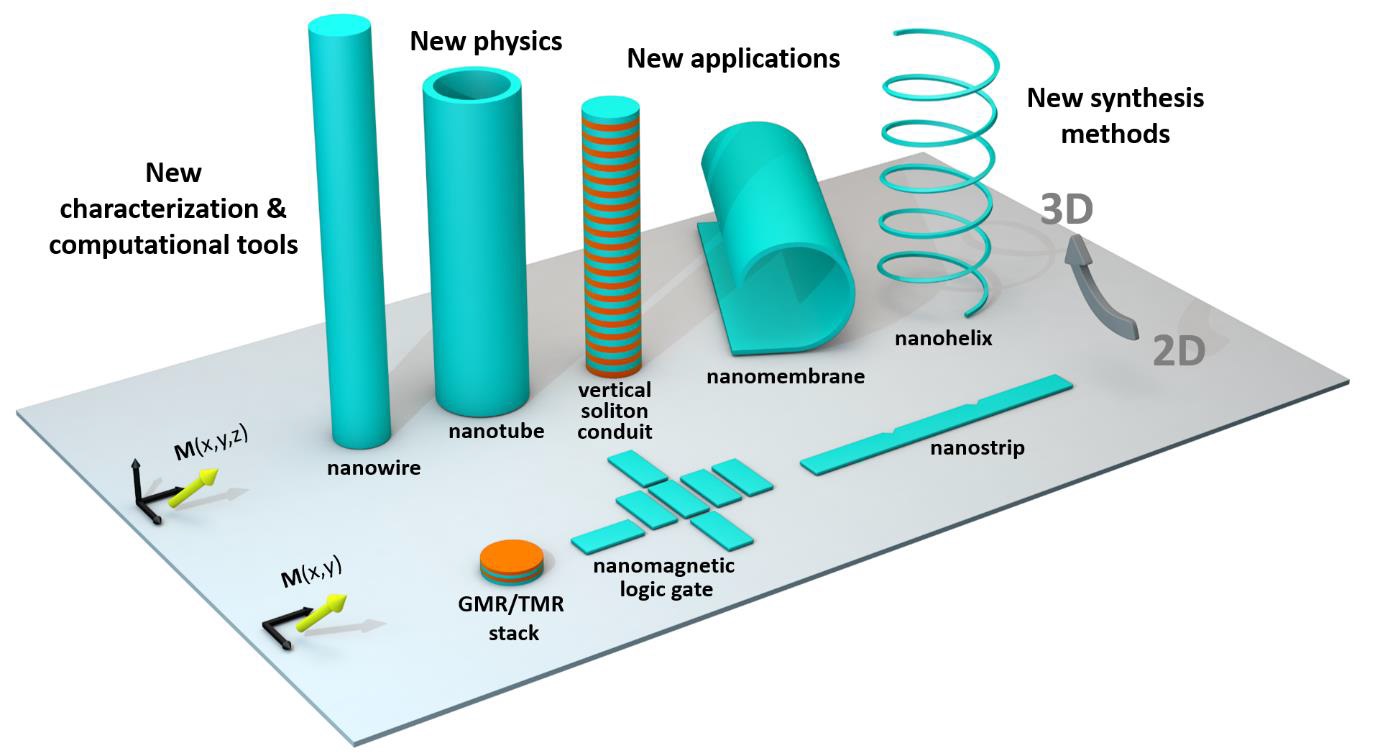}
		\caption[3D nanomagnetism]{\textbf{Towards 3D curved magnetic nanostructures}. Reprinted from~\cite{Sanz-Hernandez2018}} 
		\label{img_3D_nanomagnetism}
\end{figure}

Many of these reviews are valuable, however they are sometimes difficult to connect as written independently by different authors. Some aspects of magnetic nanowires have been covered in great detail, while others have not been reviewed. The purpose of the present chapter is to go less in depth, however provide a consistent overview. We often adopt a descriptive or handwaving approach, to allow the non-expert to get to the main point. The reader interested in a deeper insight may follow the numerous references provided. This review comes at a turning point in the field, when investigations are clearly shifting from measurements of larger assemblies, to investigations of single objects. Wire and tubes are also ideal objects to investigate the rising interest of the impact of curvature in magnetism, and more generally the emergence of three-dimensional structures\bracketfigref{img_3D_nanomagnetism}. Therefore, although several decades of investigations can now be reported, there are exciting prospects for progress and new discoveries in this field.

Still, here are a few technical remarks before starting. We mainly focus on metallic nanostructures, and only briefly mention microwires and structures from other materials (polymers). For examples of magnetic polymeric (semiconductor) nanowires consult~\citep{Ren2012}. Further, we do not cover magnetic properties of carbon nanotubes etc., for these see book \textit{Magnetism in Carbon Nanostructures}~\citep{Magnetism_in_Carbon_Nanostructures}. We will use the vocabulary \textsl{wire} to designate long one-dimensional structures with a disk (or possibly square) cross-section, opposite to the case of flat structures as fabricated by thin film deposition and patterning, which we will name \textsl{strips}. Note that in the literature some use the word \textsl{wire} also for these flat patterns. We use S.I units, except for those figures reprinted from the literature and using cgs-Gauss units. Finally, as list of notations and acronyms used is available in appendix~(\secref{sec-appendicesSymbols}, \secref{sec-appendicesAcronyms}).

\section{Fabrication} 
\label{sec_fabrication}

Experimental techniques for fabrication of magnetic nanostructures (here NWs and NTs) can be divided up according to several parameters. The first distinction is whether one uses a top-down, or bottom-up approach. The former approach requires to define the pattern to be designed. The latter approach, as the name suggests, starts with tiny building blocks (e.g. atoms, molecules) and let these assemble to form a nanostructure spontaneously.

Out of the top-down techniques, lithography is the most common. It starts with large structures and employs shrinking of these via etching, cutting, removing their parts until nanostructures remain. Lithography is typically used for the preparation of patterned thin film elements such as strips.
 While it is possible to use it for the preparation of magnetic NWs and NTs, such approach is usually not viable. Due to resolution and deep etching or coating constraints, only short structures (\eg rings instead of tubes) with somewhat large diameters have been prepared\citep{Huang2012}.  Direct writing assisted by a focused electron beam allows more flexibility, especially for vertical structures, however its throughput is low. Direct writing of nanostructures can be also in principle achieved by metalized tip of atomic force microscope~(AFM) immersed in a suitable electrolyte (precursor).

Commonly, bottom-up techniques -- mostly chemical depositions -- are preferred as they also enable synthesis of more complex structures (core-shell, non-straight vertical wires, changes of chemical composition along structure etc.), and most important, with very high vertical aspect ratio. Further, these are cost-effective, enable reasonable geometry and material (composition) control and can deliver large amounts of nanostructures during one deposition. 

In the case of the bottom-up approach, another distinction can be done according to whether or not a template giving the final structures the proper shape is used. The first ones, template-based methods, are usually preferred, still some template-less alternatives with sufficient control over the geometry exist\citep{Scott2016, Jia2005}.
 Below we will briefly discuss the main techniques used for the deposition of NWs and NTs, most of these are bottom-up techniques that employ a template. Yet certain techniques can be used both with and without template (but with different parameters). Some techniques can yield both wires and tubes alike or even core-shell structures, others can be more suitable for one nanostructure type. Before listing and discussing these techniques, we first mention the main templates of interest. Further information on the fabrication of NWs and NTs may be found in reviews from~\citet{Cao2008,Vogel2011,Ye2012,bib-VAZ2015,Tiginyanu2016}. 

\subsection{Templates}
\label{sec-fabricationTemplates}

\begin{figure}[htbp]
\centering
    \subfigure[][porous alumina]{\includegraphics[height=4.1cm]{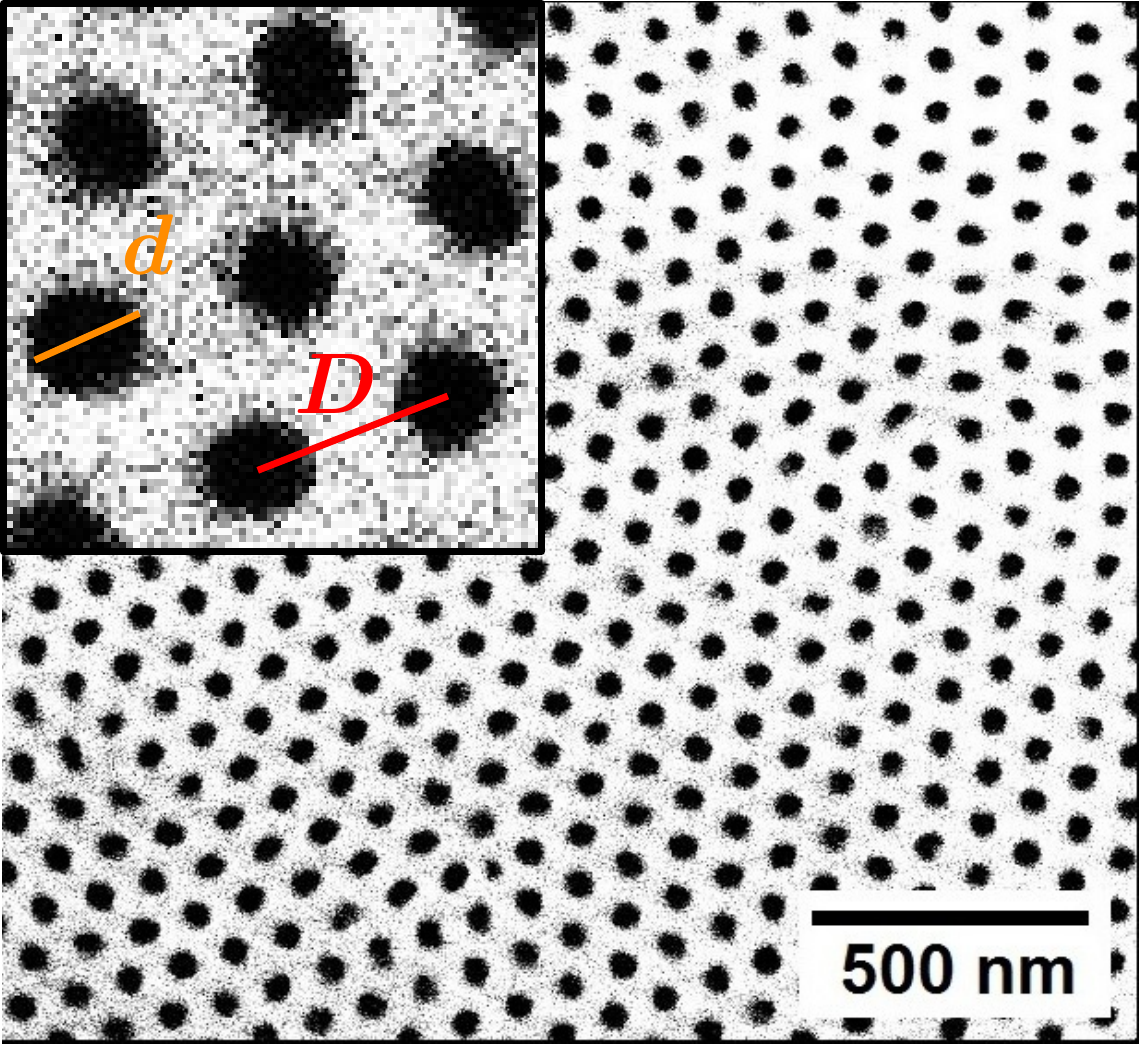}}
		\subfigure[][porous polycarbonate]{\includegraphics[height=4.1cm]{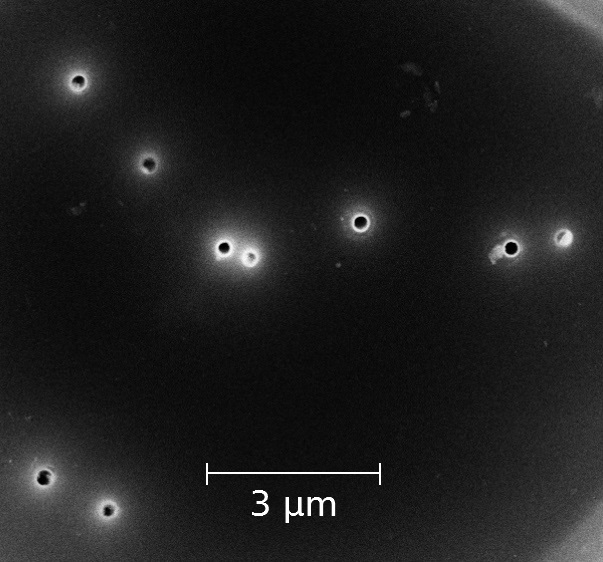}}
		\subfigure[][vertical wires]{\includegraphics[height=4.1cm]{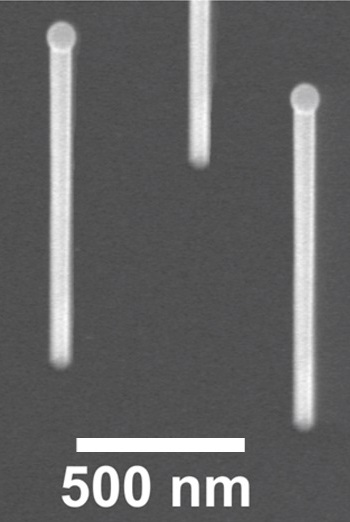}}
		\caption[Templates for fabrication]{\textbf{Scanning electron microscopy images of templates for nanowire/nanotube deposition}. (a) Nanoporous alumina (top view, empty pores); the inset shows a magnified image with a highlighted pore diameter $d$ and pitch (pore spacing) $D$. (b) Porous polycarbonate membrane (already filled with metallic tubes; otherwise the contrast is rather poor). Adapted with permission from~\citet{MS_thesis}. (c) Vertical GaAs NWs (with catalyst particle on top) for deposition of NTs and core-shell structures. Adapted with permission from \citet{Nunez2018}. Copyright 2018 American Chemical Society.}
		\label{img_templates_SEM}
\end{figure}

Various structures can be used as a template \bracketfigref{img_templates_SEM}. These include mainly porous membranes (with holes; deposition inside) and assemblies of pillars/wires (deposition on the outside surface, \ie, suitable for tube fabrication). Note that biological microtubuli\citep{Mertig1998} or other tubes (e.g. carbon NTs) enable coating of both outer and inner surface. There are other suitable bio-templates such as viruses [Tobacco mosaic virus\citep{Khan2012-TMV}]. In most cases porous membranes are preferred to pillars owing to easier manipulation, processing and measurement, of large amounts of nanostructures.

\subsubsection{Porous templates}
\label{sec-fabricationTemplatesPorous}

These include porous alumina~\citep{Masuda2005, bib-SOU2014, Losic2015-nanoporous_alumina}, mica, and ion track-etched polymeric templates~\citep{Apel2011} -- PolyCarbonate (PC), PolyEthylene Terephthalate (PET) or even Kapton (chemically resistant -- difficult to dissolve). 
Aside from track-etching, one can use also self-assembled diblock copolymer templates~\citep{Thurn-Albrecht2000}. Further, \citet{bib-WIL2017b} demonstrated that two-photon lithography using a positive photoresist can yield a complex 3D structure of pores.
 The most commonly used templates are based on nanoporous polycarbonate and alumina. Both can be either purchased, or fabricated in a laboratory.


Porous alumina \bracketsubfigref{img_templates_SEM}{a} is based on anodic oxidation of a high-purity Al sheet in acidic environment. Under proper conditions, self-organized ordered arrays of cylindrical pores perpendicular to the free surface can be obtained. As the order is poor at the surface and improves  progressively during anodization, a key development was the two-step anodization demonstrated by \citet{bib-MAS1995}: the alumina layer formed during a first anodization is dissolved, leaving an ordered corrugation at the surface of the remaining aluminum. The second anodization step happens to follow this corrugation, and yields an ordered array from the start. Note that the order, originating from stress building in the matrix, has only a middle range, over at most a few tens of~$D$. However, it it possible to use lithography as the first step, to get a long-range from a single anodization step. This has been demonstrated with interference lithography\citep{bib-JI2006} and nanoimprint lithography\citep{bib-LEE2006b,bib-WAN2008b}. Pore diameter $d$ (from a few nm to hundreds of nm) and pore spacing $D$ (pitch) can be tuned by changing the fabrication parameters\citep[anodizing voltage, type of acidic electrolyte]{bib-SOU2014}. The pore diameter and pitch ($d$, $D$) define porosity $p$, i.e. the fraction of top template surface occupied by pores (precise for template with well hexagonally ordered pores, good estimate in other cases):

\begin{equation}
p = \frac{S_{\rm pores}}{S_{\rm total}} = \frac{\pi}{2\sqrt{3}} \frac{d^2}{D^2}
\label{eq_alumina_porosity}
\end{equation}

The as-prepared ordered nanoporous alumina has about 10\% porosity\citep{Nielsch2002-porosity_rule}, but it can be adjusted by post-processing: chemical etching, increasing the pore diameter, or wall coating by, \eg, atomic layer deposition, decreasing the pore diameter. Note that after the template is filled with magnetic wires, the term \textsl{packing factor} or \textsl{packing density} is used instead of the porosity, as the filling material is now the one of interest. Note that in practice the theoretical packing density may not reflect the actual filling, as some pores could remain unfilled (e.g. due to defects, clogging of pores). In the case of tubes, one may take into account the hollow core for defining the packing density. As the pore diameter~$d$ is essentially proportional to the anodizing voltage, templates with diameter modulations (protrusions, constrictions) can be prepared, as will be further discussed in \secref{sec-fabricationEngineeredDiameterModulated}. Finally, the pore length, defining the template thickness, is set at will by the anodizing time. It ranges from typically a from hundreds of nm to tens of micrometers.

  An advantage of alumina is its suitability for the processing and characterization of arrays of magnetic structures deposited in the pores: annealing, low temperature magnetometry,\,\ldots). A disadvantage is its chemical resistivity: strong acids or bases need to be employed in order to dissolve the template and liberate nanostructures for investigation of single objects. Thus, unless magnetic structures are protected with additional layer, the chemical may also attack the magnetic material and oxidize it\citep{Apel2011}. This issue is more sever in case of magnetic tubes.


Porous polycarbonate [the process is similar for other polymers, \citet{Apel2011}] is prepared by irradiation of foils of the material with high energy ions, followed by the selective etching of the ion-tracked parts in a strong basis such as NaOH or KOH. The pore size depends on the nature of the ions and their energy, the etchant and its concentration, and the etching time. The pore density is given by the ion fluence. Typically, the resulting pore distribution is random with lower porosity compared to alumina membranes\bracketsubfigref{img_templates_SEM}{b}. Unless particles from a well-defined and collimated beam are employed (i.e. in large scale ion/particle accelerators),  the pores show a distribution of direction around the normal to the foils. Thus, even for moderate porosity and ever more pronounced for higher porosity, pores may intersect each other. On the other hand, one can exploit this to fabricate cross-linked arrays of structures. However,  for better control it is desirable to perform subsequent ion irradiation under different well-defined angles. This can lead to cross-linked orthogonal (or just tilted) pores and later also magnetic structures\citep{Gomes2016,Muench2015-interconnected,Araujo2015}. An advantage of polycarbonate membranes is that they can be very easily and rapidly dissolved in, \eg, dichloro\-methane (within seconds). Such solvent does not oxidize the fabricated magnetic nanostructures\citep{MS_thesis}. A disadvantage is that some membranes, especially commercial ones, may have a quite rough surface that translates into poor surface quality of the deposited nanostructures.

\subsubsection{Elongated nanostructures as templates}
\label{sec-fabricationTemplatesElongated}

Existing nanostructures in form of rods, nanowires, pillars (magnetic or non-magnetic) can be utilized for the deposition of tubes, while existing tubular structures can serve both for tube and wire fabrication (both sides of such template can be coated).

Such templates usually consist of \textbf{vertical rods} (arrays), grown on a suitable substrate. These can be prepared by electroplating in a porous template (with lower porosity) and dissolving the template. Another common technique is vapour-liquid-solid~(VLS) growth of semiconducting vertical wires from a catalyst, \eg, GaAs wires on Si substrate \bracketsubfigref{img_templates_SEM}{c} from a Ge droplet catalyst\citep{Ruffer2014,Nunez2018}. Related to their crystal structure and orientation, these can have a hexagonal or other non-circular cross-section.  Vertical pillars can be coated by both chemical~[\eg, electroless plating, see \citet{Cheng2007}] and physical (vapour) depositions (e.g. evaporation) under vacuum\citep{Baumgaertl2016}. An advantage of pillar templates is that one may benefit from a cleaner surface, as there are no impurities such as in case of isolated structures obtained after dissolving a porous template. This is true, \eg, for VLS-grown pillars and physical deposition performed in a row under ultra-high vacuum. Besides, one can do in principle both fabrication and investigation in-situ, inside a vacuum chamber, and avoid oxidation that is problematic especially in case of tubes with thin shells ($<\SI{10}{\nano\meter}$). Vertical pillars on a substrate are also a better starting point for micromanipulation of individual structures. A drawback is that they cannot be handled mechanically from the top side without protection, due to the fragility of the protruding wires.

An alternative method is to perform \textbf{synthesis in solution}, where suitable template structures (wires, tubes including bio-templates) are dispersed in a liquid solution and coated upon addition of suitable reagent. Typical coating method is electroless plating\citep{Mertig1998,Khan2012-TMV}.

\subsection{Common deposition techniques}
\label{sec-fabricationDeposition}

\begin{figure}[htbp]
\centering
    \includegraphics[width=0.8\linewidth]{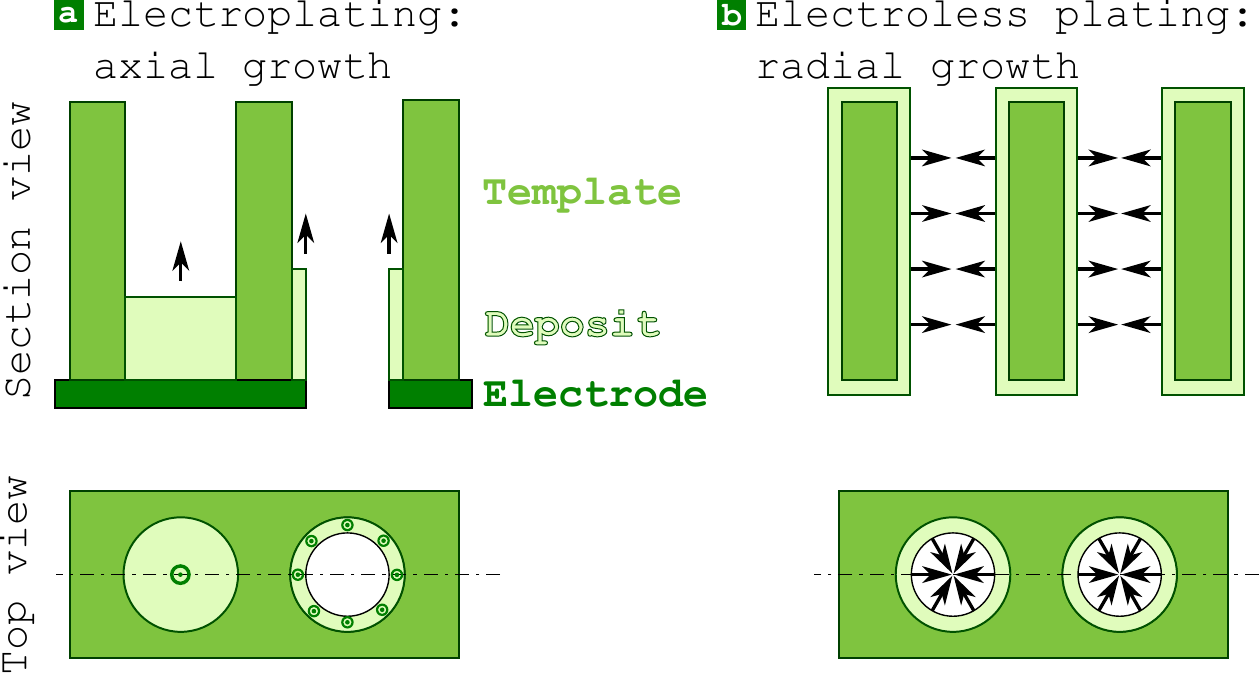}

		\caption[]{\textbf{Nanowire/nanotube growth in nanoporous templates}. (a) Axial growth in pores -- typical for electroplating where structures grow from the bottom electrode. (b) Radial growth in pores -- common for surface coating techniques such as electroless plating or atomic layer deposition. Dark blue arrows depict the growth direction inside the pores.}
		\label{img_Axial_vs_radial_deposition_in_pores}
\end{figure}

\subsubsection{Electroplating}
\label{sec-fabricationDepositionElectroplating}

Electroplating\citep{Modern_electroplating} relies on the reduction of metallic ions from an electrolyte, typically an aqueous liquid, controlled by an external source of electrons through a power supply. A common plating cell consists of at least two electrodes. Deposition takes place at the cathode (negatively-biased electrode, working electrode), while a complementary redox oxidation reaction occurs at the other electrode, called the counter electrode. For better control over the deposition, the electric potential of the cathode is often measured with respect to a third, reference electrode. This enables a better-defined shift of electrochemical potential, which can be exploited for deposition of different materials from a solution containing multiple species (\eg, Co and Cu for multi-segmented wires).

Electrodeposition is commonly done with a DC current controlled in either galvanostatic mode (fixed current), or potentiostatic mode (fixed potential). The constant-potential mode gives control over crystallography and mainly composition in alloys and segmented structures. The constant-current mode enables an easier tuning of the length of nanostructures. Indeed the growth rate is better controlled, as the amount of material is directly proportional to the passed electric charge. Also, the galvanostatic deposition setup is simpler, requiring only two electrodes. Aside from DC current, AC or pulsed (-reversed) deposition\citep{Chandrasekar2008} can be employed. It may provide a more homogeneous composition in case of alloys or other deposits with co-deposition of several elements\citep{Salem2012}, as well as close to $\SI{100}{\%}$ filling of the pores\citep{Nielsch2000}. In some instances, monocrystalline deposits can be obtained\citep{Yasui2003,bib-IVA2013,Li2008}.

Aside from pure metals, alloys and their multilayers, one can also deposit some semiconductors; electrical insulators can be prepared only upon post-processing, otherwise the deposition is self-limiting, as a conductive surface is needed to sustain the growth. Crystallography, texture and magnetic properties can be influenced by pH of the plating solution as well as changing the deposition electric potential\citep{Cortes2009}, application of a magnetic field during the growth\citep{Ge2001} or growing structures on a suitable substrate\citep{Yasui2003}.

Reference to some examples of electroplated wires and tubes are given in Tab.~\ref{tab_electroplated_nanostructures-examples}. Other examples can be found in work by \citet{Stkepniowski2014}.


{\renewcommand{\arraystretch}{1.6}
\begin{table}[htbp]
\setlength{\tabcolsep}{3pt}
\centering
\footnotesize
\caption{Examples of nanowire and nanotube depositions via electroplating.}
\begin{tabular}{lll}
\hline\hline
\textbf{} & \textbf{Nanowires} & \textbf{Nanotubes} \\ \hline
Fe & \citep{Zhang2003} & \citep{Cao2006} \\
Co & \citep{Thurn-Albrecht2000,bib-WAN2008a} & \citep{Li2008, Zhang2013-nanotubes} \\  
Ni & \citep{bib-PIT2011,Xu2008-Ni_wires} & \citep{Zhang2013-nanotubes} \\
NiCo & \citep{bib-SAM2018} & \citep{Zhang2013-nanotubes} \\
NiFe & \citep{Salem2012} & \citep{Zhang2013-NiFe} \\
CoFe & \citep{Chen2002,Ozkale2015} & \citep{Kozlovskiy2016} \\
CoNiFe & \citep{Atalay2010} & \citep{Atalay2010} \\
CoPt & \citep{Yasui2003,bib-DAH2006}  & \citep{Rozman2014} \\ \hline\hline 
\end{tabular}
\label{tab_electroplated_nanostructures-examples}
\end{table}


For deposition of wires or tubes, one uses typically nanoporous templates (alumina, polycarbonate -- electrical insulators) with a metallic layer (e.g. Au) covering the pores on one side and thus serving as the electrode where deposition is initiated [structures grow from the pore bottom, \bracketsubfigref{img_Axial_vs_radial_deposition_in_pores}{a}]. The electrode should be exposed to the plating solution only through the pores, to avoid unwanted depositions and reduced yield of nanostructures. The outer diameter of deposited nanostructures is set by the template (pore diameter) and the length is typically proportional to the deposition time (and limited by the template thickness -- pore length). Nanowires with sub-10nm diameters can be prepared as demonstrated by 5-nm-diameter nanowires from Fe\citep{Zhang2003} and Ni, Co, and Fe by \citet{bib-ZEN2002}.

In order to get hollow nanotubes instead of solid wires inside the pores, different conditions and deposition parameters should be used. Strategies to obtain tubes include starting from a porous working electrode\citep{Atalay2010,Proenca2012}, modified template pore walls\citep{Bao2001} or other particular conditions (pH, current density, over-potential). 
Tube wall (shell) thickness can be to some extent tuned as well. However, such control is in general poor compared to atomic layer deposition and electroless plating discussed below. Electroplating can yield tubes with diameters as small as 25\,nm\citep{Wang2011} and good material quality. However, wire-vs-tube growth instabilities occur\citep{Fukunaka2006} as small changes in conditions (pH, concentration, current density,\,\ldots) could be sufficient to favour growth of a solid wire, in particular in the case of smaller diameters. This can be overcome by employing a template with tubular nanoholes prepared, \eg, by deposition and controlled shrinking of polymeric NWs inside porous alumina as shown by \citet{Li2012}. However, the aspect ratio of tubes prepared this way is limited.

\subsubsection{Electroless plating}
\label{sec-fabricationDepositionElectroless}

Similar to electroplating, electroless plating, also referred to as an autocatalytic deposition, relies on the reduction of metallic ions from a liquid electrolyte\citep{Zhang2015-electroless,Modern_electroplating,ShachamDiamand2015}. Unlike electroplating, no external current source is needed as electrons for the reduction are provided by a chemical substance, so called reducing agent, added into the solution. Thus, rather simple beaker chemistry is sufficient for the deposition. Further, almost any surface can be coated, even though some (e.g. non-conductive ones) may have to be chemically modified -- using so called sensitization and activation procedures, resulting in coverage of the surface with suitable catalyst particles (Pt, Pd, Ag,\,\ldots). The plating is conformal like in the case of atomic layer deposition, and high-aspect ratio structures can be covered with the deposit [also deep pores where the growth proceeds in radial direction, \bracketsubfigref{img_Axial_vs_radial_deposition_in_pores}{b}]. However, the thickness control is not as precise as in the case of atomic layer deposition\bracketsecref{sec-fabricationDepositionALD}. A large variety of materials can be deposited: metals, alloys, metalloids, oxides\ldots

The choice of the reducing agent depends on the material to be plated, as well as on the chemical resistance of the template/substrate. Many reducing agents contain boron (\eg, dimethylamino borane) or phosphorous (sodium hypophosphite). From few up to tens of percent of these elements may be incorporated in the deposit. Material properties (such as grain size, electrical conductivity, mechanical hardness) can be influenced by changing their content. The amount of boron or phosphorous depends on the deposition process, mainly on the pH and reducing agent concentration. Almost pure metals can be obtained using formaldehyde or more often hydrazine\citep{Muench2013}.

Electroless deposition has been employed for magnetic (nano)tube fabrication from various materials such as CoNiB\citep{Schaefer2016}, NiFeB\citep{bib-FRU2017e,bib-RIC2015c}, NiB\citep{Richardson2015process}, Co and Ni\citep{Wang2006}. The above-mentioned works employed porous templates, but one can also coat arrays of pillars\citep{Cheng2007} or nanorods in a solution\citep{Rohan2008}. The technique provides good control over the tube thickness, roughly  proportional to the plating time\citep{bib-RIC2015d}. Diameters down to $\SI{100}{\nano\meter}$ have been obtained using porous templates\citep{Li2014}, and $\SI{50}{\nano\meter}$ in case of biotemplates\citep{Mertig1998}. The grain size can be decreased upon increasing the boron or phosphorous content, from of few tens of nanometers for low content, to amorphization for large content~(\eg, 10\%+)\citep[p.~122]{Watanabe2004}. Electroless plating with reducing agents free of boron/phosphorous (such as formaldehyde or hydrazine) yields large grains and in some cases also rough or even spiky surfaces\citep{Muench2013}.


\subsubsection{Atomic layer deposition}
\label{sec-fabricationDepositionALD}

Atomic layer deposition (ALD)\citep{George2010} is a special regime of chemical vapour deposition, which is self-limiting and confined to the surface. Deposition proceeds in cycles, with the sample sequentially exposed to a precursor gas, often a organometallic compound, and then to a second reactant, such as water vapour, oxygen, hydrogen\,\ldots The reaction chamber is purged by an inert gas such as nitrogen or argon in-between expositions, to remove the excess precursor. This way the reaction is self-limiting, both precursors react only at the surface of the sample, and sufficient time can be provided for diffusion. Therefore, ALD provides a conformal coating of high aspect ratio pores without clogging their inlet\citep{bib-DAU2007}, and similarly rods\citep{Ruffer2014}. Thus it is very suitable for nanotube fabrication \bracketsubfigref{img_Axial_vs_radial_deposition_in_pores}{b}. The technique gives excellent control over the NT shell thickness~(also called tube wall thickness), which is simply proportional to the number of ALD cycles. The typical deposition rate is $\SI{1}{\AA/\mathrm{cycle}}$  or even lower. ALD commonly yields oxides, \eg, Fe$_2$O$_3$. So far only, only oxides (NiO, CoO$_x$) that were reduced after the deposition or during ALD with an extra third precursor\citep{bib-DAU2007,Ruffer2014,Pereira2016}, have been used for the preparation of magnetic nanotubes. In case of Fe oxides, Fe$_2$O$_3$ is either studied directly, or it can be reduced to Fe$_3$O$_4$\citep{Bachmann2009,Albrecht2011}. Notice, however, that these reduced materials are quite granular from the structural as well as magnetic point of view, as revealed by X-ray magnetic imaging\citep{bib-KIM2011b}.

Aside from oxides, one can deposit a very large variety of different materials by ALD\citep{Miikkulainen2013}. This includes rather pure metals such as noble metals\citep{Elliott2010}, and transition metals such as Fe, Co, Ni, and Cu when using molecular hydrogen as the second reactant instead of water/oxygen\citep{Lim2003}. Ferromagnets prepared in this way could exhibit better magnetic properties than reduced oxides.

As will be mentioned below, thanks to its sequential and conformal features, ALD is very suitable for the fabrication of multilayered and core-shell structures.


\subsubsection{Focused electron beam induced deposition}
\label{sec-fabricationDepositionFEBID}


Focused electron beam induced deposition [FEBID, \bracketfigref{img_Utke2008-FEBID}]\citep{Utke2008,Huth2012} allows the direct writing of nanostructures with the electron beam of a scanning electron microscope equipped with a gas injection system that doses a gaseous precursor.

The focused electron beam dissociate the precursor adsorbed on a substrate. Thus, metal is deposited at the beam location and to a lesser extent also around, with a spatial resolution of a few tens of nanometers. Carbon is also embedded in the deposit, coming from the decomposition of the metallorganic precursors. Without further purification, such as annealing in hydrogen or laser-assisted deposition\citep{Lewis2017}, the metallic content can be lower than 50\%. With purification it can reach purity above $\SI{95}{\%}$.

\begin{figure}[htbp]
\centering
    \includegraphics[height=5cm]{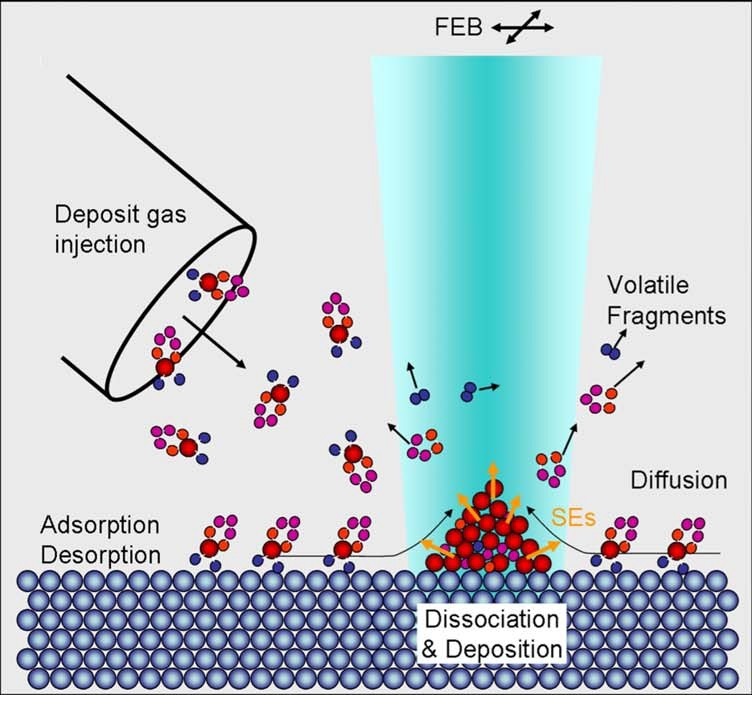}
		\caption[]{\textbf{Scheme of focused electron beam (FEB) induced deposition}. SEs stands for secondary electrons that are mainly responsible for dissociation of a precursor and the deposition. Reprinted with permission from~\citet{Utke2008}. Copyright 2008, American Vacuum Society.}
		\label{img_Utke2008-FEBID} 
\end{figure}

FEBID enables the deposition of Co\citep{Lau2002} and Fe. Ni-containing deposit can be prepared as well, but so far the quality and composition is poorer compared to Fe and Co deposits, as pointed out in a review on FEBID-grown magnetic nanostructures by \citet{De_Teresa2016-review}. One can also prepare a non-magnetic metallic deposit such as Ag\citep{Hoflich2017}, Au, Cu\citep{Esfandiarpour2017}, Pt~\citep{Lewis2017}. These can serve either as a protection, electrical contacts, or for injection of spin current through the spin-Hall effect in case of Pt.

Magnetic deposits are typically magnetically soft with small crystallites embedded in a carbon-rich matrix. One can prepare wires lying on a surface, vertical nanowires\citep{Cordoba2016} and core-shell structures, \eg,  Co wires covered with Pt\citep{Pablo-Navarro2016}. Further, one can to some extent adjust the wire diameter during the growth\citep{Pablo-Navarro2017} and create bends\citep{bib-FRU2018} or even helices\citep{Fernandez-Pacheco2013}. Even with non-negligible carbon content, the materials (especially Co deposit) is good enough to provide rather uniform magnetization in domains\citep{Wartelle2018,Pablo-Navarro2018} and also domain walls can be nucleated and trapped in wires with bends (hook-shaped) as demonstrated by \citet{Wartelle2018}.


\subsubsection{Electrospinning}
\label{sec-fabricationDepositionElectrospinning}


Electrospinning is a technique commonly used for the fabrication of polymeric nanofibres, with basis a sol-gel reaction. A precursor solution or sol-gel is being fed through a metallic nozzle with a small aperture, high voltage of several kV is applied between the nozzle and a collector electrode (typically rotating/spinning drum/cylinder). High voltage in this geometry leads to the formation of so-called Taylor cone. With the help of (strong) electric field a fibre is formed from the precursor, and collected on the electrode.

Aside from polymeric fibres with magnetic particles, one can also obtain metallic and oxide fibres via the combination with sol-gel chemistry and postprocessing\citep[thermal annealing, calcination]{Khalil2013}. Electrospinning enables almost industrial fabrication of very long free-standing wires and their networks/clusters, as well as preparation of shorter structures\citep[\figref{img_Sakar_BFO_electrospinning}]{Sakar2016}. In certain cases, one can also get hollow fibres\citep[nanotubes]{Eid2010}, core-shell and nested structures\citep{Mou2010}.

\begin{figure}[htbp]
\centering
    \includegraphics[height=6cm]{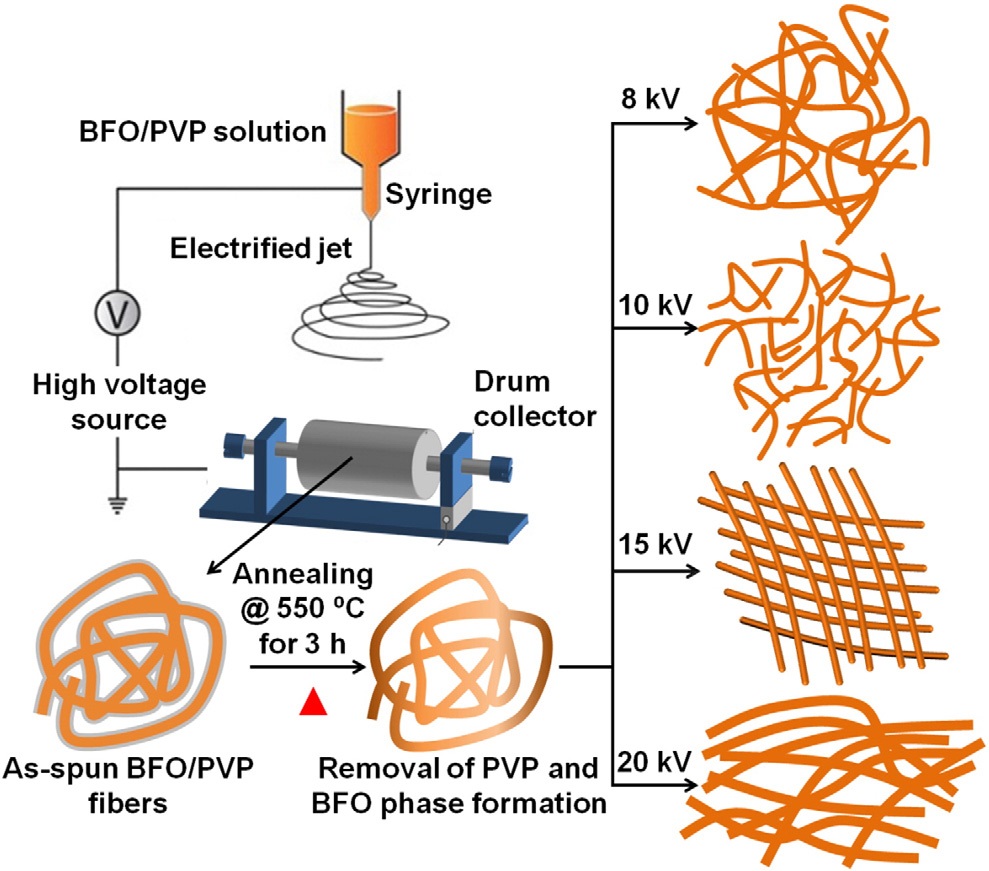}

		\caption[]{\textbf{Applied voltage-dependent electrospinning formation of 1D nanostructures of BiFeO$_3$}. Reprinted from~\citep{Sakar2016}, Copyright 2016, with permission from Elsevier.}
		\label{img_Sakar_BFO_electrospinning}
\end{figure}

A non-exhaustive list of examples of prepared magnetic nanowires from various materials follows: Fe, Co, and Ni (diameter around 25\,nm)\citep{Wu2007}, Yttrium iron garnet \citep[characterized with ferromagnetic resonance]{Jalalian2011}, CoNi\citep{Barakat2010}, FePt\citep{Zhang2015}, and BiFeO$_3$\citep{Sakar2016}.







\subsubsection{Other techniques}
\label{sec-fabricationDepositionOther}

Magnetic NTs and NWs can be prepared via sol-gel deposition~\citep{Bahuguna2016}, employing templates. Sol-gel synthesis is concerned with preparation of colloidal solution with nm-sized particles (sol) and its conversion to gel and finally to a solid material. Unlike a simple preparation of aqueous solution, preparation of the sol often requires several steps with mixing different chemicals. Often, an alcaline salt of a metal to be deposited is mixed with an alcohol and additives in order to obtain alkoxides that serve as a precursor for the gel. Precipitation of colloidal particles and later solvent removal for gelation is often assisted by heating. The deposition technique yields mostly oxides [e.g. CoFe$_2$O$_4$~\citep{Ji2003}] or other complex compounds with some of them displaying multiferroic properties [BiFeO$_3$~\citep{Javed2015}] and/or antiferromagnetic ordering [FeTiO$_3$~\citep{Khan2016}]. Post-processing such as annealing and reduction in an hydrogen atmosphere, provides metallic nanostructures, such as Fe~\citep{Xu2008}, Ni or CoFe~\citep{Hua2006}. Sol-gel is also being employed in the electrospinning method.


\paragraph{Rolling thin sheets} Tube-like structures can be obtained via letting strained bilayered thin sheets roll by themselves. The strain determines the radius of curvature, while the length and number of multilayers in the roll are defined through lithography\citep{Mendach2008,Streubel2015}. Such structures have rather micrometric diameters. The advantage is the readily integration at precise locations on a surface. A disadvantage is their limited range of radius and the fact that they form "Swiss-rolls" rather than perfectly connected tubes.

\paragraph{Spinodal decomposition} 
Phase separation of materials forming different \linebreak phases, sometimes immiscible, is a standard route for producing materials. It can be applied to form wires, where the direction order is provided through a thin-film growth process. \citet{Mohaddes2004} prepared arrays of single crystal Fe NWs inside an oxide matrix by thermal decomposition and phase separation in a source perovskite thin-film. Magnetic columns~(nanowires) with diameters below 5\,nm have been prepared by phase separation in epitaxially grown matrices, \eg, Co in CeO$_2$\citep{Schio2010}, or Mn-rich GeMn in Mn-poor matrix of GeMn\citep{bib-JAM2006}. Bifunctional materials such as artificial multiferroics can also be prepared\citep{bib-MOH2004}.

\paragraph{Chemical vapour deposition} CVD allows the growth of vertical wires at a surface, often through the VLS process~(Vapour-Liquid-Solid) and a catalyst forming a eutectic with the growing material. Ferromagnetic Fe$_{1.3}$Ge NWs were grown on sapphire and epitaxially on graphite (few-layer graphene or highly-oriented pyrolitic graphite) using chemical vapour transport\citep{Yoon2011}. The same technique was employed for the preparation of magnetic Si-based NWs containing ferromagnetic metal and/or Mn, such as Fe$_{1-x}$Mn$_x$Si\citep{Hung2012-FeMnSi_NWs}. Aside from semiconductor-based nanowires, one can also prepare single-crystalline Ni NWs\citep{Chan2012-Ni_CVD_NWs}, Co\citep{Kim2015-CVD_Co_NWs}, and alloys NiCo, CoFe, and NiFe; aside from Ni and Co\citep{Scott2016}.

\paragraph{Microwires -- glass-coated melt spinning} 
Another worth-mentioning method is glass-coated melt spinning, or so-called modified Taylor-Ulitovsky method, see \citet{Larin2002}. It has used for the fabrication of rapidly solidified amorphous microwires with glass-cladding for decades. The quenching provides two aspects. First, amorphous materials of complex composition can be obtained, suitable to achieve very soft magnetic properties. Second, radial strain may exist, and can be used to control the distribution of magnetization in the wires, either axial, radial or core-shell. Recently, metallic cores (nanowires) with diameters down to $\SI{100}{\nano\meter}$ have been prepared. However, the total diameter including the cladding is still micrometric\citep{bib-OVI2014}. Further information on the microwires can be sought, \eg, in book \textit{Magnetic nano- and microwires}\citep[chap. 7]{bib-VAZ2015}.



\subsection{Engineered structures}
\label{sec-fabricationEngineered}

Some phenomena and applications require not only straight NWs and NTs, but rather more complex or modified structures with additional functionality and/or alternation of magnetic properties along the structures. These can be achieved by:

\begin{itemize}
	\item change of geometry: diameter modulation, wire-tube segments, notches, bends 
	\item change of material: segments with different chemical composition, varying microstructure through a local treatment (doping, irradiation with ions, laser,\,\ldots)
	\item core-shell structures, multilayered tubes
	\end{itemize}
\noindent Examples of some experimentally-prepared engineered cylindrical nanostructures are given in Fig.~\ref{img_engineered_structures}. Geometry and material change can be used for modulating the energy profile for domain walls, \eg, the creation of artificial pinning sites, definition of bits for the race-track memory. Periodic change of magnetic properties can be exploited for the preparation of magnonic crystals, used for the manipulation of spin-waves, in analogy with photonics crystal. Further, one could obtain complex structures integrating magnetic and optically active parts; multi-segmented/multi-layered structures can serve as sensors utilizing magnetoresistence effects, \ie, the combination of magnetic and non-magnetic parts. Core-shell structures, multilayered tubes and their vertical arrays can be used as 3D curved analogues of planar (2D) multilayers that form the basis of current spintronics. Therefore, one can obtain spin valves and interfaces with heavy metals such as platinum for spin-Hall-effect-assisted~(SHE) manipulation of magnetization in the adjacent magnetic tube/nanowire.

\begin{figure}[htbp]
\centering

\begin{minipage}{0.75\linewidth}
\centering
\subfigure[][]{\includegraphics[height=1.5cm]{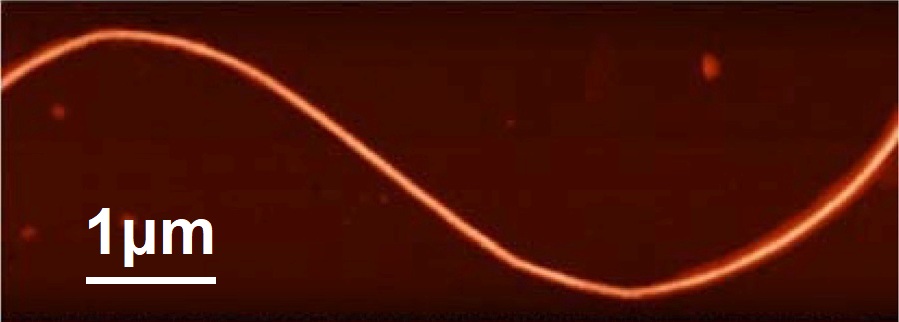}}
	\subfigure[][]{\includegraphics[height=1.5cm]{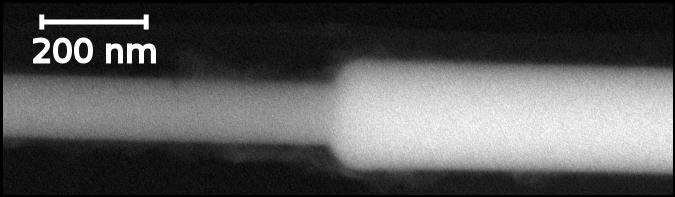}}\\
	\subfigure[][]{\includegraphics[height=1.9cm]{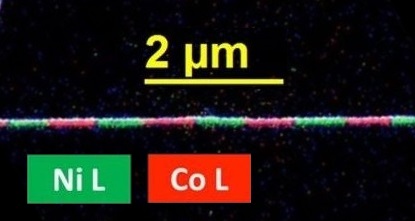}}
	\subfigure[][]{\includegraphics[height=1.9cm]{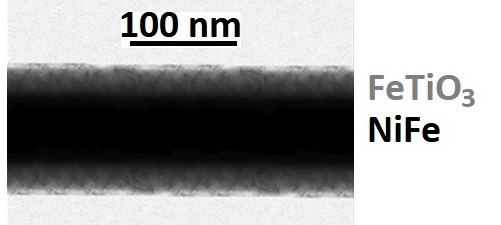}}
\end{minipage}
\begin{minipage}{0.24\linewidth}
\subfigure[][]{\includegraphics[height=4.2cm]{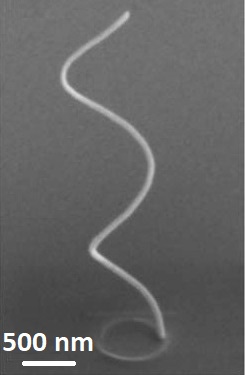}}
\end{minipage}

	\subfigure[][]{\includegraphics[width=1\linewidth]{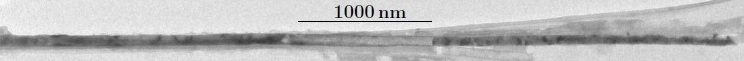}}
	
	\caption[Examples of engineered cylindrical structures]{\textbf{Examples of engineered cylindrical structures}. (a) Bent Ni cylindrical nanowire lying on a Si surface. Reprinted from~\citet{bib-FRU2016c}, with the permission of AIP Publishing. (b) NiCo diameter-modulated nanowire. Reprinted with permission from~\citep{MS_thesis}. (c) Chemical map of a nanowire with alternating Co and Ni segments. Adapted with permission from \citet{bib-IVA2016b}. Copyright 2016 American Chemical Society. (d) Permalloy-FeTiO$_3$ core-shell nanowire. Adapted from~\citep{Khan2016} with permission of The Royal Society of Chemistry. (a)-(c) prepared by electroplating in nanopores and (d) sol-gel deposition (shell) combined with electroplating (core). (e) Reprinted by permission from Springer Nature: \citet{Fernandez-Pacheco2013}, copyright 2013. (f) Transmission electron microscopy of Ni wire-tube-wire nanoelement. Adapted from \citet{MS_thesis}.} %
	\label{img_engineered_structures}
	
	
\end{figure}


\subsubsection{Nanostructures with diameter modulation}
\label{sec-fabricationEngineeredDiameterModulated}


Cylindrical nanostructures with diameter changes along their length \bracketsubfigref{img_engineered_structures}{b} have been prepared by selective etching of wires with alternating composition~\citep{Liew2011}, or with more versatility, by coating or filling of a suitable template such as nanoporous alumina with varying diameters~\citep{bib-PIT2011}. Porous alumina templates are very suitable for this purpose, as the pore diameter can be tuned via various way. A first way is to stop anodization at specific times, and enlarge~(resp. reduce) the pore diameter through etching~(resp. ALD), or combinations of these between successive steps of anodisation. For instance, ALD of silica can be used to decrease the pore diameter of a first pore segment and protect it from further chemical etching during the chemical widening of further uncapped segments\citep{bib-SAL2018b,bib-FRU2018b}. The advantage of this method is the nearly unlimited freedom and good control in terms of segment length and variation of diameter. The disadvantage is the limited number of segments that can be fabricated, both for the sake of time and compatibility of the verious steps. A second way is changing the voltage during anodization\citep{bib-LEE2010b,Losic2015-nanoporous_alumina}, as the natural diameter $d$ of the pores depends on voltage. The advantage is the easy implementation and high throughput, as a complex structure may be obtained in a continuous run of anodization. A disadvantage is that the pitch $D$ also depends on voltage, so that a prolonged change of voltage results in pore branching or closing. Thus, only pulses of voltage may be applied, resulting in a limited longitudinal length of the modulations, and an only partial control on their geometry. Here are some examples of resulting modulated magnetic structures: Iron oxide nanotubes with modulated diameters prepared by ALD in a multi-step modulated template\citep{bib-PIT2009}, nanowires prepared in similar templates by electroplating\citep{bib-IGL2015,Bran2016}. Diameter changes can be also prepared by so called coaxial nanolithography~\citep{Ozel2015} that combines pore widening, intercalated between steps for deposition of different nanowire segments.





\subsubsection{Segmented structures}
\label{sec-fabricationEngineeredSegmented}

Here we report processes to fabricate nanowires with alternating segments, sometimes called multilayered nanowires. Such structures had been extensively prepared by electroplating and explored in the nineties for investigation of the giant magnetoresistance (GMR), using the model system Co/Cu\citep{Piraux1994,Blondel1994}. The interest has been revived by further studies of spin transfer torques\citep{Huang2009}, and more recently for domain wall motion studies, for the definition of bits for a 3D race-track memory\citep{bib-BER2017,Bochmann2017}. These feature alternating hard and soft magnetic segments such as Co/Ni\bracketsubfigref{img_engineered_structures}{c}, or the combination of magnetic/nonmagnetic sections such as Co/Au for magnetic nano-oscillators\citep{bib-BRA2017}.

The method of choice for the fabrication of multisegmented structures is electroplating in nanoporous membranes. A first way is to exchange the plating solution for each segment. This is a flexible however time-consuming method. A second way is to use a solution containing all elements of the various segments and take advantage the fact that these elements have different reduction potentials. At moderate electric potential~(more positive), only the more noble metal is plated~(such as Au, Cu). For a potential applied below all deposition potentials, the less-noble ferromagnet (Co, Fe) is reduced. However, the noble metal is being deposited to some extent as well. In order to get a magnetic segment, one chooses a concentration of the more noble metal (Au, Cu) in the solution much lower than that of the less noble ferromagnetic metal. This so-called kinetic control decreases the contamination of magnetic segments by the noble metal. So, this second way is easier to implement, however, less versatile. Aside from wires, multisegmented tubes\citep{Davis2006} were prepared by electroplating as well for the GMR studies. Magnetoresistance of segmented structures is further discussed in section~\ref{sec_magnetoresistance}


\subsubsection{Wire-tube nanoelements}
\label{sec-fabricationEngineeredWireTube}

Instead of modifying the chemical composition, one can modulate magnetic properties by alternating wire~(solid) and tubular~(hollow) segments. Such wire-tube structures \bracketsubfigref{img_engineered_structures}{f} have been considered mainly in micromagnetic simulations\citep{Neumann2013,Neumann2015,Espejo2015,bib-SAL2013b,Salazar-Aravena2014}.

Experimentally, such structures have been prepared by direct electroplating in nanopores or by a related so-called coaxial lithography\citep{Ozel2015}. Most works report only one transition, where a tube segment is formed during the initial stage of electroplating\citep{Dryden2016} and its tube wall thickness gradually increases until the growth continues with a solid wire. This was shown, \eg, in case of Co-Pt nanostructures \citep[diameter 200\,nm, tube length about 1\,$\micro$m]{Arshad2014}. A theoretical kinetic model~\citep{Philippe2008} for electrodeposition of tubes, wires and even wire-tube elements, was experimentally demonstrated by deposition of Co nanostructures with multiple well-defined but very short wire-tube segments of length $\leq 30$\,nm and diameter of 60\,nm. Sharp multiple Ni wire-tube transitions with longer segments were shown in~\citep{MS_thesis}; length of tube segments was sometimes even several $\micro$m. In addition, tubular segments were present also in the centers of nanostructures \bracketsubfigref{img_engineered_structures}{f}, not only at the ends. However, the growth could not be well-controlled -- many deposited structures were just solid wires. A similar behaviour was reported by \citet{Dryden2016} for Co, but only with one tube-wire transition located at the end. As shown by both contributors, these structures are very fragile and tend to break at the wire-tube transition.

To conclude, only a few experimental works on wire-tube elements exist and controlled growth is still challenging, especially in the case of longer tubular segments~($\geq 1\micro$m, mostly much shorter segments have been realized), smaller diameters and sharp transitions. While there are several theories for the formation of wire-tube structures, the precise mechanism that could explain all experiments and allow a better control is still unclear.


\subsubsection{Core-shell structures}
\label{sec-fabricationEngineeredCoreShell}


There are two main fabrication routes for multilayered tubes and core-shell wires: coating (sputter-deposition, evaporation,~\ldots) of vertical pillars/wires and deposition inside porous templates combining different chemical methods.

The former method enabled the fabrication of structures such as core-shell nanowire spin valves\citep{Chan2010}, CoO (10\,nm)/Co (5\,nm)/Cu (5\,nm)/Co (5\,nm), deposited through sputter deposition around the chemical vapour-deposited Ni NWs. Vertical and fully 3D core-shell nanostructures can be grown also by focused electron beam-induced deposition, such as Co NWs with a Pt shell\citep{Pablo-Navarro2016}.

The latter, the chemical filling of porous templates, is more common. It can provide a higher density of structures in arrays compared to physical coating of pillars. On the other hand, the fabrication is more challenging for (pore) diameters below 100\,nm in diameter, due to the high aspect ratio inducing limitations of diffusion. Some examples using a combination of chemical methods are given below.

\paragraph{Sequential electrochemical steps} Co/NiO/Ni~(inner tube)\citep{Chen2011}; Ni shell + Cu core\citep{Chen2015}.

\paragraph{Electrochemical and atomic layer deposition} Fe$_3$O$_4$~(ALD shell)/SiO$_2$/Ni (electroplated core)\citep{Chong2010,bib-KIM2011b}.
	
\paragraph{Electroless plating} Ni/Co and Ni/CoNiFe multilayered tubes\citep{Rohan2008}; 
Ni shell deposited on Cu nanowires in solution\citep{Rathmell2012}. 
	
\paragraph{Sol-gel and electrochemical deposition}
FeTiO$_3$ (antiferromagnetic shell) + Ni or Ni$_{80}$Fe$_{20}$ core\citep{Khan2016}; BiFe$_{0.95}$Co$_{0.05}$O$_3$ (multiferroic shell) + permalloy core\citep{Javed2015}; Cr$_2$O$_3$ (antiferromagnetic shell) + Ni or Fe core\citep{Irfan2017}. 
	
\paragraph{Electrochemical co-deposition with phase separation} Ni shells + Cu core\citep{Wang2005,Liu2008,Li2010}.

\paragraph{Coaxial lithography} Core-shell NWs, also with rather short segments from different materials, have been prepared by a so-called coaxial lithography presented by Ozel and coworkers~\citep{Ozel2015}.

So, core-shell nanowires and multilayered nanotubes can be prepared by a variety of approaches, mostly by combination of bottom-up chemical methods in porous templates. Many heterostructures with an interesting potential for magnetism have been prepared, but in most cases no magnetic characterization has been performed, except for magnetometry on arrays of structures. The only exception are the tubular spin valve\citep{Chan2010}, and a pioneering work by \citet{bib-KIM2011b} with magnetic imaging using polarized X-rays on Fe$_3$O$_4$(shell)/SiO$_2$/Ni(core). Unfortunately, the iron oxide shell prepared by ALD was rather granular and displayed an irregular distribution of magnetization.

\subsection{Summary of wire/tube fabrication}
\label{sec-fabricationSummary}

Deposition of NWs and NTs is usually done via electrochemical deposition in nano\-porous templates that provide the desired geometry. Wires are typically prepared employing electroplating, whereas tubes are easier to prepare via atomic layer deposition or electroless plating (coating of a surface). An alternative template for nanotube deposition is an ensemble of vertical pillars/rods that can be coated by both chemical and physical deposition techniques. There exist also template-less chemical depositions, but generally with larger deviations from desired shape/geometry. An alternative is the direct writing of structures using focused electron beam and a gas precursor (focused electron beam-induced deposition), that can yield wires, tubes, and core-shell structures. Yet, despite several improvements, the material quality (carbon and other contaminants from precursors) can be still an issue, and the throughput remains low.

The combination of a tailored template (typically nanoporous alumina), and, \eg, electroplating, can yield structures with changes of diameter along their length. Further, alternating segments from different materials or wire-tube segments can be prepared. Core-shell structures, equivalents of multi-layered thin films/strips, can be deposited either using a single technique~(atomic layer deposition, electroless plating, electroplating,\,\ldots), or their combination, also with physical depositions (coating of magnetic pillars).

Even though the bottom-up fabrication routes are preferred, for device fabrication (or just in case of sample preparation for electrical measurements), one usually combines both top-down and bottom-up approach. This provides us with nanowires and nanotubes with various geometry and material composition. These can be prepared in many cases also in a reasonably controlled way with sufficient material quality that is necessary for nanomagnetic and spintronics studies (still quality of interfaces in some cases could be an issue and may require further optimization).

\section{Magnetic properties}
\label{sec-Magnetism}

In the following we discuss together the cases of wires and tubes. Indeed, most of their features are similar, at least qualitatively. Often, the case of wires was considered first, later generalized to tubes by adding the extra degree of freedom of the inner radius. Some aspects are well established and both theory and experiments are largely documented. This the case of dipolar interactions in dense arrays of wires. Other aspects are mostly or uniquely described by theory and only emerging experimentally, for instance domain-wall dynamics or magnonics, respectively. In all cases, theory and experiments are brought together in the same discussion.

\subsection{Micromagnetism in the cylindrical geometry}
\label{sec-MagnetismCylindrical}

In this paragraph we outline some specific aspects of magnetization textures in the cylindrical geometry. We first introduce our notations, then review some specific aspects in the cylindrical geometry. Finally, we detail a few test cases.

\subsubsection{Notations for micromagnetism}
\label{sec-MagnetismCylindricalNotations}

We remain in the framework of the continuous theory of micromagnetism, in which the magnetization texture is described by a continuous field of magnetization. We use lower case, upper case and script letters for dimensionless, volume density and volume-summed quantities, such as magnetization vector field~$\vect m(\vect r)$ or~$\vect M(\vect r)$ and total moment~$\vectMoment$. Similarly, we write magnetic energies $e$, $E$ in $\SI{}{\joule\per\meter\cubed}$ and $\mathcal E$ in $\SI{}{\joule}$. We consider the following volume densities of energy:

\begin{eqnarray}
  \Ez &=& -\muZero\vect M \dotproduct \vect H\label{eqn-zeemanEnergy}\\
  \Eex &=& A\left({\vectNabla{\vect m}}\right)^2 \label{eqn-exchangeEnergy}\\
  \Ea &=& Kf(\theta,\phi) \label{eqn-anisotropyEnergy}\\
  \Ed &=& -\frac{1}{2}\muZero\vect{M} \dotproduct \vectHd  \label{eqn-dipolarEnergy}
\end{eqnarray}
$\Ez$ is the Zeeman energy, with $\vect H$ the external magnetic field. $\Eex$ is the exchange energy, with $A$ the exchange stiffness. $\Ed$ is the dipolar energy~(also called magnetostatic). $\Ea$ is an anisotropy energy, which may be of various origins: magnetocrystalline, magnetoelastic, interface. $\theta$ and $\phi$ are two angles defining the direction of magnetization. For a second-order local uniaxial anisotropy we write: $\Ea=K\sin^2\alpha$, $\alpha$ being defined against a given local direction. All these energy densities are expressed in $\SI{}{\joule\per\meter\cubed}$.

We write $\DipolarExchangeLength=\sqrt{2A/\muZero\Ms^2}$ the dipolar exchange length, with $\Ms$ the spontaneous magnetization. When dipolar and exchange are the only two relevant energies, such as in the case of a soft magnetic material, the behavior in any given system is universal, provided that all length are scaled against $\DipolarExchangeLength$, and external magnetic field is measured against~$\Ms$.

The width of any type of wall is written phenomenologically $\pi\Delta_\mathrm{W}$, with $\Delta_\mathrm{W}$ being called the wall parameter. The expected formula for $\Delta_\mathrm{W}$ depends on the situation, where dominated by anisotropy, dipolar or other types of energies. We write and call $\Kd=(1/2)\muZero\Ms^2$ the dipolar constant.

The geometry of wires and tubes will be described with: length~$L$, radius~$R$ (or, outer radius for tubes), inner radius of tubes $r=\beta R$, $t=(1-\beta)R$ the wall thickness for tubes. Length and outer radius normalized with $\DipolarExchangeLength$ are $\ell$ and $\rho$.

\subsubsection{Exchange energy in the cylindrical geometry}
\label{sec-MagnetismCylindricalSpecific}

We use the cylindrical system $(\rho,\varphi,z)$ for spatial coordinates, to reflect the structural axial symmetry. The unit vectors for this coordinate system are $\unitvect \rho$, $\unitvect\varphi$ and $\unitvect z$. Magnetization is expressed with spherical coordinates $(r,\theta,\phi)$. The unit vectors for the spherical coordinate system are $\unitvect r$, $\unitvect\theta$ and $\unitvect\phi$, which we define locally such that $(\theta;\phi)=(\pi/2,0)$ stands for the direction of the wire/tube radius\bracketsubfigref{img_cylindricalSphericalCoordinates}{a,b}. The magnetization vector field $\vect m=\unitvect r$ reads, upon projection in the cylindrical system:

 \begin{equation}\label{eqn-magnetizationCylindrical}
   \vect m=\cos\theta\,\unitvect z + \sin\theta\cos\phi\,\unitvect\rho + \sin\theta\sin\phi\,\unitvect\varphi,
 \end{equation}
with $\theta(\rho,\varphi,z)$ and $\phi(\rho,\varphi,z)$ two functions defined over space. Below are examples of distributions of magnetization:
\begin{itemize}
  \item Uniform axial magnetization: $\theta(\rho,\varphi,z)=\pi(1\pm1)/2$
  \item Azimuthal magnetization, circular left or right~(also named orthoradial):\\$\theta(\rho,\varphi,z)=\pi/2$; $\phi(\rho,\varphi,z)=\pm\pi/2$
  \item Transverse magnetization, \eg for a wire saturated with a transverse magnetic field:  $\theta(\rho,\varphi,z)=\pi/2$; $\phi(\rho,\varphi,z)=-\varphi$
\end{itemize}

\begin{figure}[thbp]
\centering\includegraphics[width=110.579mm]{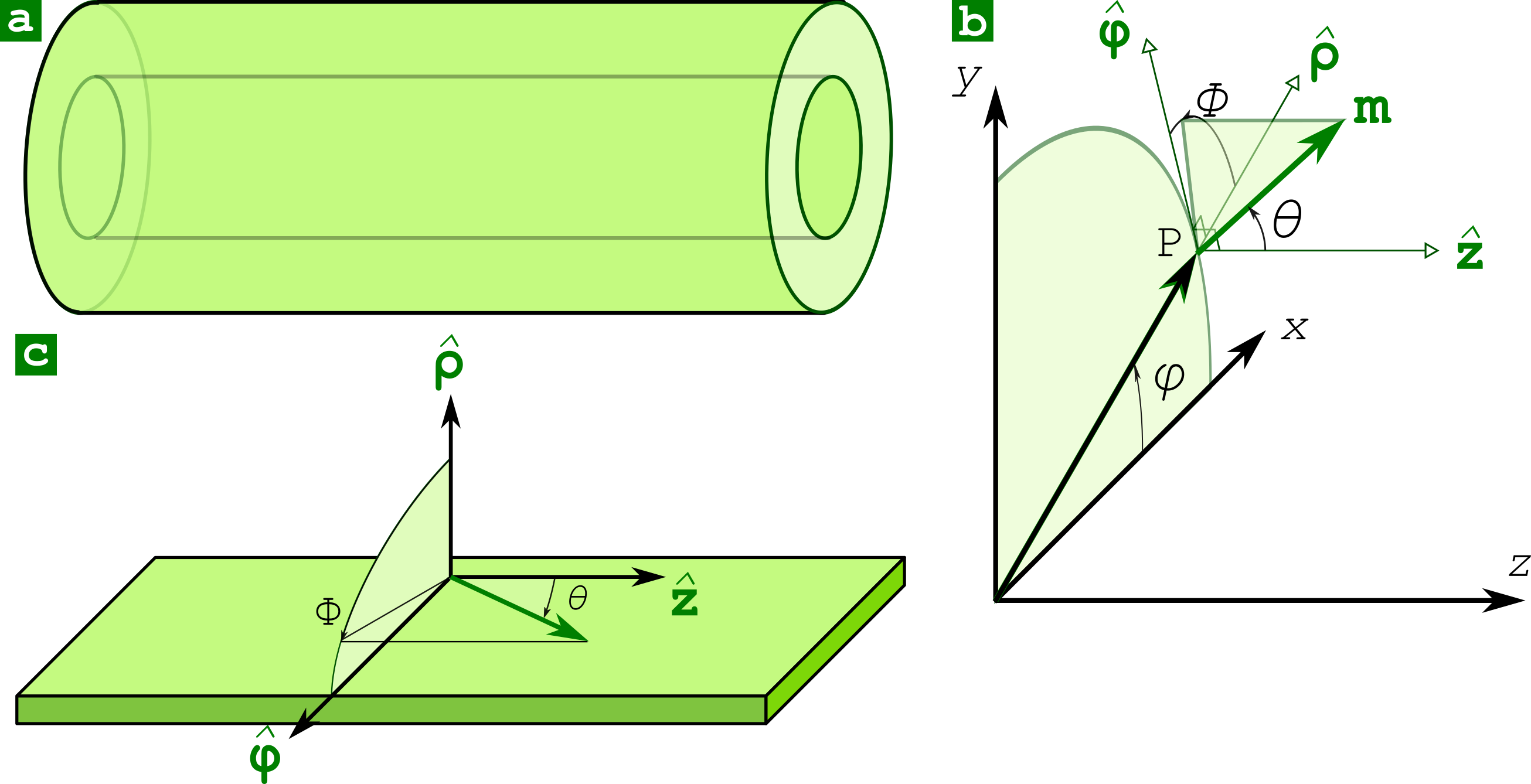}
\caption{(a)~Wire or tube to be described (b)~Points in space in the wire or tube such as P here, are defined with cylindrical coordinates $(\rho, \varphi, z)$. At this point, magnetization $\vect m$ is defined with spherical coordinates $(1, \theta, \phi)$ defined in the local axes $(\unitvect z, \unitvect\rho, \unitvect\varphi)$ (c)~The tube or wire surface is unrolled into a slab. The axes defining the spherical coordinates are now the same everywhere.}\label{img_cylindricalSphericalCoordinates}
\end{figure}
This convention of rotating frame is used by some authors\citep{bib-LAN2010}, while others chose a fixed frame\citep{bib-VIL1995,bib-SAX1998}. One should be careful, as the choice leads to different formulas for exchange energy. Indeed, when deriving \eqnref{eqn-exchangeEnergy} from \eqnref{eqn-magnetizationCylindrical}, not only terms $\linepartial{\theta}{j}$ and $\linepartial{\phi}{j}$ contribute~($j=\rho,\varphi,z$), but also the spatial variation of unit vectors~$\linepartial{\unitvect i}{j}$~($\unitvect i=\unitvect\rho,\unitvect\varphi,\unitvect z$). One finds\citep{bib-LAN2010}:

\begin{multline}\label{eqn-exchangeCylindrical}
   \Eex=A\left[ { \left({\fracpartial\theta \rho}\right)^2 +\frac{1}{\rho^2}\left({\fracpartial\theta \varphi}\right)^2 + \left({\fracpartial\theta z}\right)^2}\right.\\
   \left.{+ \sin^2\theta\left({\fracpartial\phi \rho}\right)^2 +\frac{\sin^2\theta}{\rho^2}\left({1+\fracpartial\phi \varphi}\right)^2 + \sin^2\theta\left({\fracpartial\phi z}\right)^2 } \right]
 \end{multline}
We examine consequences of this form of exchange in the next paragraph. Note that \citet{bib-LAN2010} did not explicitly mention the radial variation, as they were considering tubes with thin walls. Other works considered only cylindrical symmetry\citep{bib-DAN1995}, or on the contrary extended the formulation for elliptical cylindrical coordinates\citep{bib-SAX1997}. Equations for exchange on surfaces with arbitrary curvature were also mentioned\citep{bib-SAX1997} or derived explicitly\citep{bib-GAI2014}, and applied to cases such as the Moebius ring\citep{bib-PYL2015}. Finally, while \eqnref{eqn-exchangeCylindrical} handles angles, exchange has also been expressed in the most general 2D curved case based on local components\citep{bib-SHE2015b}, which here would be $m_z$, $m_\rho$ and $m_\varphi$. The latter form makes it easier to draw a parallel with the formalism of the Dzyaloshinskii-Moriya interaction~(DMI). A review of magnetism in curved geometries was made by \citet{bib-STR2016}.

\subsubsection{Examples}
\label{sec-MagnetismCylindricalExamples}

We examine here the consequences of \eqnref{eqn-exchangeCylindrical}. In this course we consider distributions of magnetization with sometimes non-zero dipolar energy. We will ignore it, as well as consider a material with no magnetocristalline anisotropy, to highlight the role of exchange. This means that in a real system the energetics is more complex than described below. Nevertheless, the discussion suitably describes magnetism at 2D surfaces in their asymptotic limit of zero thickness\citep{bib-STR2016}.

While all terms are reminiscent of exchange in cartesian coordinates, the fifth term on the right-hand side may seem puzzling at first sight:

\begin{equation}\label{eqn-exchangeCylindricalAzimuthal}
   A\frac{\sin^2\theta}{\rho^2}\left({1+\fracpartial\phi \varphi}\right)^2.
 \end{equation}
It is non-zero and thus implies a cost even in the case $(\theta,\phi,z)\equiv(0,0,0)$. This is a consequence of the choice of a local frame to express spherical coordinates and the projection in the cylindrical frame, so that uniform $m_\rho$ or $m_\varphi$ ~(both related to $\sin\theta$) do not imply uniform magnetization~$\vect m$, as $\unitvect\rho$ and $\unitvect\varphi$ are not uniform. It is handy to analyze the consequences by hypothetically unrolling the surface of the wire or the tube into a flat and long strip of width $2\pi R$\bracketsubfigref{img_cylindricalSphericalCoordinates}{c}. In this new system, the frame for the former spherical coordinates is now uniform, so that \eqnref{eqn-exchangeCylindricalAzimuthal} looks like a contribution to energy mixing all three components/directions in space. We see below on simple examples that, in this planar context, this is reminiscent of a magnetic anisotropy or the Dzyaloshinskii-Moriya interaction, depending on the situation. For the sake of illustration, we consider the case of a tube with very thin wall~($\beta\simeq1$), and radius~$R$.

\begin{figure}[thbp]
\centering\includegraphics[width=126.245mm]{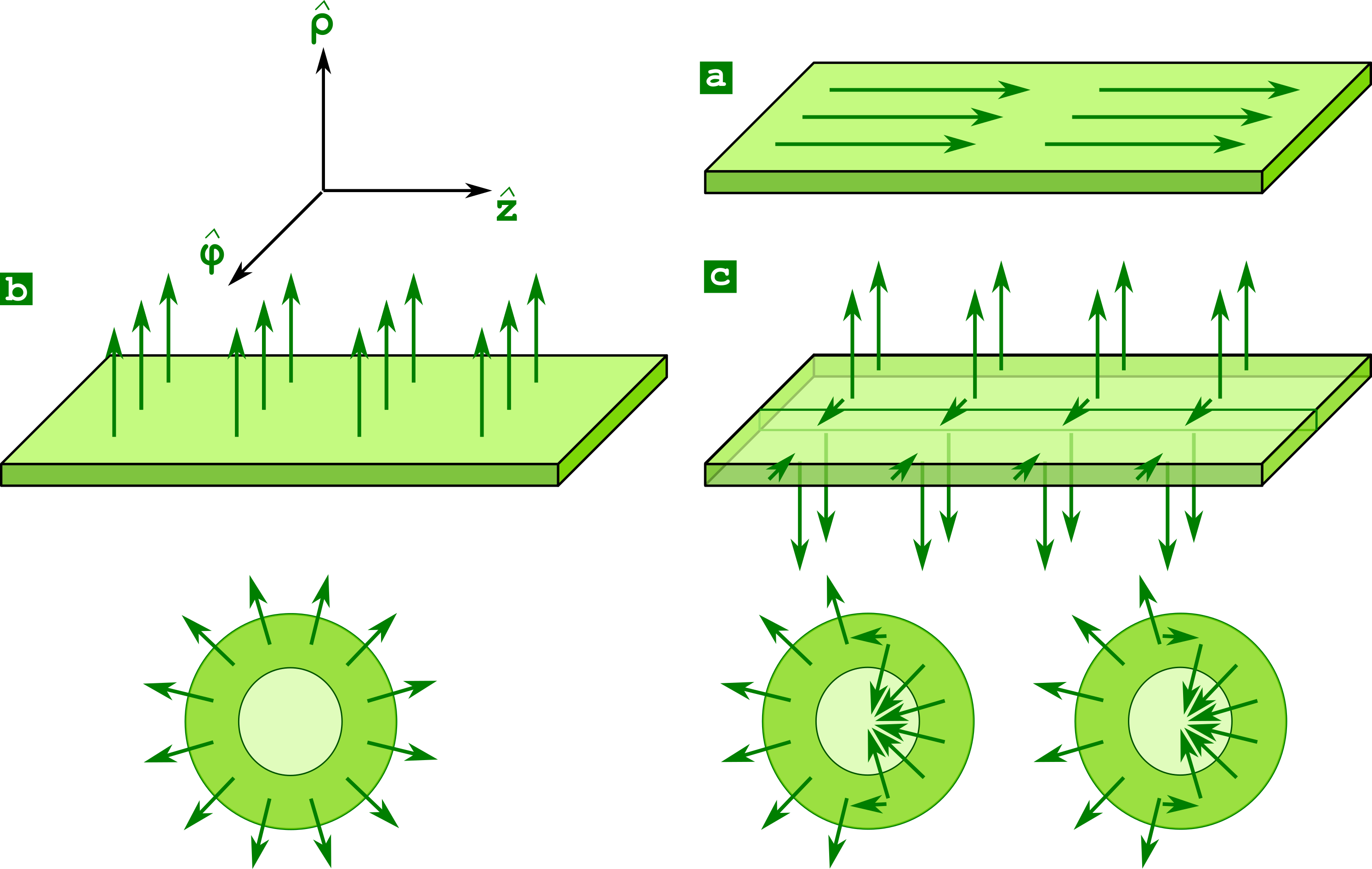}
\caption{Examples of distributions of magnetization at the surface of wires or tubes. (a)~Uniform axial magnetization~(unrolled view) (b)~Outward radial magnetization~(radial and cross-sectional views) (c)~Two domains with inward and outward radial magnetization, separated by two longitudinal domain walls~(unrolled and cross-sectional views). The chirality of the N\'{e}el walls from the bottom-left view, allows to reduce the exchange energy compared to the chirality from the bottom-right view.}\label{img_magnetization-configurations}
\end{figure}

\paragraph{Uniform magnetization}
\label{sec-MagnetismCylindricalExamplesUniform}

Uniform magnetization implies $\phi=-\varphi$\bracketsubfigref{img_magnetization-configurations}a. Consequently, \eqnref{eqn-exchangeCylindricalAzimuthal} cancels and exchange energy is zero, as expected for uniform magnetization.

\paragraph{Curvature-induced anisotropy}
\label{sec-MagnetismCylindricalExamplesAnisotropy}

Consider radial\bracketsubfigref{img_magnetization-configurations}b or azimuthal magnetization, characterized by $\theta=\pi/2$ and $\phi=(0,\pm\pi/2)$, respectively, or any other intermediate situation with $\phi=\textrm{cst}$. \eqnref{eqn-exchangeCylindricalAzimuthal} simply reads $(A/R^2)\sin^2\theta$. This is equivalent to a uniaxial anisotropy of magnitude $K=(A/R^2)$, favoring axial magnetization. To set numbers, for $A=\SI{1e-11}{\joule\per\meter}$, $\Ms=\SI{10^6}{\ampere\per\meter}$ and $R=\SI{20}{\nano\meter}$, one gets $\Ku=\SI{2.5E4}{\joule\per\meter\cubed}$ or in terms of anisotropy field: $\muZero\Ha=\SI{50}{\milli\tesla}$. The consequence, of course trivial, is that exchange favors axial and thus uniform magnetization. As this also happens to generate no dipolar energy in the limit of infinitely-long wires and tubes, this is in general the ground state. Another consequence is that, in the presence of a microscopic source of anisotropy favoring azimuthal\citep{bib-FRU2017e,bib-ZIM2018} or radial magnetization, a hard-axis loop or ferromagnetic resonance spectrum needs to be corrected by this value to extract  the microscopic anisotropy energy.

\paragraph{Curvature-induced Dzyaloshinskii-Moriya interaction~(DMI)}
\label{sec-MagnetismCylindricalExamplesDMI}

Consider again radial magnetization imposed by a strong local perpendicular anisotropy, however, with the tube split longitudinally in two domains\bracketsubfigref{img_magnetization-configurations}c. The longitudinal domain walls between the domains may be of either Bloch or N\'{e}el type\citep{bib-HUB1998b}, each with two possible directions. However, due to the finite extent of the wall, rotation is less~(resp. more) than $\pi$ depending on the direction in the N\'{e}el wall, while it is exactly $\pi$ for a Bloch wall. Thus, exchange favors N\'{e}el walls of a given chirality related to the direction of magnetization in the neighboring domains~(this comes at the expense of the wall dipolar energy, like for the thin film case, which in practice needs to be considered in addition). Note that the topology is the same as the onion state, rotation all moment by $\pi/2$ around $\unitvect z$\citep{bib-SUN2014}.  These situations are analogous to the DMI in ultrathin films with perpendicular magnetization\citep{bib-HEI2008}, and promoting high-speed domain-wall motion\citep{bib-THI2012,bib-EMO2013}. Let us put figures on this situation. For a wall parameter $\Delta_\mathrm{W}$ one can show in a few steps starting from \eqnref{eqn-exchangeCylindricalAzimuthal}, that the resulting difference of energy is $8\pi A\Delta_\mathrm{W}/R^2$. If one makes the parallel with the DMI energy $\pi D$ of a wall\citep{bib-HEI2008}, with $D$ a DMI parameter, one gets an effective value:

\begin{equation}\label{eqn-exchangeCylindricalDMI}
   D=8\pi A\frac{\Delta_\mathrm{W}}{R^2}.
 \end{equation}
As an example, for $\Delta_\mathrm{W}=\SI{5}{\nano\meter}$ and $A=\SI{1e-11}{\joule\per\meter}$, $D=\SI{0.13}{\milli\joule\per\meter\squared}$ for $R=\SI{100}{\nano\meter}$, and $D=\SI{3.1}{\milli\joule\per\meter\squared}$ for $R=\SI{20}{\nano\meter}$. The DMI coefficients are in the same range for ultrathin-films\citep{bib-BEL2016b}, which shows that static chiral effects may be quite significant in tubes.

The discussion above remains valid for walls of smaller extent, for example a localized bubble with $m_\rho=-1$ inside a domain with  uniformly $m_\rho=1$. The parts of longitudinal domain walls are favored as N\'{e}el type with opposite chirality. So, such bubbles should be chiral, and reminiscent of skyrmionic bubbles\citep{bib-BER2018b}. Let us draw a final link between cylinders and DMI in ultrathin films. DMI requires absence of inversion symmetry, which in thin films is achieved using two different materials with high and different spin-orbit coupling for bottom and top layers\citep{bib-MOR2016}. In a tube, the breaking of symmetry arises from the curvature, positive on one side and negative on the other. We will see in \secref{sec-MagnetismDWsPropagation} and \secref{sec-MagnetismSpinWaves} that this has important consequences on magnetization dynamics, which through the Landau-Lifshitz equation is also a chiral phenomenon.

Note that the possibility of material- or interface-related DMI in tubes has also been considered by theory\citep{bib-GOU2016}, with impact on domain-wall dynamics like for thin films. However, so far no suitable material has been reported experimentally.

\paragraph{Bends and modulations of diameter}
\label{sec-MagnetismCylindricalExamplesBends}

Above, we considered the case of wires and tubes with translational structural invariance along their axis. These display curvature along one direction only: the azimuth. Curvatures along two directions occur in several situations, which have been realized experimentally and considered by theory: longitudinal modulations of diameter\citep{bib-ALL2009}, and non-straight wires, \ie presenting bends\citep{bib-FER2013,bib-STR2016,bib-FRU2017b}. The general treatment of curvature is then required, with impact on the statics and dynamics of domains and domain walls\bracketsecref{sec-MagnetismDWs}, as well as in magnonics\bracketsecref{sec-MagnetismFMRAndMagnonics}.

\subsection{Dipolar interactions in dense arrays of wires}
\label{sec-MagnetismDipolar}

\subsubsection{Motivation and general considerations}
\label{sec-MagnetismDipolarGeneral}

For a long time, reports on the magnetic properties measured on single wires and tubes have remained scarce, appart from a few pioneering works\citep{bib-WER1996,bib-BEE1997,bib-EBE2000,bib-VIL2002}. Most measurements were made on macroscopic assemblies, kept in their template used for the growth, using resonance or magnetometry techniques. The properties of individual objects and of the constituting material had to be inferred from such measurements, and thus disentangled from the effect of inter-objects dipolar interactions. Thus, understanding dipolar interactions in arrays has been an early and major topic for magnetic nanowires and nanotubes.

Dipolar interactions and their understanding is a general concern in magnetic materials. They are both important and complex to tackle, due to their long range compared with, \eg, exchange interactions. It is important for the following discussion, to define what \textsl{long range} means. Dipolar fields may be calculated based on magnetic charges, of volume type $-\Div \vect M$ and surface type~$\vect M\dotproduct \vect n$. For largely-uniformly-magnetized systems with a three-dimensional outer shape and size~$L$, the charges are of surface type and their quantity thus scales like~$L^2$. The $1/r^2$ decay of resulting dipolar fields exactly compensates the decay. More precisely, what matters is the solid angle through which charges are seen from a given point in the system. This underpins the concept of demagnetizing factors: demagnetizing effects depend on the sample shape, not on its size.

Dipolar fields in low-dimensional systems are \textsl{short-ranged}\citep{bib-FRU1998}. Besides calculation, this can be understood handwavingly in thin films as most of the stray field escapes in the free space rather than interact with magnetization. Another way to view this, is that the solid angle from which a part of the film is seen from another part, decays rapidly at distances significantly larger than the film thickness. So, in alumina and polymer foils serving as templates for the growth of wires and tubes, of thickness typically a few tens of micrometers, the lateral range of dipolar fields is of the order of $\SI{100}{\micro\meter}$. This involves a very large number of objects, making the use of simplifying, however, trustful models crucial. In particular, we stress that approaches based on pinpoint dipoles or simulations with a small number of objects as sometimes used, may give trends, however not quantitatives figures. In the following we consider separately uniformly-magnetized arrays and the case of hysteresis loops, as the modeling of the underlying phenomena are significantly different.

\subsubsection{Magnetic anisotropy}
\label{sec-MagnetismDipolarAnisotropy}

Magnetic anisotropy can be measured from uniformly-magnetized states, through ferromagnetic resonance~(FMR) or monitoring a hard-axis loop, \ie, displaying no hysteresis. In the former case one extracts an anisotropy field, including both the external and internal field, itself reflecting the anisotropy of energy. In the second case the saturation field, or more generally the area above the loop, is related to the difference between final and initial energy, which again is magnetic anisotropy. The several contributions to magnetic anisotropy are: (1)~microscopic contributions, mostly magnetocrystalline anisotropy, depending on composition, crystalline structure and texture; (2)~the single-object shape anisotropy, be it wire or tube, depending on length and radii, and possibly axial modulations of radii or material;  (3)~inter-object dipolar anisotropy, depending on object shape and inter-object distance.

In the following we review the modelling of the second and third contributions, the dipolar effects. This boils down to determining the demagnetizing factors of a composite system. While general theories exist for this\citep{bib-DOB2009}, we focus here on the specific case of arrays of parallel objects, starting from the simpler case of wires. We call in-plane~(ip) the direction parallel to the template surface, \ie transverse to the wires axes, and out-of-plane~(oop) the axial direction. We like to avoid using \textsl{parallel} and \textsl{perpendicular} as sometimes done, because of the possible confusion between the reference to the wire axis or to the plane of the template. $\theta$ is the angle between magnetization and the wire axis. We consider wires with a diameter $d$ and inter-axis distance $D$ in ordered alumina membranes, yielding the packing factor~$p=[\pi/(2\sqrt{3})]/(d/D)^2$. This model was first described by \citet{bib-ENC2001} to analyze ferromagnetic resonance data, following a pioneering work of \citet{bib-NET1990} regarding dipolar fields in a composite medium for magnetic tape recording.

\begin{figure}[thbp]
\centering\includegraphics[width=127.999mm]{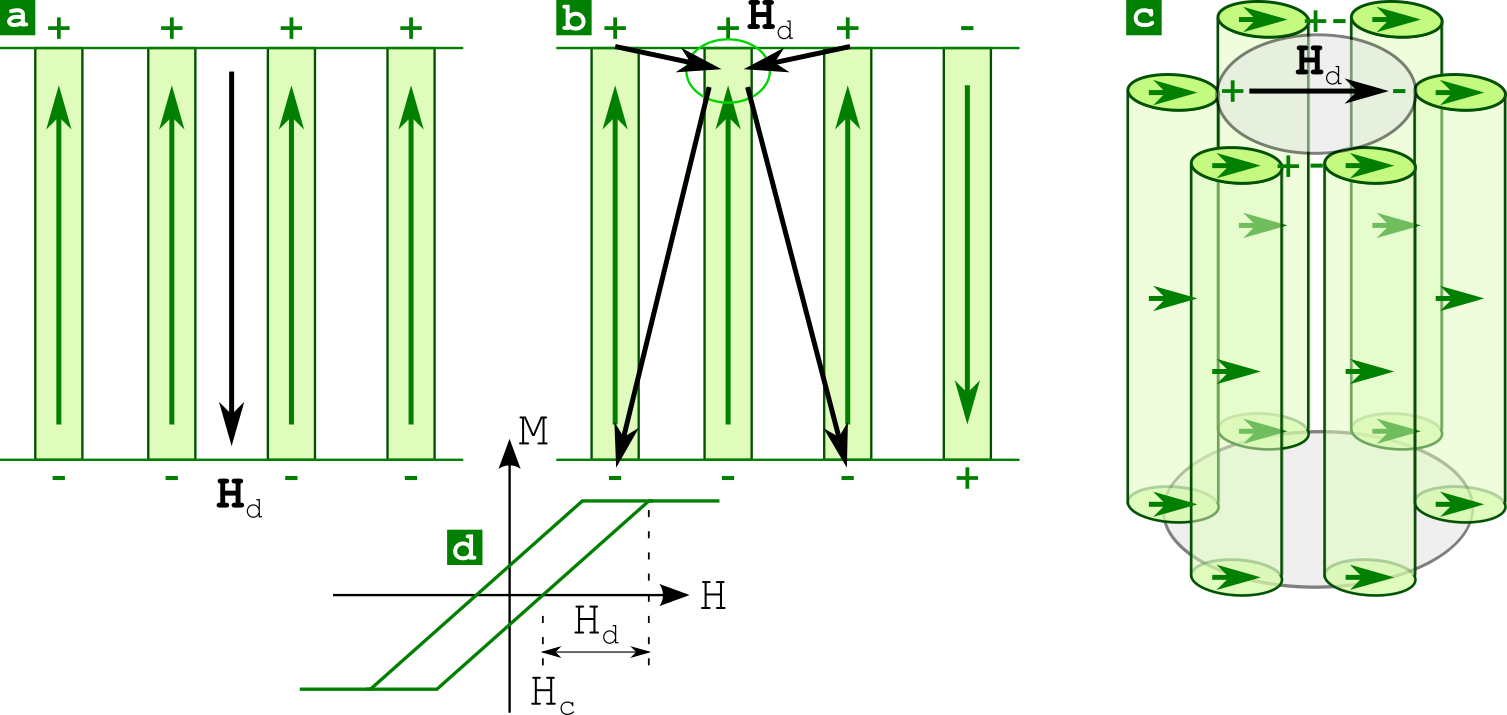}
\caption{\textbf{Schematics of dipolar fields in arrays} (a)~Array uniformly-magnetized along the positive vertical direction, giving rise to a dipolar field, which is destabilizing, and in mean field is exactly along the opposite direction (b)~Partly-reversed array. The array of a nucleation volume is highlighted at the end of a wire. Charges from the opposite side may be modeled in mean field, while those from the same side cannot as short-ranged, and their effect is lower due to the projection angle (c)~Wire or tubes magnetized across their axis, give rise toa stabilizing field modeled with a cylindrical Lorentz cavity}\label{img_dipolar-fields-in-array}
\end{figure}

The inter-wire energy can be taken into account in mean-field, assuming a medium of magnetization $p\Ms$. The oop component of dipolar field arises from the charges located at both ends of the wires, like for a thin film: $\vect H_\mathrm{oop}=-p\Ms\cos\theta\;\unitvect z$\bracketsubfigref{img_dipolar-fields-in-array}{a,b}. For the sake of simplicity we neglect here the self-demagnetizing factor of each cylinder, given their long length, although the calculation may be extended with no formal difficulty to finite-length objects. The ip component of dipolar field has two contributions. The first contribution is the self dipolar field for each cylinder, with strength $-(\Ms\sin\theta)/2$ following the demagnetization factor $N=1/2$ transverse to the axis. The second contribution is inter-wire fields, approximated as the Lorentz-cavity field for a cylindrical hole in a medium of magnetization $p\Ms$: $\Hin=+(\Ms p\sin\theta)/2$\bracketsubfigref{img_dipolar-fields-in-array}c. The total density of dipolar energy $\Ed=-\muZero\vectM\dotproduct\vectHd$ reads, letting aside a constant value:

\begin{equation}\label{eqn-dipolarEnergyAAM}
   \Ed=\Kd\left({\frac{3p-1}{2}}\right)\cos^2\theta.
 \end{equation}
Dipolar energy is as usual of second-order. Let us consider three specific cases:
\begin{itemize}
  \item $p=0$ describe the case of non-interacting wires. Accordingly, \eqnref{eqn-dipolarEnergyAAM} predicts a uniaxial oop easy axis with strength $\Kd/2$, set by the transverse demagnetizing factor~$N=1/2$ of individual wires. Such low-porosity matrices have been investigated experimentally, immediately informing of single-wire properties, letting aside their distribution\citep{bib-DEM2002b,bib-EBE2001b}.
  \item $p=1$ describe the hypothetical case of a matrix fully filled with magnetic material. This would be a continuous thin film. Consistently, \eqnref{eqn-dipolarEnergyAAM} predicts an easy plane with strength $\Kd$, determine byd the oop demagnetizing factor os a thin film~$N=1$.
  \item $p=1/3$ is the cross-over value, for which a uniformly-magnetized matrix behaves isotropically.
\end{itemize}
The above remains a valid description of hysteresis loops performed with an ip field. In that case the population of up and down wires balances at remanence, so that the oop component of dipolar field vanishes, and the ip loop is linear, with a saturation field $\Ms(1-p)/2$. Measuring this field is a rather reliable way to determine $\Ms$ if $p$ is known, or the reverse. The model may be applied to disordered arrays, especially to polycarbonate foils, considering the average packing factor. However, this also implies the broadening of ip resonance peaks and the rounding of ip hysteresis loops, due to the distribution of inter-wire distances, and thus of inter-wire dipolar couplings. Finally, the modelling is also valid for tubes, with two specific aspects. First, the packing factor is reduced compared to a wire, due to the hollow part. Second, while the transverse demagnetizing factor of a tube is also $1/2$~(because $2N_\mathrm{ip}+N_\mathrm{oop}=1$, and $N_\mathrm{oop}\approx0$), the internal field inside the tube material is highly non-uniforme, so that again ip FMR spectra are wide and ip hysteresis loops are expected to be rounded.

\subsubsection{Impact on magnetization switching}
\label{sec-MagnetismDipolarSwitching}

We now consider the case of oop hysteresis loops, and restrict the discussion to wires and tubes essentially uniformly-magnetized along their axis~(the case and meaning of angular-dependent hysteresis loops is covered in \secref{sec-MagnetismDWsNucleation}). Due to the long aspect ratio of wires and tubes, this applies, in particular, to soft magnetic materials. Recoil curves show that the hypothesis of uniformly-magnetized structures and sharp switching is valid\citep{bib-FRU2011g}. Starting from an oop-magnetized state, wires/tubes tend to switch before their intrinsic switching field is reached, under the influence of the matrix self-dipolar field\bracketsubfigref{img_dipolar-fields-in-array}a. On the reverse, upon coming close to saturation along the opposite direction, the matrix has a stabilizing effect on the wires/tubes not yet switched. For intermediate situations the internal field is opposite and proportional to the normalized  moment of the sample~$\vect m$, in a mean-field approach. This induces a slanting of the global loop, even if each wire/tube has a square loop. In theory, the full width of the slanting equals twice the interaction field for a fully magnetized matrix\bracketsubfigref{img_dipolar-fields-in-array}d. In practice, we expect that this is a lower bound, due to the contribution of the switching field distribution of individual objects\citep{bib-FRU2011g}. Consistently, experiments show that widening~(resp. narrowing) the wires diameter for a given pitch~$D$, increases~(resp. decreases) the slanting\citep{bib-VAZ2004,bib-FRU2011g}. If the packing factor is low enough, then the full remanence of the loops may be preserved\citep{bib-NIE2002}. Interactions can be measured directly using Henkel plots for wires\citep{bib-ZHE2000,bib-FOD2002} and tubes\citep{bib-VEL2014}, or the more elaborate First-Order Reversal Curve~(FORC) method\bracketsecref{sec-MagnetismTechniquesMagnetometry}, a technique that allows to extract independently the distribution of individual switching fields, from the interactions\citep{bib-PIK1999b}. FORC has been used both for wires\citep{bib-ROT2011} and for tubes\citep{bib-PRO2013,Proenca2014}. We review below the various approaches that have been proposed to model quantitatively the interaction field.

A simple thought is to model the array as a 2D network of pinpoint dipoles, each holding the moment of an entire wire/tube\citep{bib-FOD2003,bib-GHA2011}. Although this reproduces an increased interaction with larger packing factor, this cannot work quantitatively. Indeed, we have seen in \secref{sec-MagnetismDipolarGeneral} that we expect significant interaction over a large number of wires/tubes, while on a 2D array the interaction is short-ranged. Some authors introduced an effective coefficient to correct for this\citep{bib-FOD2003}, however this correction is not known \apriori, and it leaves the range of interaction incorrect.

A step ahead is to perform micromagnetic simulations of an ensemble of wires\citep{bib-HER2001,bib-LAV2012}. However, as mentioned in \secref{sec-MagnetismDipolarGeneral}, the lateral range of interactions extends up to several times the length of the wires, implying a large number of wires, which cannot all be considered in the simulation. Thus, again trends are grasped, however in practice this approach cannot be fully quantitative.

A significant improvement of the modelling is through considering the top and bottom charges of the array to estimate the average dipolar interaction in mean field, similar to the approach in the previous section\citep{bib-WAN2008a}. An interaction field $-p\Ms\vect m$ is predicted, with $\vect m$ the normalized moment of the array,. This remains valid for tubes, taking in to account for the reduction of packing factor due to the hollow core\citep{bib-VEL2014}. The agreement with experiments is much improved, although the experimental slanting is often lower than the model by a few tens of percent\citep{bib-ZEN2000}, while, on the reverse, we expect a larger distribution in experiments, as pointed out above. Besides, it is intriguing that theory does predict that coercive field is independent of interactions, while in experiments it slightly decreases for increasing interactions\citep{bib-VAZ2004}. The reason for these slight discrepancies and the solution for a better modelling are the following.

The above considers the \textsl{average} demagnetizing field in the entire array. However, as we will see in \secref{sec-MagnetismDWsNucleation}, nucleation for magnetization switching occurs at the end of the wires/tubes, while the average field reflects rather the interior of the array, at mid-height of the wires/tubes. At the wires/tubes ends, the end charges of the neighbors cannot be taken into account in mean field. A model has been proposed to take this into account\citep{bib-FRU2011g}, considering separately the neighboring charges, from the one of the opposite surface of the template\bracketsubfigref{img_dipolar-fields-in-array}b. The range of the latter is a few times the membrane thickness, as discussed earlier, while due to the projection angle the range of the former is a few times~$D$. In the end, the latter is $-(p/2)\Ms\vect m$, while the former is lower, resulting from a 2D summation over the neighbors. To perform this summation, the model requires to assume phenomenologically the longitudinal length of the nucleation volume, to compute the projection angle of the demagnetizing field in this very volume. Assuming a length of the order of the wire diameter provides a very good fit with the experiments\bracketfigref{img_dipolar-fields-in-array-fits}. Besides, as the demagnetizing field at the nucleation site may be different in strength and in direction~(\ie, not parallel to the pore axis) compared with the average value, this picture also explains why in experiments the coercive field may depend on the interactions, as reproduced by micromagnetic simulations\citep{bib-HER2001}. Finally, a consequence is that one does not expect identical ip and oop for $p=1/3$, \textsl{although the corresponding saturated states are isotropic} (With this model in mind, an interesting contribution was made by \citet{bib-LAN2009b}, calculating in the entire space the dipolar field induced by nanotubes having various types of distribution of magnetization). Note, however, that despite the good agreement, nucleation volumes are a phenomenological concept, and a choice needs to be made for their size, so that the model becomes clearly non-rigorous, although providing an improved agreement. This means, it is a better choice to consider fitting ip loops, if the objective is extract geometrical or magnetization data of the system. In the end, a confident fit requires that both ip and oop loops are fitted suitably with the same material and geometrical parameters.

\begin{figure}[thbp]
\centering\includegraphics[width=128.441mm]{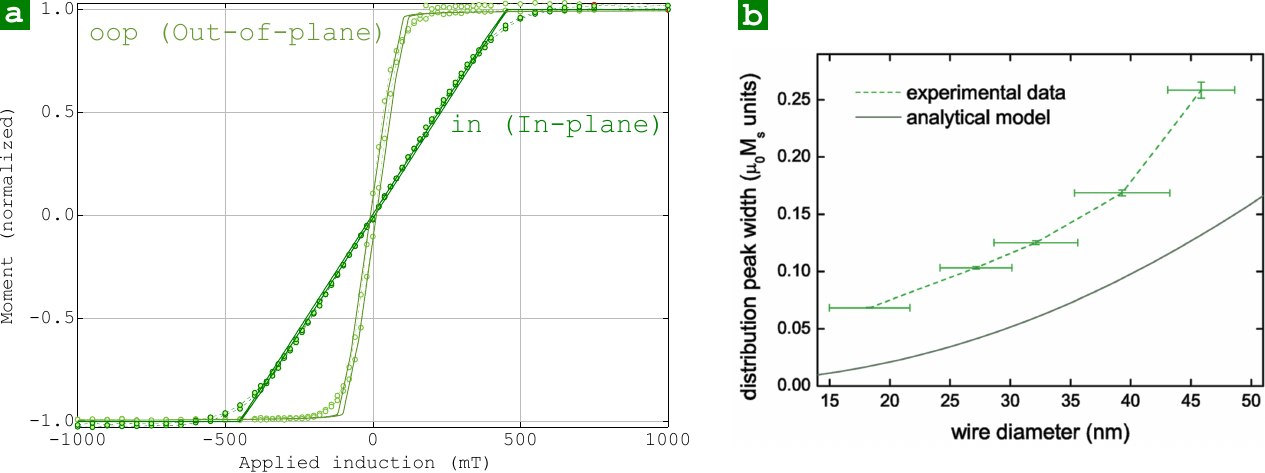}
\caption{\textbf{Modeling of interactions in arrays} (a)~Fit of ip and oop hysteresis loops of trisegmented wires with diameters $\SI{200}{\nano\meter}$/$\SI{150}{\nano\meter}$/$\SI{200}{\nano\meter}$, segment length $\SI{10}{\micro\meter}$ each, and pitch $\SI{410}{\nano\meter}$. Symbols show experiments, with lines for fits with magnetization as the sole adjustable parameter, yielding $\Ms=\SI{822}{\kilo\ampere\per\meter}$. Note that the model is slightly more elaborate than the one presented in the text, taking into account the end curling and predicting the coercive field with no adjustable parameter\citep{bib-FRU2018b}. (b)~Experiments~(symbols) and model~(line) for the interaction field versus diameter of Ni wires for a constant pitch $D=\SI{105}{\nano\meter}$, normalized to magnetization. The two curves are parallel one with another, the difference being interpreted as the intrinsic switching field of the wires. Adapted from \citet{bib-FRU2011g}, copyright 2011, with permission from AIP publishing.}\label{img_dipolar-fields-in-array-fits}
\end{figure}

Finally, wires with modulated diameters have also been measured as assemblies\citep{bib-ROT2011,bib-ARZ2017,bib-SAL2018}. This case calls for further refinements of the modelling, proposed by, \eg, \citet{bib-FRU2018b}. There is probably no model suitable for all cases, given the various phenomena that may occur at modulations of diameter: domain-wall pinning, nucleation, curling\bracketsecref{sec-MagnetismSingleSoft} etc. Global measurements have also been reported for segmented wires\citep{bib-BRA2017b}. It is clear that a lower packing density of magnetic segments induces a reduced dipolar interaction, although the changes of dipolar effects versus individual switching field distributions have not been disentangled.


\subsection{Magnetic domains}
\label{sec-MagnetismSingle}

\subsubsection{Magnetic ordering}
\label{sec-MagnetismSingleOrder}

Two types of effects may be expected to affect the magnetic ordering of nanowires and nanotubes. First, the material may be affected by the synthesis method or stress from the template, giving rise to a different magnetization or even magnetic order, compared to its bulk counterpart. Second, finite-size effects may be sizable, inducing a faster thermal decay of magnetization compared to the bulk.

Metals are quite robust against strain and finite-size effects, so that magnetization and the type of magnetic order is not different from the bulk. This is in general the case for electroplated materials. However, there exist cases where an extreme strain may change this, such as for $\mathrm{Ge}_{1-x}\mathrm{Mn}_x$ epitaxial nanocolumns embedded in a Ge matrix and resulting from a spinodal decomposition\citep{bib-JAM2006}. In that specific case, an alloy not even existing in the bulk phase diagram is stabilized, displaying magnetization and room-temperature ordering. Another case of drastic alteration is when the composition of the material is changed by incorporation of various species during the synthesis. This is the case, \eg, for electroless-plated 3d metals with dimethylamine borane as reducing agent. This may induce incorporation of boron, whose content sensitively depends on the pH\bracketsecref{sec-fabricationDepositionElectroless}. This can induce amorphisation of the material, and also reduce magnetization, especially for Ni due to the already close-to-filled 3d shell\citep{bib-RIC2015}. Incorporation of Phosphorus can also result from another common reducing agent: sodium hypophosphite\citep{Watanabe2004}.

As regards thermal decay of magnetization, in thin films, deviations from the bulk behavior are evidenced for thickness below a few nanometers\citep{bib-GRA1993}. Thus, given the larger surface-over-volume ratio in low dimensions, an effect may be expected in nanowires until a diameter around~$\SI{10}{\nano\meter}$, due to the lower dimensionality than for thin films. This was indeed seen for Co, Ni and Fe electroplated in alumina\citep{bib-ZEN2002}. Fitting magnetization with the Bloch law $M(0)(1-B\,T^\lambda)$ yields $B$ and $\lambda$ respectively larger and smaller than in the bulk: for Ni wires with diameter of $\SI{8}{\nano\meter}$, $\lambda=1.2$ instead of 1.5, and $B$ is increased by one order of magnitude. The difference  in $\lambda$ may reflect the change of spin-wave spectrum in low dimension. A similar investigation of Co wires embedded in a $\mathrm{Ce}\mathrm{O}_2$ matrix confirms the much faster decay, however this time with still $\lambda=1.5$\citep{bib-VID2009}.

\subsubsection{Soft magnetic materials}
\label{sec-MagnetismSingleSoft}

When modelling macroscopic arrays of parallel wires and tubes in the previous section, we assumed that each individual object is essentially magnetized uniformly up or down along its axis at any time. The reason is that dipolar energy favors alignment of magnetization parallel to the longest direction of an object. Numbers can be put on this using demagnetizing factors, which have been computed for cylinders\citep{bib-ROW1956,bib-RHO1962,bib-AHA1983}. The axial demagnetizing factor becomes negligible when the length exceeds by far the diameter, so that axial magnetization may be expected if uniform magnetization is assumed. Consistently, the first measurements on individual wires confirmed the longitudinal magnetization from sharp magnetization switching\citep{bib-WER1996} and anisotropic magnetoresistance\citep{bib-VIL2002}. Note that, due to the hollow part, the end charges for tubes are less than for wires in the case of axial magnetization, which translates in even lower axial demagnetizing factors.

A more general phase diagram was derived analytically as a function of the geometry of shorter cylinders, considering axial uniform magnetization, transverse uniform magnetization, and also curling around the axis as a strongly non-uniform magnetization distribution~(the vortex state)\citep{bib-MET2002}\bracketsubfigref{img_phase-diagrams}a. For sufficiently large diameters, the vortex state is favored against axial magnetization even for slightly elongated cylinders, thanks to the closure of magnetic flux and despite the cost of exchange energy. Note the proximity of the dash-dotted lines with the axial-versus-vortex full line, showing that the vortex fills the entire cuylinder. \citet{bib-LEE2007d} derived the energy of various states for short tubes. However, they did not consider the uniform curling state, but rather a higher-energy state including a domain wall at mid-height. A first realistic phase diagram for tubes with rather thin walls was derived by \citet{bib-SUN2014}. Only the axial magnetization and uniform azimuthal magnetization were considered~(also called curling, or vortex state). The latter allows to avoid end charges and thus dipolar energy, at the expense of exchange energy. Which one is the ground state depends on the ratio:

\begin{equation}\label{eqn-energyTubes}
   \gamma=\frac{R^2}{\pi\DipolarExchangeLength^2}\frac{t}{L}\left[{\ln\left({\frac{8R}{t}}\right)+\frac32}\right],
 \end{equation}
with $L$ the tube length and $t=(1-\beta)R$ the tube wall thickness. $\gamma=1$ determines the boundary between the axial and curling ground states, the former being ground state for $\gamma<1$. Transverse and onion states, also considered, are stable only in the case of a transverse applied field. Note that the curling state is more favorable in tubes than in wires for a given $L/R$ ratio, due to the $1/R^2$ dependence of exchange energy, avoided towards the axis in the tube case. The transition between these two states has been confirmed experimentally by \citet{bib-WYS2017}.

\begin{figure}[thbp]
\centering\includegraphics[width=127.856mm]{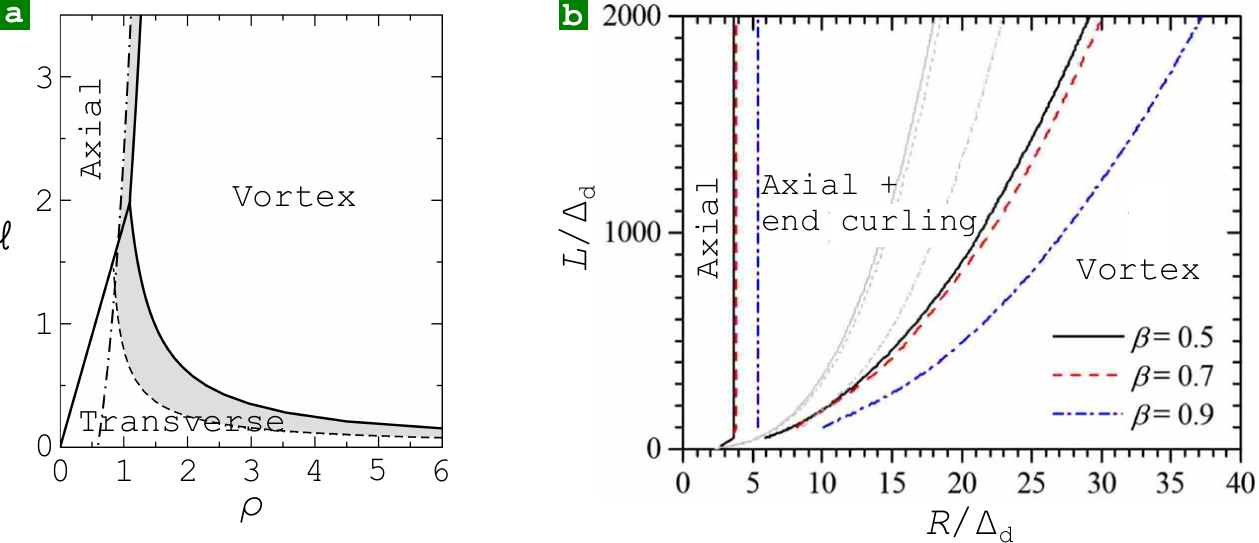}
\caption{\textbf{Analytical phase diagrams in cylinders and tubes} States of lowest energy in short cylinders~(\ie, thick disks): vortex~(curling around the axis), uniform transverse magnetization, uniform axial magnetization. $\rho=\sqrt{4\pi}R/\DipolarExchangeLength$ and $\ell=\sqrt{4\pi}L/\DipolarExchangeLength$, with $R$ and $L$ the cylinder radius and length, respectively. Full lines are iso-energy lines. The dashed line shows the boundary beyond which the vortex state is unstable. The dash-dotted line shows the size of the vortex core. Adapted from \citet{bib-MET2002}, copyright 2002, with permission from Elsevier. (b)~States of lowest energy for tubes of various wall parameter $\beta=r/R$: uniform axial magnetization, essentially axial magnetization plus end curling, global curling (vortex state). The thinnest lines stand for the transition directly from the uniform axial to global vortex state, such as considered by \citet{bib-SUN2014}. Reprinted with permission from \citet{bib-LAN2009}. Copyright 2009 by the American Physical Society.}\label{img_phase-diagrams}
\end{figure}

The curling state in tubes, although non-uniform, may be called global because it can be described by a single degree of freedom: the uniform $m_\rho=1$. However, a state of lower energy may be achieved by releasing the constraint of translational invariance. The object may remain largely uniformly axially magnetized, with rotation of magnetization occurring only close to the ends, to spread part or all of the end charges inside the volume in the form of volume charges. This allows to significantly reduce magnetostatic energy, occurring only at the object ends, while not paying a cost in exchange energy proportional to the full object length. This feature is widely known in micromagnetism as an end state\citep{bib-HUB1998b}. In the phase diagrams mentioned above, this increases the domain of stability of the mostly-uniformly-magnetized axial state, as calculated by \citet{bib-LAN2009}~\bracketsubfigref{img_phase-diagrams}b. As a rule of thumb, note that for $\beta<0.7$ the lines are nearly superimposed, so that these are expected to be also largely valid for wires. Also, it is confirmed that the consideration of end domains extends the range of stability of the largely axial state. Finally, it is below a diameter around $7\DipolarExchangeLength$ that no end curling occurs, which is consistent with micromagnetic simulations\citep{bib-FRU2015b}.

Which kind of non-uniformity occurs in the end domains depends on the geometry of the system. This was first simulated in wires by \citet{bib-ARR1979b}, showing that curling occurs for $R$ significantly larger than $\DipolarExchangeLength$. Thus, these end domains have been named \textsl{localized curling} by some\citep{bib-ZEN2002}. Later, simulations revealed that curling remains relevant down to $2R\approx7\DipolarExchangeLength$ for wires\bracketsubfigref{img_curling-and-C}c, below which the exchange energy becomes too costly, and rotation in the end domain occurs rather uniformly across the cross-section\citep{bib-HIN2000,bib-HER2002a}\bracketsubfigref{img_curling-and-C}{a-b}. In the latter case, end domains may occur only close to the coercive field, not at remanence, for low diameter. This case may be called a transverse end domain. Simulations and models have been extended later to tubes\citep{bib-ESC2007c,bib-LAN2009}. Trends are similar, except that transverse end domains are less favorable, due to the dipolar field they induce inside the hollow core. In all cases, the length of the end domains is expected to increase linearly with the diameter for $2R\simeq5-10\DipolarExchangeLength$\citep{bib-USO2006}, and quadratically for larger diameter\citep{bib-FRU2015c}. Experiments confirmed the existence of end domains for wires, through the decrease of remanence for increasing diameter\citep{bib-CHI2002}, reduced remanence at the surface probed by magneto-optical Kerr effect and compared with global magnetometry\citep{bib-WAN2008a}, quantitative modeling of MFM contrast\citep{bib-VOC2014}. End domains were observed directly by shadow-XMCD-PEEM\citep{bib-FRU2015c} and electron holography\citep{bib-CAN2015} in wires and in tubes\citep{bib-WYS2017,bib-FRU2017e}. The variation of length of these end domain under magnetic field, prelude to nucleation and magnetization switching, has been imaged by MFM\citep{bib-FRU2018b}. We will see in \secref{sec-MagnetismDWsNucleation} that the type of end domain determines the type of domain wall entering the object upon nucleation.

\begin{figure}[thbp]
\centering\includegraphics[width=128mm]{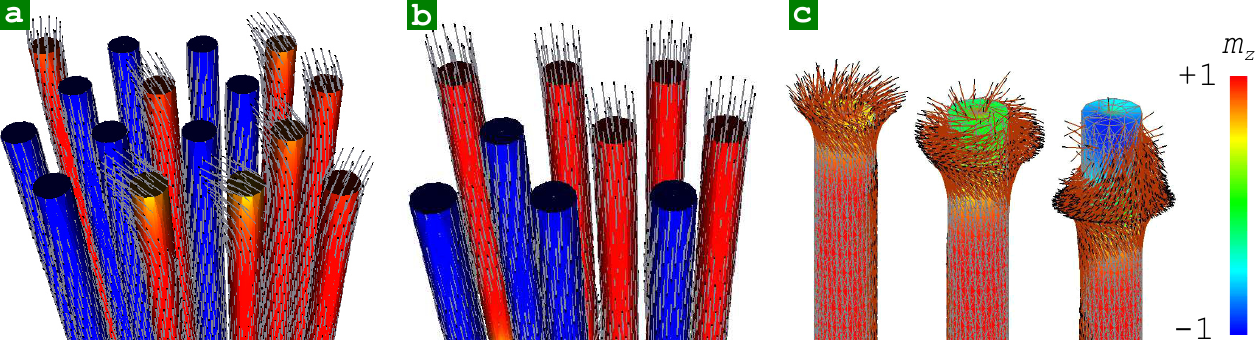}
\caption{\textbf{End states in wires}. (a)~zero-field  and (b)~near-coercive field states of Ni wires with diameter $\thicknm{40}$. Reprinted from \citet{bib-HER2001}, with the permission of AIP Publishing (c)~From a curling end domain to the nucleation-propagation process of a BPW, in a Ni wire with diameter \thicknm{60}.
The color codes the axial component of magnetization. Adapted from \citet{bib-HER2002a}, copyright 2002, with permission from Elsevier. We thank R.~Hertel for sharing the original and color versions of all figures.}\label{img_curling-and-C}
\end{figure}

A fine point about end curling is the following. It has been claimed rather early in tubes\citep{bib-WAN2005c}, and since then both in wires and tubes\citep{bib-LEE2007d,bib-BET2008,bib-CHE2011b}, that curling at opposite ends is lower in energy with an opposite winding, on the basis of micromagnetic simulations. At first sight this seems intuitive, as antiparallel alignment is of lower energy for moments along parallel lines. However, the rigorous calculation of dipolar energy shows that the two states are degenerate. Indeed, both magnetic charges $\rho$ and magnetic potential $\phi$ are identical for opposite windings, so that the dipolar energy is identical, as it volume density $-\rho\phi$ is the same. Even worse, for short wires the two curling zones are slightly coupled by exchange, so that parallel curling is of slightly lower energy, as shown early by \citet{bib-USO2006}. \citet{bib-WYS2017} showed recently that the long-standing discrepancy betweeen these statements comes from the difference of energy in simulations as an artifact related to the finite-size of the discretization mesh, probably inducing some artificial stray field, while a smooth surface wouldn't. So, the difference asymptotically vanishes for a finer mesh. Note that thie argument does not hold for non-circular cross-sections, so that in these, opposite windings may be the ground state. Still, it remains that opposite windings tend to occur in all cases in micromagnetic simulation because of the chirality of the Landau-Lifshitz equation. Indeed, this sets initially magnetization into motion along opposite azimuths, under the effect of the opposite radial components and thus torques of demagnetizing fields at both ends of the object\citep{bib-USO2006}.

\subsubsection{Anisotropic magnetic materials}
\label{sec-MagnetismSingleAnisotropic}

In the previous section, only dipolar and exchange energy determined the magnetic state of a wire. The addition of a local anisotropy contribution of magnetocrystalline or magnetoelastic origin gives rise to a larger variety of other magnetic states.

\begin{figure}[thbp]
\centering\includegraphics[width=128mm]{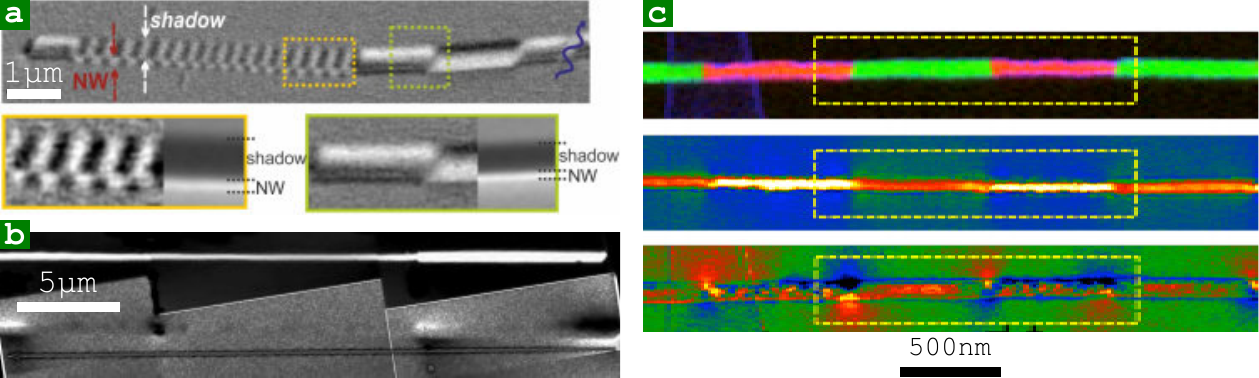}
\caption{\textbf{Magnetization state of various types of wires}. (a)~XMCD-PEEM image of a $\mathrm{Co}_{65}\mathrm{Ni}_{35}$ wire with diameter \thicknm{120}. The magnetization state comprises both regions with stripe domains of regular period~(bottom-left blow-up), and a global curling state~(bottom-right blow-up). Reprinted with permission from \citet{bib-BRA2017}. Copyright 2017 by the American Physical Society. (b)~AFM~(top) and MFM~(bottom) images of a uniformly-magnetized $\mathrm{Co}_{40}\mathrm{Ni}_{60}$ wire, consisting of three segments of diameters $\SI{200/150/200}{\nano\meter}$. Contrast due to the stray field arises both at the wire ends, at the modulations of diameter, and at domain walls. Sample courtesy S.~Bochmann, J.~Bachmann. (c)~Transmission Electron Microscopy image of a segmented [Co/Ni]$_n$ wire, with chemical contrast~(top, Ni appearing with three segments of lighter color, green in the electronic version), and axial~(middle) and transverse~(bottom) components of magnetic induction, using the DPC (differential phase contrast) method in the Foucault mode. Reprinted with permission from \citet{bib-IVA2016b}. Copyright 2016 American Chemical Society.}\label{img_magnetizationStateOfWires}
\end{figure}

A polycrystalline material with an anisotropic texture of the grains does not display new directions of magnetization. However, it can be affected by local distributions of anisotropy. The macroscopic state may then still be largely uniformly-magnetized along the axis, however with a smaller remanence even for single objects, due to a spread of local directions of magnetization. The impact will be larger for a material with significant magnetocrystalline anisotropy, such as hcp Co. This has been reported for, \eg,  Co-rich CoNi alloys\citep{bib-SER2016}. ALD magnetic materials are rather granular, so that tubes synthesized this way usually display rounded hysteresis loops\citep{bib-DAU2007}, reflecting a rather irregular pattern of domains at the microscopic scale\citep{bib-KIM2011b}. The size of the grains, especially compared with the wire diameter or tube radius\citep{bib-SKO2000b,bib-SAM2018}, is expected to have have a different impact because averaging in the random anisotropy model depends on dimensionality\citep{bib-HER1992}. Materials with an isotropic response however with a high microscopic magnetic anisotropy may be obtained this way, \eg based on CoPt or FePt\citep{bib-DAH2006}.

In the case of electroplated wires\bracketsecref{sec-fabricationDepositionElectroplating}, some synthesis conditions may yield a texture of the grains. The associated magnetocrystalline anisotropy may best be probed by ferromagnetic resonance\bracketsecref{sec-MagnetismFMRAndMagnonics}. If this anisotropy favors the axial direction, then the dipolar shape effect is reinforced and the behavior described for soft magnetic materials remains largely valid\citep{bib-HEN2001}. An increase of coercivity and possibly remanence may be expected. If the resulting anisotropy rather favors directions transverse to the axis, then radically new distributions of magnetization can be observed. First, magnetization may point in a direction transverse to a wire axis, and form large domains. The situation is then analogous to rather thick films of magnetic materials with perpendicular anisotropy, where magnetocrystalline anisotropy and demagnetizing effects compete. This, measured by the quality factor $Q=\Ku/\Kd$ with $\Ku$ being uniaxial anisotropy coefficient, may lead to the formation of so-called stripe domains: this is domains with a well-defined period and alternating up and down directions of magnetization\citep{bib-KOO1960,bib-MUR1966,bib-MUR1967}. Here the dimensionality is different, however the same result may occur. This way of partly closing the flux often competes with the existence of global curling states~(also called vortex states), in which magnetization curls around the wire axis. These have been reported for, \eg, CoNi alloys\citep[\subfigref{img_magnetizationStateOfWires}a]{bib-BRA2017}, single-crystalline hcp Co\citep{bib-VIL2009,bib-IVA2013,bib-CHE2016b}, single-crystal CVD Ni wire\citep{bib-KAN2018}. Magnetic anisotropy may also come from strain, for wires embedded in a solid matrix. This may be the case in an alumina matrix\citep{bib-LEI2011b}, or in the case of epitaxial strain for wires arising from spinodal decomposition\citep{bib-SCH2015b,bib-SCH2016}.

Tubes may give rise to another specific situation, that of azimuthal easy direction of magnetization. This is thought to result from the interplay of inverse magnetostriction with the uniaxial curvature. It has been observed in rolled sheets\citep{bib-STR2014} as well as in electroless tubes\citep{bib-FRU2017e} and angle-coated-wire tubes\citep{bib-ZIM2018}, with typical diameters a few micrometers and a few hundreds of nanometers, respectively. In both cases, the change to a non-magnetostrictive material, or one with an opposite sign of magnetoelastic coefficient, drives magnetization parallel to the tube axis. The situation is reminiscent of the case of glass-coated microwires, where strain plays a key role in determining an azimuthal anisotropy~\cite[chap. 7, 8, 12, 13, 17, and 19]{bib-VAZ2015}. These materials are interesting experimentally, as azimuthal magnetization is the situation in which the non-reciprocity of spin wave propagation has been predicted\citep{bib-OTA2016}, see \secref{sec-MagnetismSpinWaves}

\subsubsection{Modulated and 3D structures}
\label{sec-MagnetismSingleEngineered}

The first type of wire engineering, is diameter modulation. In the case of axial magnetization, the change of cross-section induces a local density of magnetic charges, demagnetizing and stray fields. The sign of the charges is positive~(resp. negative) if the diameter decreases~(resp. increases) along the direction of magnetization\bracketsubfigref{img_magnetizationStateOfWires}b. This effect is evidenced clearly with MFM for single modulation\citep{bib-PIT2011,bib-IGL2015, bib-FRU2016c}, constriction\citep{bib-CHA2012}, protrusions\citep{bib-BER2016}. The charges are shared between surface and volume type, and may induce curling at the transition from the thick to the thin part, similar to a wire end.

The second type of a modulated wire, is composition modulation, or in other terms: segments along the axis. The case of long segments of soft magnetic material is simple: magnetization remains essentially axial, with possibly curling at the transition between segments. This is valid for either two types of magnetic materials\citep[\subfigref{img_magnetizationStateOfWires}c]{bib-IVA2016b}, or one magnetic and the other one non-magnetic\citep{bib-GRU2017,bib-BRA2018}. In case the segments display a transverse shape or magnetocristalline anisotropy, each may display a type of domain on its own, \eg with alternation of longitudinal and curling domains\citep{bib-BER2017}. More complex cases can arise with the interplay of transverse anisotropy of either magnetocrystalline or shape origin, and give rise to metastable states involving transverse, axial or curling magnetization\citep{bib-BEL1999,bib-REY2016}. A general phase diagram has been proposed by \citet{bib-LEI2009}. Proposals have been made to use well-defined sequences of exchange-coupled layers for propagating information along vertical wires in a ratchet style, however so far this has been demonstrated on PVD films\citep{bib-LAV2013}.

Finally, we are witnessing the emergence of three-dimensional structures made of segments of straight or bent wires, synthesized by FEBID or two-photon lithography\bracketsecref{sec-fabricationEngineered}. The current status is the development of methodologies to monitor magnetism at the nanoscale in three dimensions. Demonstrations have been made of local hysteresis loops using focused MOKE\citep{bib-FER2013} or micro-Hall probes\citep{bib-MAM2018}. More complex structures made by two-photon lithography and coated with magnetic material raise the issue of branching between different segments, and call for advanced microscopy. Tomography is being developed on such structures. The first demonstrations were made at the rather unconventional 3d K edge, allowing a longer penetration depth of X-rays\citep{bib-DON2015}. The large focus depth of SEMPA proved to be valuable, however in that case some parts of the sample are shaded\citep{bib-WIL2017b}.

\subsection{Magnetic domain walls}
\label{sec-MagnetismDWs}

In the previous section we discussed the types of domains existing in wires and tubes. Here we focus on walls separating such domains. Besides describing their inner magnetization texture, the interest in these lies in the possibility to move them under the stimulus of a magnetic field or spin-polarized current. This involves fundamental aspects of magnetization dynamics, however lies in the background of possible applications in information and communication technologies. They have been addressed extensively by theory and simulation since pioneering works around 2000, see two excellent reviews by \citet{bib-THI2006,bib-THI2008}. Experiments are emerging since a couple of years. Note that domain walls and their dynamics happen to be qualitatively similar in wires and tubes. Accordingly, we discuss both cases together in the following section.

\subsubsection{Types of domain walls}
\label{sec-MagnetismDWsType}

The direction of magnetization will be monitored in spherical coordinates, $z$ being the wire/tube axis, and $\varphi$ referring to an azimuth. We first consider soft magnetic materials. In these the domains are axially-magnetized, so that domain walls are intrinsically charged, with for wires, $Q=\pm2\pi R^2$ in the head-to-head and tail-to-tail cases. There exists two types of domains for these, depicted on  \figref{img_tvw-bpw} and described below.

We call the first one the transverse-vortex wall~(TVW), for reasons explained below. It was first described independently by \citet{bib-NIE2002,bib-FOR2002} in wires. Most authors name this wall a transverse wall, as its first feature is a component of magnetization transverse to the wire/tube axis~(see \subfigref{img_tvw-bpw}e for tubes). Its simplest description is thus a one-dimensional model, with $\theta(z)$ ranging from 0 to~$\pi$. For wires with diameter larger than $\approx7\DipolarExchangeLength$ the wall is no more one-dimensional, however magnetization curls both around the transverse component and along the wire/tube axis\bracketsubfigref{img_tvw-bpw}{a,c}. This allows to decrease the magnetostatic energy of the wall. These dual features of transverse component and curling structure motivate the use of the name transverse-vortex wall, which is further described for wires elsewhere\citep{bib-FRU2015b}. The combination of transverse and vortex features does not seem to have been document to tubes yet.
\begin{figure}[thbp]
\centering\includegraphics[width=128mm]{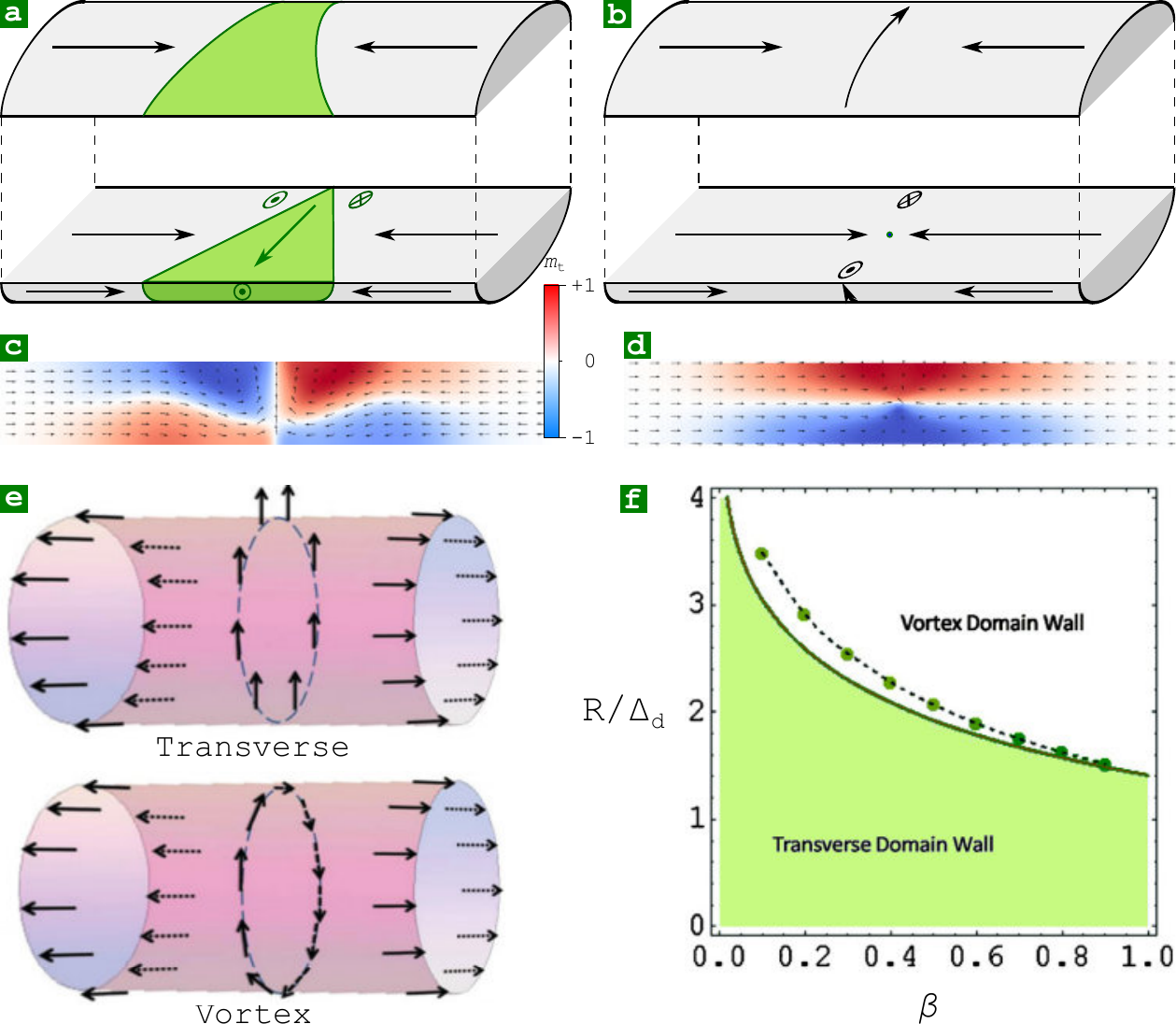}
\caption{\textbf{Transverse-vortex and Bloch-point/vortex walls in wires and tubes} Sketches for (a)~the transverse-vortex wall~(TVW) and (b)~the Bloch-point wall~(BPW) in a wire. The wire is split along the length for clarity. Reprinted with permission from \citet{bib-FRU2014}. Copyright 2014 by the American Physical Society. Cross-section of micromagnetic simulations of $\thicknm{80}$-diameter permalloy wires, for (c)~the TVW and (d)~the BPW. The arrows display the direction of magnetization, while color stands for the component of magnetization perpendicular to the cross-section. Adapted from \citet{bib-FRU2015b}, copyright 2015, with permission from Elsevier\finalize{copyright Woodhead, \citet{bib-FRU2015b}}. (e)~Sketches for the transverse wall and vortex walls in tubes, counterparts of (a) and (b). (f)~Phase diagram of domain walls in tubes versus tube wall thickness $\beta=r/R$ and normalized tube outer radius. Adapted from \citet{bib-LAN2010}, with the permission from AIP publishing.}\label{img_tvw-bpw}
\end{figure}

At this stage it is useful to describe the TVW using topology, and highlight its generality. The transverse feature means that there exist two areas at the wire/tube outer surface, such as $\vect m\dotproduct\vect n=\pm1$. The curling features arise by a continuous transformation of this transverse feature, so that its topology is unchanged, and remains valid for both circular and square cross-section. This topology also applies to the well-known transverse wall~(TW) and vortex wall~(VW) in thin flat strips. For the TW the transverse component lies in-the-plane, while for the VW it lies perpendicular-to-the-plane, consisting of the vortex core. Curling is largely developed for the latter, however not for the former, unless the strip is thick.

We call the second wall in wires the Bloch-point wall~(BPW). It has been described first by \citet{bib-FOR2002,bib-HER2002a}. Its main feature is curling around the wire axis and its invariance upon azimuthal rotation\bracketsubfigref{img_tvw-bpw}{b,d}, so that it may be described by a one-dimensional model for $m_\varphi$. The radial component $m_\rho$ is weak, magnetization being only slightly tilted outwards~(resp. inwards) by the head-to-head positive~(resp. tail-to-tail negative) charge of the wall. Owing to these boundary conditions, magnetization may be neither axial, nor in any transverse direction at the very core of the wall, so that it must lie undefined. This peculiarity is a Bloch point~(BP)\citep{bib-FEL1965}, the only singularity predicted in micromagnetism\citep{bib-BRA2012}. This point gives its name to the wall, as coined by \citep{bib-THI2006}, and used by others\citep{bib-KIM2013}, besides us. It is a common misconception that the BPW gives rise to no stray field, related to the azimuthal flux closure and close-to-absence of surface charges, and thus should not be visible with MFM\citep{bib-IVA2013}. The head-to-head total charge $2\pi R^2\Ms$ must be conserved; it is only spread along the wall in the form of volume charges $-\linepartial{m_z}{z}$. Another misconception is that the BPW is chiral. It is not, a plane perpendicular to the wire axis is a plane of symmetry. Both circulations of the wall are equivalent and degenerate in energy.

A key feature of the BPW is that there is no point at the wire surface with $\vect m\dotproduct\vect n=\pm1$, so that it is clear that its topology is distinct from the one of the TVW. The BPW is called by some a vortex wall, because of the curling around the axis. However, given its topological inequivalence with the VW in strips, we consider that it can be ambiguous to name it this way. The singularity for the vector field $\vect m$ is a serious issue for micromagnetism, which by essence is a continuous theory. Its artificial interaction and thus pinning with the mesh in simulations is known\citep{bib-THI2003}. Multiscale codes describing down to atomic moments have been proposed to tackle this\citep{bib-LEB2012,bib-HAN2014,bib-LEB2014,bib-EVA2014}. However, while this is clear progress, they consider Heisenberg spins, so that the constraint of uniform magnetization is not fully lifted, or consider an effective penalty for exchange through LLG-Bloch equations\citep{bib-GAR1997b,bib-GAL1993}. Modeling the BP within band magnetism remains an open challenge, which may require combining micromagnetism with density functional theory.

The first experimental observation of the TVW was reported in Ni wires by \citet{bib-BIZ2013} using electron holography, soon followed by \citet{bib-FRU2014} reporting both the TVW and the BPW in permalloy wires, using shadow XMCD-PEEM. The first report confirmed the transition from the symmetric TW~(one-dimensional) to the curling one~(TVW), comparing wire diameters $\SI{55}{\nano\meter}$ and $\SI{85}{\nano\meter}$. Both confirmed that the TVW, although metastable above diameter $7\DipolarExchangeLength$, are found in practice, possibly alongside BPWs.

When the outer boundary conditions of the BPW are applied to a tube, one obtains a DW fundamentally distinct from the TVW, because of different topology. This wall is often call a vortex wall for tubes. It is similar to the BPW, except that there is no BP, due to the hollow nature of the tube\bracketsubfigref{img_tvw-bpw}e. Again, care must be taken to avoid confusion with other vortex walls.

The energetics of the BPW~(or vortex wall for tubes) versus the TVW have been simulated, leading to phase diagrams for square\citep{bib-THI2006}, circular and arbitrary rectangular\citep{bib-FRU2015b} cross-sections. For circular and square wire cross-section, the BPW is of lower energy than the TVW wall diameter~(or square side) higher than $7\DipolarExchangeLength$. The latter report shows that BPWs exist in a wide range of geometries, not only for circular or square cross-section. Note that, similar to first-order transitions, these two states may exist for a given geometry. One has the lowest energy and the other one is metastable, however in practice does not transform spontaneously because of the energy barriers. Also, the BPW does not exist for an arbitrarily narrow wire, because the associated energy would be too large\bracketeqnref{eqn-exchangeCylindricalAzimuthal}. In tubes, the phase space in which the TVW is of lowest energy versus the vortex wall, is reduced, owing to the associated stray field in the hollow part in the case of transverse magnetization. The TVW is notably unfavored for increasing $\beta$, \ie, for thin tube wall. \citet{bib-LAN2010} derived the following approached formula for the boundary:

\begin{equation}%
\label{eqn-wallsEnergyTubes}
   \rho=2\sqrt{\frac{\ln(1/\beta)}{1-\beta^2}},
 \end{equation}
where $\rho=R/\DipolarExchangeLength$. \subfigref{img_tvw-bpw}f shows this boundary, based on this formula and compared with micromagnetic simulations. \citet{bib-LOP2012} discuss the impact of axial and cubic anisotropy on the phase boundary.

The case of walls for azimuthal magnetization in tubes has been considered only rarely in simulations. In the case of a soft magnetic material, it was considered in short tubes displaying end curling of opposite sign\citep{bib-LEE2007,bib-BET2008}. It was considered, more generally, for tubes displaying azimuthal domains, such as rolled sheets\citep{bib-STR2014}, electroless tubes\citep{bib-FRU2017e} or some angle-coated-wire tubes\citep{bib-ZIM2018}. We name the walls after their classification in thin films, considering the surface of the tube as an unrolled sheet. The latter considered cases of N\'{e}el walls~(axial core) and Bloch walls~(radial core) and discussed their energetics, while all former reported the occurrence of cross-tie walls\citep{bib-HUB1958,bib-MID1963}, however not discussing whether or not they are the ground state. DWs between azimuthal domains in tubes have been described in various works\bracketfigref{img_DW_types_for_orthorad_domains-schemes}, however, a phase diagram such as the one on \subfigref{img_tvw-bpw}f is still lacking. Experimentally, N\'{e}el walls were claimed by \citet{bib-WYS2017}, while Bloch walls were observed by us\citep{bib-FRU2017e}. There is no necessary contradiction, as the materials and geometries are different, which could result in two different ground states fro the walls. So far, there is no experimental report of cross-tie wall.

\begin{figure}[htbp]
\centering
    \includegraphics[width=128mm]{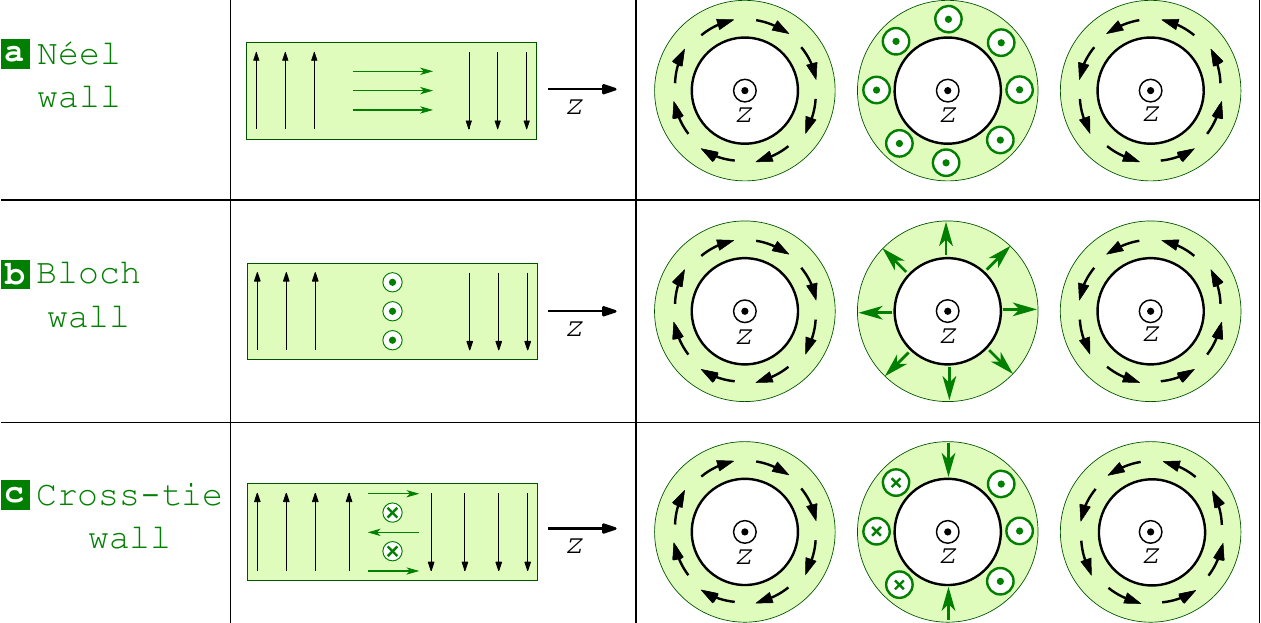}
\caption[Sketch of domain walls in nanotubes with azimuthal domains]{\textbf{Sketch of domain walls in nanotubes with azimuthal domains}. Central column: unrolled views of the tube shell. Right column: cross-sections of the tube shell. The tube axis lies along $z$. Adapted with permission from~\citep{MS_thesis}.}%
\label{img_DW_types_for_orthorad_domains-schemes}
\end{figure}

Domain walls are usually associated with a width $\delta=\pi\Delta_\mathrm{W}$ and an energy, which for wires and tubes may be expressed as a total energy~$\mathcal{E}$, or energy per unit surface $\sigma$ upon normalization with the cross-section. $\delta$ is predicted and simulated to equal asymptotically $\pi\sqrt{2A/\Kd}$ towards zero wire diameter, associated with the dipolar energy density $\Kd/2$ for transverse magnetization\citep{bib-THI2006,bib-FRU2015b}. For radius much larger than $\DipolarExchangeLength$, a simple analytical model balancing exchange and dipolar energy predicts that $\delta\sim R^2/\DipolarExchangeLength$, which is confirmed by numerical simulations in wires with both square and circular cross-section, and equally true for both TVW and BPW\citep{bib-FRU2015b}. The same model predicts that $\mathcal{E}\sim AR^2/\DipolarExchangeLength$. For intermediate size, $\delta$ is expected to scale roughly linearly with~$R$, given the triangular shape of the transverse wall. An analytical model was proposed to refine the wall width in this regime\citep{bib-HER2015b}. Walls have been imaged by several groups and various microscopy techniques. Experimentally, the increase of wall width with wire diameter is clear\citep{bib-FRU2016c}, although it has not been studied quantitatively in a systematic manner. In tubes, analytical modelling predicts that the width of TVWs steadily decreases for increasing~$\beta$, while the width of BPWs reaches a maximum for a given $\beta$\citep{bib-LOP2012}. These authors also discuss the impact of axial and cubic anisotropy on the wall width.

\subsubsection{Nucleation of domain walls}
\label{sec-MagnetismDWsNucleation}

Magnetization reversal is a central topic in magnetism. In large and especially in bulk systems, the processes involved are complex. The model of coherent rotation of magnetization of \citet{bib-STO1948,bib-STO1991} highlights a link between coercivity and magnetic anisotropy, however fails to reproduce reality quantitatively. It is an experimental evidence that, even if a system is mostly uniformly-magnetized at rest, magnetization reversal does not proceed uniformly\citep{bib-FRU1999b}. Instead, it involves processes such as nucleation of small reversed domains, before their possible expansion through domain-wall motion. In bulk materials, these processes cannot be monitored in detail, so that they are described with a phenomenological theory\citep{bib-GIV2003}.

 The situation is much simpler in nanowires and nanotubes, which are mostly one-dimensional. Here, we restrict the discussion to wires and tubes, whose ground-state at remanence is axial magnetization. Letting aside the possible role of structural defects, a natural locus for nucleation is the end of the wire/tube. Indeed, it is the locus of magnetic charges\citep{bib-EBE2000}, so that the demagnetizing field is the highest, assisting the external field in rotating magnetization\citep{bib-BRA1999}. This phenomenon we already described in \secref{sec-MagnetismSingleSoft}, as the reason for the formation of end domains. Upon increase of the strength of magnetic field applied antiparallel to axial magnetization, these end domains expand and eventually give rise to the nucleation of a domain wall, whose propagation along the wire/tube means magnetization switching, thus determining coercivity. The energetics of this expected localized nucleation has been examined theoretically\citep{bib-SKO2000b}, and its relevance confirmed by viscosity measurements highlighting the volumes involved in nucleation events\citep{bib-ZEN2000,bib-VID2012}.

\begin{figure}[htbp]
\centering
    \includegraphics[width=128mm]{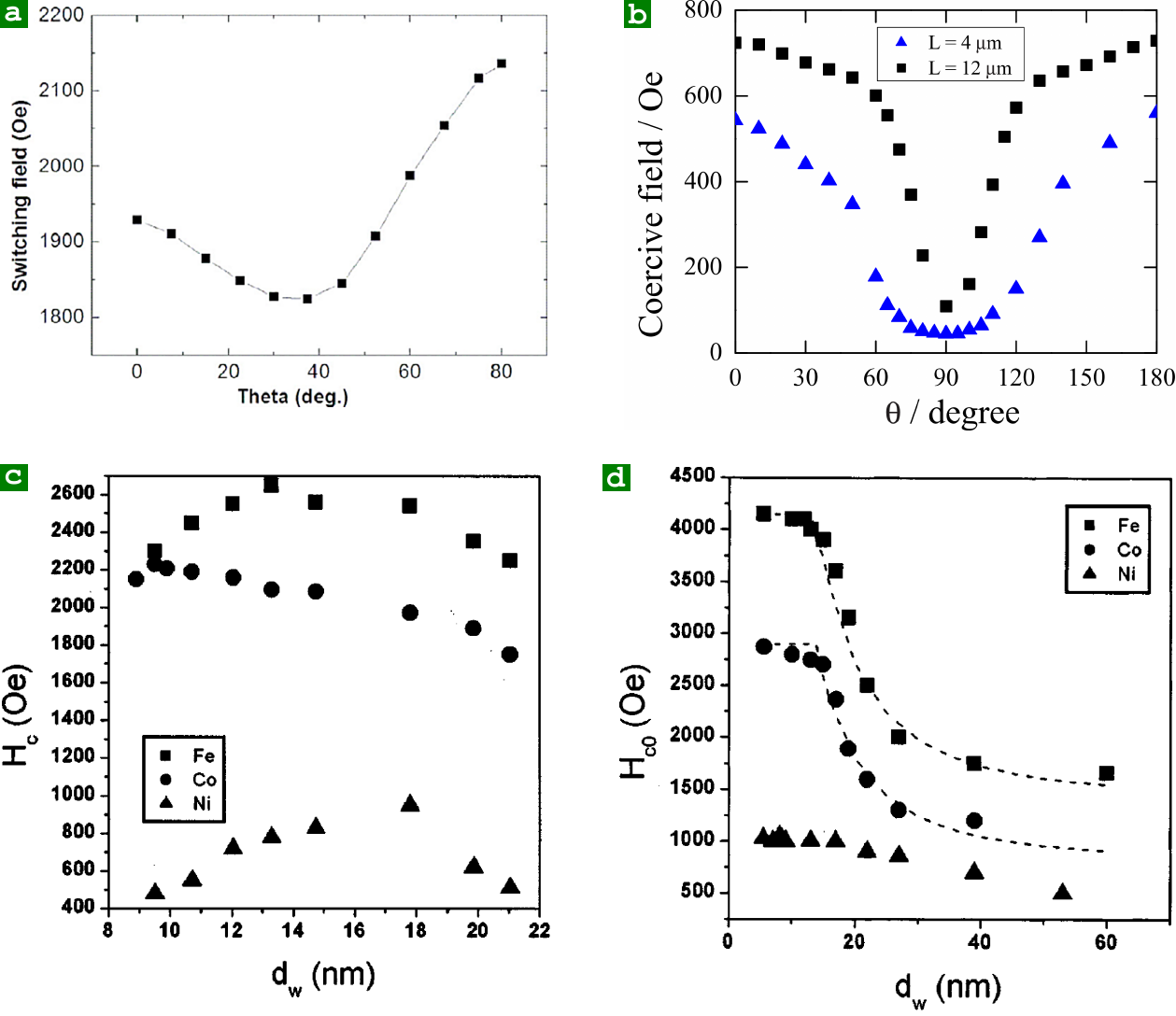}
\caption{\textbf{Experimental switching field in nanowires}. Angular variation of: (a)~the switching field, measured by counting individual switching events by MFM at the surface of an array of Co nanorods with diameter $\thicknm{30}$, pitch $\thicknm{100}$ and length $\thicknm{300}$\citet{bib-WAN2008a}. \copyright IOP Publishing.  Reproduced with permission.  All rights reserved. (b)~Angular variation of the coercive field, measured through global hysteresis loops of an array of Ni wires with diameter $\thicknm{50}$, pitch $\thicknm{100}$, and two lengths as indicated. In both cases, $\vect H$ is parallel to the axis for $\theta=\angledeg{0}$. Reprinted from \citet{bib-LAV2009}, with the permission of AIP Publishing. (c)~Coercive field versus wire diameter for Fe, Co and Ni wires, measured at room temperature, for magnetic field applied along the axis. (d)~the same over a different range of diameters, measured at low temperature~($\tempK4$). The dashed mine is a phenomenological fit with a $1/R^2$ laa and a plateau cut-off. Reprinted with permission from \citet{bib-ZEN2002}. Copyright 2002 by the American Physical Society.}%
\label{img_switching-field}
\end{figure}

We mentioned earlier that end domains can be of either transverse or curling type\bracketfigref{img_curling-and-C}. Theory\citep{bib-BRA1999} and simulation\citep{bib-FER1997b,bib-FOR2002,bib-HER2002} predict that, through the preservation of the symmetry of this non-uniform texture of magnetization, a transverse end domain leads to the nucleation of a TVW, while a curling end domain leads to the nucleation of a BPW~(or vortex wall in the case of a tube)\citep{bib-LAN2007,bib-USO2008}. In analytical theories of magnetization reversal such as Stoner-Wohlfarth and curling\citep{bib-FRE1957}, the effect of tilting the applied field with respect to the easy axis, is considered. Processes such as coherent rotation, curling, or propagation of an existing domain wall\citep{bib-BEC1932,bib-KON1937}, all have a specific signature in the angular variation of the switching field. Nucleation at wires and tubes ends as well, has been investigated by theory and simulation as a function of angle of applied field\citep{bib-ALL2008,bib-LAV2009}. They show that transverse nucleation, occurring at rather small radius, is analogous to the case of uniform rotation, being characterized by a sharp maximum of nucleation field along the axis, \ie, parallel to the wire/tube axis. To the contrary, curling nucleation, occurring at large radius, is characterized by a soft plateau or even a shallow minimum of nucleation field along the axis\citep{bib-LAN2007}. In the cross-over regime of radius, axial field tends to favor curling nucleation, while transverse field tends to favor transverse nucleation. In tubes with small radius, consistent with the energetics of TVW versus vortex walls, thinner tube walls favor curling nucleation, while thicker walls favor transverse nucleation\citep{bib-ESC2008}. As a consequence, the measurement of the angular variation of nucleation field has been used to infer which end domain occurs, and which wall type is nucleated, without using magnetic microscopy. A pioneering experimental work was angular measurements of a single Ni nanowire using the micro-SQUID technique\citep{bib-WER1996}. A decade later, following the development of theories for the angular variation of nucleation at the end of wires and tubes, several groups used measurements over arrays to guess the nucleation mode in individual wires\citep{bib-LAV2009}, evidencing the transition from transverse to curling nucleation as a function of diameter\citep{bib-GHA2011}, wall thickness\citep{bib-ESC2008}, or angle of applied field\citep{bib-LAV2012}. A fine point is the following. Close to the transverse direction, one should make the difference between the nucleation field\bracketsubfigref{img_switching-field}a and the coercive field\bracketsubfigref{img_switching-field}a, whose meaning and mean of measurement sharply differ\citep{bib-STO1948}: the switching field is an irreversible jump of magnetization, a concept relevant for single objects and measured my monitoring these individually; the coercive field is defined \apriori on any hysteresis loop, as the field for which $\vect M\dotproduct\vect H=0$. This is a material-science and rather global quantity, in the sense that there may be no special event associated with the coercive field in a single particle. Most experimental reports concern coercivity, which is a macroscopically-measurable quantity, while only a few investigations are able to measure the nucleation field close to the hard axis\citep{bib-WER1996,bib-WAN2008a,bib-WAN2009}, which is a single-object quantity. Note, however, that nucleation can be extracted from FORC measurements, which access irreversible events\citep{bib-ALI2016}. An extensive review of experimental results was made by \citet{bib-IVA2013b}.

 Another variation of interest is the nucleation field versus the diameter, and tube wall thickness for tubes. Models for estimating the length of end domains have been made\citep{bib-ZHU2002b,bib-USO2008}, and magnetic microscopy confirmed the variation of end domains with applied field\citep{bib-FRU2018b}. The variational solving of these models against the applied field, provides a value for the nucleation field, which is expected to decrease with the wire radius, typically like $1/R^2$. This is qualitatively similar to the curling theory, however, with an ad hoc coefficient. Note that this $1/R^2$ scaling law may be found by a simple model\citep{bib-FRU2018b}. As regards tubes, for very thin wall dipolar fields are negligible, so that nucleation is only hampered by exchange energy involved in curling\bracketeqnref{eqn-exchangeCylindricalAzimuthal}, predicting nucleation field varying again with $1/R^2$. This nucleation field increases with the wall thickness like $1/\beta$, arising from the integration of exchange over the tube thickness\citep{bib-USO2008}. \citet{bib-IVA2013b} provides a review of experimental nucleation fields for various materials and diameters. However, the above-mentioned theories do not consider thermal activation, which may play a role in the experiments. Its effect is expected to be more important for smaller activation volumes, so, for wires and tubes with smaller diameter and/or wall thickness. Consistently, \citet{bib-ZEN2002} showed that, at room temperature, the nucleation field decreases for diameters below typically $\SI{10-20}{\nano\meter}$\bracketsubfigref{img_switching-field}c. This is an effect of thermal activation and not of poor structure of low-diameter wires, because the nucleation field keeps increasing for these smaller diameters at low temperature\bracketsubfigref{img_switching-field}d. The latter measurement were well reproduced by a $1/R^2$, with a cut-off plateau at low diameter. This plateau, reached in a regime were the domain-wall width is constant and solely determined by the transverse cylindrical shape anisotropy $\Ms/2$\citep{bib-THI2006}, is roughly $\Ms/3$. It has not been investigated, whether this value is intrinsic and due to the absence of exchange at the wire end~(an end domain, compared with a regular domain wall}, or is smaller than $\Ms/2$ and determined by defects.

Other specific cases have been considered for nucleation. For instance, some have reported the impact of wire ends different from a perfect cylinder, such as with a dendritic shape\citep{bib-VID2015}. The impact of an Oersted field has also been considered to assist nucleation, as the curling mode has the same symmetry as the Oersted field\citep{bib-OTA2015}. Nucleation as a function of (short) wire length has also been investigated\citep{bib-FRU2017}, with application to the realization of a ratchet in multisegmented wires with an axial gradient of segment length\citep{bib-BRA2018}.

\subsubsection{Motion of domain walls}
\label{sec-MagnetismDWsPropagation}

The development of theory and simulation of DW motion in magnetic nanowires and nanotubes has been very active since fifteen years, through several tens of research articles. These now provide a comprehensive view of the phenomena to expect, many specific to cylindrical structures. Experimental realizations involving domain walls and their motion in wires and tubes are only emerging, and do not allow yet to support the theoretical claims. Accordingly, in the following we first draw the panorama of theoretical work, before switching to experiments.

We will consider DW wall motion under the stimulus of either an external magnetic field applied along the wire/tube axis, or a spin-polarized current. DW motion is intrinsically of precessional nature, following the Landau-Lifshitz-Gilbert equation, generalized with the so-called adiabatic and non-adiabatic spin torques\citep{bib-THI2008}:

\begin{equation}
  \label{eqn-precessionLLG}
  \diffdot{\vect m}=-\gamma_{0} \vect m\crossproduct \vect H + \alpha\vect m\crossproduct\diffdot{\vect m}-\left({\vect u\dotproduct \vectNabla}\right)\vect m
    +\beta \vect m\crossproduct\left[{(\vect u\dotproduct\vectNabla)\vect m}\right].
\end{equation}
$\gamma_{0}=\muZero|\gamma|$~($\gamma$ is the gyromagnetic ratio), $\alpha\ll1$ the damping parameter, $\beta\ll1$ the non-adiabatic coefficient, and $\vect u$ a velocity related to the spin-polarized current: $u=v_\mathrm{d} P (n_\mathrm{e}/n_\mathrm{s})$, with $v_\mathrm{d}$ the drift velocity, $P$ the spin polarization ratio of the current, $n_\mathrm{e}$ the volume density of conduction electrons, and $n_\mathrm{s}$ magnetization expressed in unit of volume density of Bohr magnetons. With these notations, $J=e\,n_\mathrm{e} v_\mathrm{d}$ is the charge current, with $e<0$.  So, we can also write: $u=P(J\muB/e\Ms)$. The first spin torque term expresses the transfer of angular momentum from the conduction electrons, to magnetization. This is the so-called Slonczewski term\citep{bib-SLO1996}. The second spin torque term acts like a field term expressing the exchange between magnetization and a non-adiabatic fraction of electrons with their direction of spin not yet aligned with magnetization. $\vect H$ is the total effective field, including external field, plus effective terms related to anisotropy, exchange and dipolar energy.

Although it is the most common one, this form of the Landau-Lifshitz equation is not practical to derive the time evolution of magnetization, which appears on both sides of the equation. Instead, we will make use of the so-called solved form:

\begin{equation}
  \label{eqn-precessionLLGsolved}
  \diffdot{\vect m}=-\frac{\gamma_{0}}{1+\alpha^2} \vect m\crossproduct \vect H - \frac{\alpha\gamma_0}{1+\alpha^2}\vect m\crossproduct\left({\vect m\crossproduct\vect H}\right) - \frac{1+\alpha\beta}{1+\alpha^2}\left({\vect u\dotproduct \vectNabla}\right)\vect m
    +\frac{\beta-\alpha}{1+\alpha^2} \vect m\crossproduct\left[{(\vect u\dotproduct\vectNabla)\vect m}\right].
\end{equation}
The solved form is strictly equivalent to \eqnref{eqn-precessionLLG}, however the prefactors of the spin torques are changed, and notably the second one. Thus, the physical meaning of non-adiabaticity is changed from one equation to the other, where $\beta-\alpha$ would be named the non-adiabatic coefficient in the solved form.

\begin{figure}[htbp]
\centering
    \includegraphics[width=128mm]{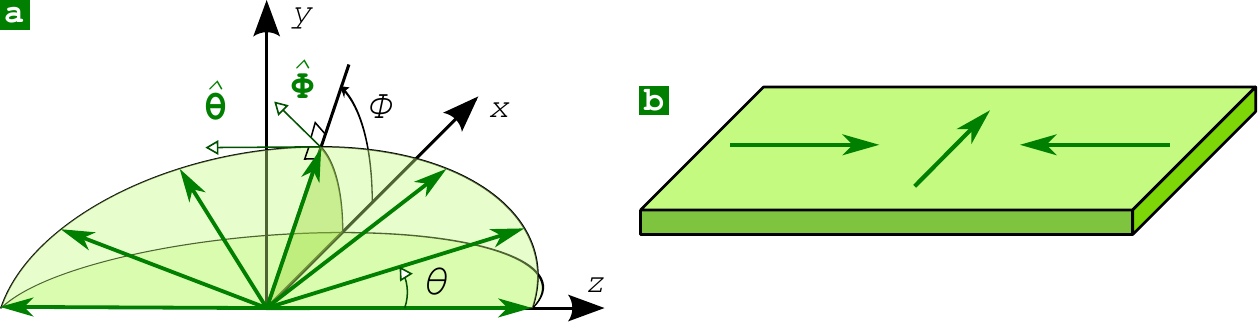}
\caption{\textbf{Notations for domain-wall motion}. (a)~$(\theta;\phi)$ coordinates to describe the magnetization direction inside a domain-wall, in the one-dimensional model. (b)~One-dimensional schematic for a head-to-head domain-wall.}%
\label{img_wall-motion-notations}
\end{figure}

We consider a head-to-head DW, with no loss of generality, as a tail-to-tail DW can be obtained by time-reversal symmetry. One can use the so-called one-dimensional model\citep{bib-MAL1979} to describe simply the physics, in which the domain wall is considered as a rigid spin texture characterized by a position $q$ and an angle $\phi$\bracketsubfigref{img_wall-motion-notations}{a,b}. Its relevance stems from the largely symmetric shape of DWs in wires and tubes, which allows one to reduce them to a single angular degree of freedom, besides their position. Micromagnetic simulation provides a more quantitative picture, if needed. The physics of the TVW and the BPW are drastically different, and need to be considered separately.

The situation of the TVW is simple and well documented in wires. $\phi$ shall be defined in the laboratory frame so that uniform $\phi$ across the axis implies uniform transverse magnetization, contrary to the general notation of \figref{img_cylindricalSphericalCoordinates}. The advantage of \eqnref{eqn-precessionLLGsolved} is that each term on the right hand side translates into either $\diffdot\phi$ or $\diffdot\theta$, so that the features of motion are solved at once. The situation under magnetic field was first described by numerical micromagnetics, independently by \citet{bib-FOR2002,bib-FOR2002b} and \citet{bib-NIE2002,bib-HER2002a}. A comprehensive review, both with simulations and analytics, is provided by \citet{bib-THI2006}. Under field $\Ha$, as $\alpha\ll1$, the leading effect is precession around the axis, with angular frequency $\diffdot{\phi}=|\gamma_0| \Ha$. To set ideas, $|\gamma/2\pi|=\SI{28}{\giga\hertz\per\tesla}$ for spin moments. So, were it just for this term, $\diffdot \theta=0$, so the DW would not move forward. Only the (weak) damping term contributes to wall motion, reading $\diffdot{\theta}=-\alpha\gamma_0 \Ha/(1+\alpha^2)$. The angular motion can be converted in speed through the wall profile $\linediff{\theta}{z}=1/\Delta$: $\diffdot{q}=\alpha\gamma_0 \Ha \Delta/(1+\alpha^2)$. To set ideas, for $\muZero \Ha=\SI{10}{\milli\tesla}$, $\Delta=\SI{50}{\nano\meter}$ and $\alpha=0.01$: $\diffdot q\approx\SI{1}{\meter\per\second}$. This is a very low standard for wall motion, against what is achieved in thin films\citep{bib-BEA2005,bib-THI2006}. The case of current-driven motion was investigated only later\citep{bib-YAN2010,bib-WIE2010}, although with no surprise as it was already very well understood by the 1d model\citep{bib-THI2007}. Only the third and fourth terms contribute in \eqnref{eqn-precessionLLGsolved}. The Slonczewski term contributes purely to forward motion, while the field-like term contributes purely to azimuthal rotation: $\diffdot q=u(1+\alpha\beta)/(1+\alpha^2)$ and $\diffdot\phi=(u/\Delta)(\beta-\alpha)/(1+\alpha^2)$. As $(\alpha,\beta)\ll1$, in the end one has: $\diffdot\phi\approx(\beta-\alpha)(u/\Delta)$ and $\diffdot q\approx u$. To set ideas, for $P=0.7$, $J=\SI{1e12}{\ampere\per\meter\squared}$, $\Ms=\SI{800}{\kilo\ampere\per\meter}$, one has: $u\approx\SI{50}{\meter\per\second}$. As an overview, the physics is much simpler than that of wall motion in thin flat strips, owing to the rotational invariance around the wire/tube axis. This yields an absence of inertia, and a mobility valid for all fields. It is the counterpart of the mobility in strips above the Walker field, characterized by azimuthal precession. So, one can consider that the Walker field of the TVW in wires exists and equals zero field. The phenomenology is largely unchanged in wires with square cross-section\citep{bib-THI2006}, except that the angular velocity varies periodically\citep{bib-USO2007}, reflecting the configurational anisotropy\citep{bib-COW1998c,bib-COW1998b} of the core of the TVW when either parallel to the edge or to the diagonal of the cross-section.

{\renewcommand{\arraystretch}{1.6}%
\begin{table}
  \capstart
  \begin{center}
    \caption{\label{tab-wallMotion}Angular and forward domain wall speed under magnetic field $\Ha$ and current~$u$, in the limit $(\alpha,\beta)\ll1$. Results for the BPW/vortex wall are below a possible Walker field.}
    \begin{tabular}{cccc}
      \hline\hline
      Wall type & Stimulus & $\diffdot{\phi}$ & Speed $\diffdot q$ \\
      \hline
      TVW & Field & $|\gamma_0| \Ha$  & $\alpha\gamma_0 \Ha \Delta$ \\
        & Current & $(\beta-\alpha)\frac{u}{\Delta}$ & $u$ \\
       BPW & Field & N.A. & $\approx\gamma_0 \Ha \Delta/\alpha$ \\
       & Current & N.A. &  $\approx u\beta/\alpha$ \\
      \hline\hline
    \end{tabular}
  \end{center}
\end{table}
}

\begin{figure}[htbp]
\centering
    \includegraphics[width=128mm]{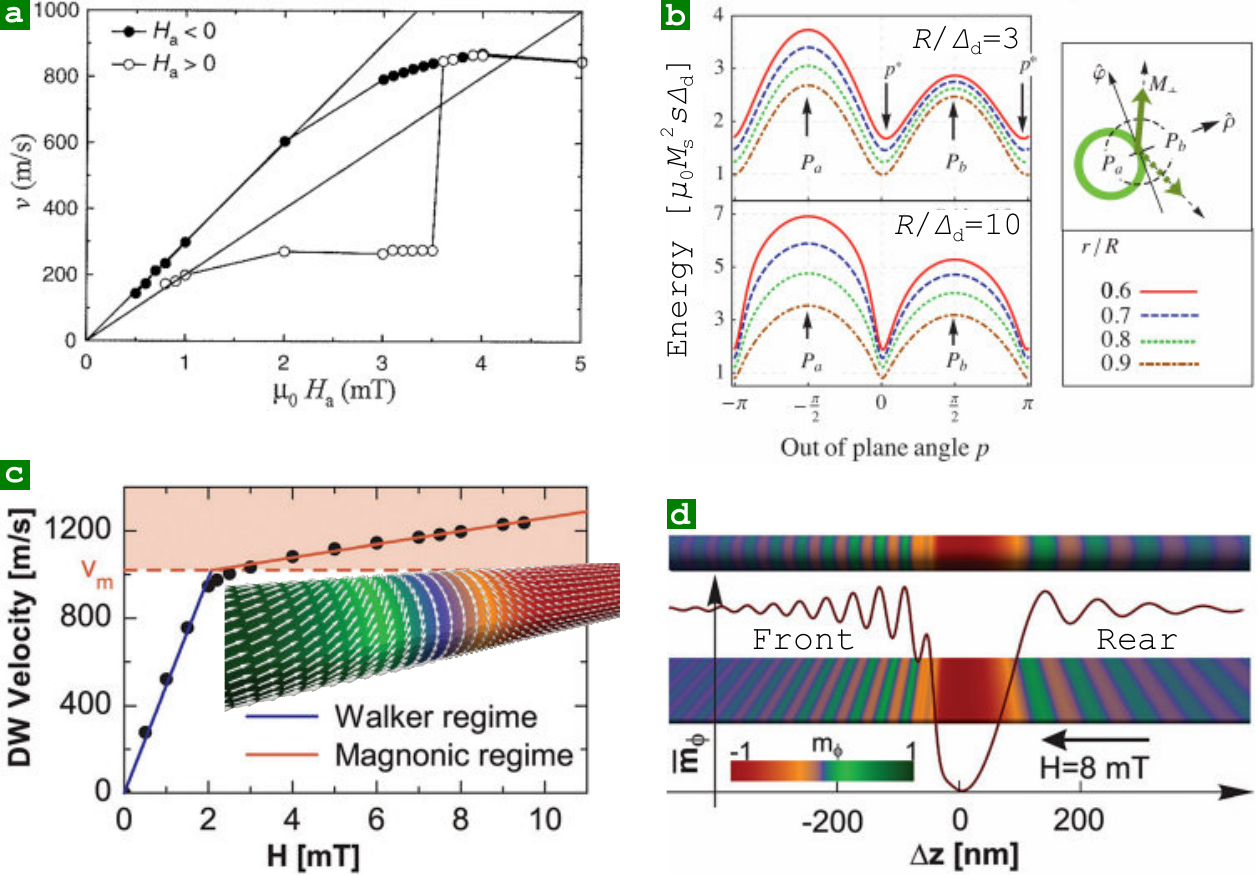}
\caption{\textbf{Simulations of domain-wall motion}. (a)~Simulated speed of a BPW in a permalloy wire with diameter $\thicknm{40}$, for a given initial circulation and depending on the sign of applied field. Reprinted by permission from \citet{bib-THI2006}, Copyright 2006. (b)~Analytical modelling of the energy of a vortex wall in a tube, frozen in a given magnetization azimuth $\phi$~(the authors used the notation: $p=\phi$). Reprinted from \citet{bib-OTA2013}, Copyright 2013, with permission from Elsevier. (c)~DW speed versus (negative) applied field for a head-to-head vortex wall right-handed with respect to the applied field, in a permalloy nanotube with $R=\thicknm{30}$ and $r=\thicknm{20}$. (d)~3D and unrolled views of the spin waves building up at the front and the rear of the DW, in the magnonic regime.  Reprinted from \citet{bib-YAN2011b}, with the permission of AIP Publishing.}%
\label{img_wall-motion}
\end{figure}

We now turn to the case of the BPW in wires, with its counterpart vortex wall in tubes. Pioneering works were made under applied field and in wires, independently by \citep{bib-FOR2002,bib-FOR2002b} and \citep{bib-HER2002a}. They pointed at the very high speed of BPWs, increasing with applied field, lower damping coefficient\citep{bib-FOR2002} and larger diameter\citep{bib-FOR2002b}. \citet{bib-WIE2004a} noticed that, over a large range, the mobility of the BPW scales like~$1/\alpha$. The reason for the high speed is the following. Magnetization is largely along the azimuthal direction in the central plane of the BPW. Precession around $\Ha$ induces rotation of magnetization around the axis, towards the radial direction. Thus, unlike the TVW, the BPW gets deformed, giving rise to a possibly large effective field of exchange and dipolar origin. The wall then moves under the stimulus of this internal field, analogous to the physics of precession-driven switching of nanomagnets\citep{bib-BAC1999}. This situation may be described by the 1d model, with $\phi$ describing the rotational invariant azimuthal angle of magnetization, such as introduced in \secref{sec-MagnetismCylindricalNotations}. The mention of this radial tilt of magnetization leads us to another effect, outlined in a second stage in wires\citep{bib-THI2006}, then in tubes with a similar phenomenology however also supported by analytical modelling\citep{bib-LAN2010,bib-YAN2012,bib-OTA2013}. At rest, magnetization in the central plane of the BPW is not purely azimuthal, due to the charge of the wall. It is tilted radially outwards for a head-to-head wall, and inwards for a tail-to-tail wall. So, depending on the direction of $\Ha$, precession will tend to reinforce or go against the radial tilt existing at rest. Internal fields are higher in the latter case, resulting in a higher mobility\bracketsubfigref{img_wall-motion}a. Internal fields are lower in the former case, resulting in a lower mobility.  This lifts the degeneracy between the two possible circulations of the BPW, which were equivalent at rest. In other words, the BPW gets a chiral character under dynamics, in relation with the direction of applied field, and intrinsic chirality of the LLG equation. This selection of circulation under dynamics is another manifestation of the curvature-induced chirality: the inner and outer surface of a tube are not equivalent, nor are the inward and outward radial directions in a wire, due to the inward or outward dipolar field arising from the core of the wall. The favored circulation is right with respect to the direction of propagation, \ie, given by the right-hand or corkscrew rule.  Above a certain threshold of field the torque arising from $\Ha$ may even switch the circulation. Then, the situation becomes identical to the latter case, upon time-reversal symmetry, and the mobility is again high\bracketsubfigref{img_wall-motion}a. This rotation process is similar to the Walker breakdown in thin strips. Two cases occur: in the case of thin-walled tubes, which resemble rolled thin films, precession keeps going, radially inward and then outward, considerably slowing down the wall. For a fixed ratio $\beta\lesssim1$, the Walker field is smaller for larger-radius walls. In thick-walled tubes or in wires, the internal field is so large when it is opposed to the torque created by the external field, that further rotation cannot occur and magnetization gets locked at a given angle $\phi$, in a steady-state regime\bracketsubfigref{img_wall-motion}a. This situation may be viewed as a 'once-only' Walker breakdown, in which $\phi$ has shifted by $\approx\pi$. The energetics of the vortex wall versus its azimuth is shown in \subfigref{img_wall-motion}b, in the general case of a tube. The thinner the wall tube is, the closer are the situations for outward and inwared radial magnetization, which the DW needs to reach for a Walker process. The asymmetry increases for smaller tube radius or larger relative tube wall thickness. The phenomenology is very similar when driven under current, see the overview by \citet{bib-WIE2010} for BPWs in wires, and by \citet{bib-OTA2012} for vortex walls in tubes. In both cases the 1d model may be applied handwavingly, where the internal exchange+dipolar field acts like the demagnetizing field for thin strips\citep{bib-WIE2004a}. This allows to get an order of magnitude of the wall speed, which is analogous to the case of thin films: $\diffdot{q} = \gamma_0 \Ha \Delta/\alpha$ for the field-driven case, and $\diffdot{q}=u\beta/\alpha$ for the current-driven case\citep{bib-WIE2010}. In particular, no motion is expected for $\beta=0$. While the above is a general phenomenon related to geometry, the selection of circulation with DMI was also investigated theoretically, with an impact on DW motion\citep{bib-GOU2016}. However, the effect is probably much weaker than the dynamics one. Wall motion has been outlined for tubes with azimuthal magnetization, under the stimulus of an ac \OE rsted field\citep{bib-BET2008}. Wall speed of a few tens of meters per second was found under applied field of magnitude circa $\SI{10}{\milli\tesla}$, which is much lower than for the vortex wall. However the wall speed varies non-monotonously with field frequency, which points at possible inertial and damping effects. This regime is also related to the giant magneto-impedance effect, covered in the next section\bracketsecref{sec-MagnetismFMR}. This was further refined in \citet{bib-JAN2018}: starting from a simple N\'{e}el wall between two domains with azimuthal magnetization, the transition to the Walker regime involves azimuthal instabilities of the N\'{e}el wall, ending in the nucleation of vortex-antivortex pairs end the transformation into cross-tie walls. These are shown to be intrinsically unstable under \OE rsted fields through the azimuthal motion of vortices and antivortices, so that this highlights the transition towards the Walker regime. Note that, contrary to the case of strips, the periodic boundary conditions do not set a threshold field for the motion of vortices and antivortices through a transverse demagnetizing field. This transition sensitively depends on the axial applied field, having an impact on the DW width and internal energy.

Finally, there exists yet another specific phenomenon associated with BPWs in wires and vortex walls in nanotubes. \citet{bib-WIE2004a} mentioned than BPWs moved by field emit spin waves for low damping. Soon after, \citet{bib-THI2006} noticed that the speed of BPWs moved by field reaches a plateau around $\SI{1}{\kilo\meter\per\second}$ for fields above a few milliteslas\subfigref{img_wall-motion}a. The same was shown under current\citep{bib-WIE2010}. The underlying phenomenon was explained quantitatively by \citet{bib-YAN2011b}\subfigref{img_wall-motion}c, considering this time vortex walls in nanotubes: the moving DW is pumping energy into spin waves. This happens only above a given threshold of wall speed, which coincides with the minimum of the phase velocity curve $v_\varphi(k)=\omega/k$ of the spin waves in the tubes. Above this threshold, spin waves progressively build-up coherently at the front and the back of the moving wall, with two opposite $\vect k$ vectors of different magnitude\subfigref{img_wall-motion}d. A handwaving way to understand this, is to consider the wall in the fixed laboratory frame, not in the moving frame. For a instant, a magnetic moment in the core of the wall is precessing at an angular frequency $\omega$, liable to excite spin waves. The spin wave with propagation speed identical to that of the wall builds up coherently over time, by the way explaining that it is the phase velocity that matters. The energy injected in the spin waves does not contribute to accelerating the wall, whose speed therefore reaches a near plateau. This process was confirmed theoretically for BPWs\citep{bib-AND2014b}.

In the previous paragraphs we discussed separately the cases of the two types of walls, because all simulations predicted no dramatic changes in the case of straight wires and tubes. Only an axial decrease of diameter had been reported liable to induce a transformation from BPW to TVW\citep{bib-HER2004,bib-ALL2009}, the former becoming unstable at small diameter. It was shown recently experimentally and reproduced by simulation that the picture is more complex for larger diameters: above a given threshold of field, TVWs transform into a BPW after a fraction of nanosecond\citep{bib-FRU2018}. While the reverse transformation has not been predicted, instabilities were reported at large driving field, such as the nucleation of magnetic droplets with a pair of Bloch point and anti-Bloch point at the droplet surface\citep{bib-HER2004a}, or the spiralling of the BP around the wire axis\citep{bib-HER2016}. It is not clear whether this is an intrinsic physics associated with the BP, or an artefact due to the difficulty to model BPs in micromagnetism, despite the emergence of multiscale codes\citep{bib-LEB2012,bib-AND2014b,bib-LEB2014,bib-EVA2014}. The pinning of the BP on a micromagnetic or atomic mesh is indeed a recognized effect, leading to a threshold field for wall motion\citep{bib-PIA2013,bib-KIM2013,bib-SHE2014,bib-HER2015}, and modulation of speed associated with the mesh spatial period\citep{bib-THI2003,bib-USO2007}, the latter author reporting a decrease of these effects for increasing wire diameter.

While theory considered magnetization textures at will in model systems, experiments require ways to image, nucleate, control the walls through magnetic field and current. For this, individual wires are collected by dissolution of the template, and spreading on a grid or surface. We summarize below the experiments related to wall motion, possibly supported by further simulation and theory.

A first issue in experiments is to prepare domain walls. When axial magnetization is concerned, the techniques developed for flat strips cannot be applied easily to wires: saturation with magnetic field transverse to the strip in bends\citep{bib-TAN1999}, current lines with an \OE rsted field\citep{bib-HAY2006}, or injection pads\citep{bib-THO2005}. Besides, unless a material has strong pinning sites, which makes it less prone to be a model system, magnetization reversal under axial field consists of nucleation of a DW at the end of the wire, and fast propagation and annihilation at the other end. Thus, the remanent state, even following a demagnetization procedure, does not exhibit domain walls. Specific strategies need to be deployed in the case of wires and tubes. Using the natural bending of very thin wires at surface is efficient to mimick bends in flat strips\citep{bib-FRU2016c}, however the process is poorly controlled. Another way is to use modulations of diameter to create potential barriers with a view to keep the wall in a given area\citep{bib-FRU2018b}. Yet another way is to proceed to demagnetization with a magnetic field applied transverse to the wire/tube axis\citep{bib-BIZ2013,bib-FRU2014,bib-FRU2016c,bib-BRA2016}, which seems to be the most efficient method. To the contrary, it seems that domain walls nucleate spontaneously towards remanence, whatever the direction of previous saturation, in the case of tubes with azimuthal magnetization\citep{bib-FRU2017e}.

A second issue, is that domain walls in real systems feel an energy landscape related to microstructural and geometrical defects, inducing pinning. The Becker-Kondorski one-dimensional model was introduced in the early days of magnetism, to describe the microscopic process of coercivity\citep{bib-BEC1932,bib-KON1937}. In essence, it is a very good start for walls in wires and tubes. Based on this, \citet{bib-IVA2011} made an analytical description of, \eg, pinning resulting from local variations of diameter, or averaging of random anisotropy grains, similar to Herzer's model\citep{bib-HER1990}. A pioneering experiment of quasistatic field-driven DW motion was reported by \citet{bib-EBE2000}, on $\SI{35}{\nano\meter}$-diameter hcp Co wires. A distribution of pinning sites was evidenced, with propagation fields in the range $\SI{50-100}{}\milli\tesla$. The significant pinning is understandable, as hcp Co has a rather large magneto-crystalline anisotropy. This is why several groups turned to $\mathrm{Ni}_x\mathrm{Fe}_{1-x}$\citep{bib-FRU2016c} and $\mathrm{Ni}_x\mathrm{Co}_{1-x}$\citep{bib-VEG2012} alloys, which are magnetically soft for $x\approx0.8$\citep{bib-OHA2000}. \citet{bib-FRU2016c} performed a statistical analysis of pinning sites in $\mathrm{Ni}_{80}\mathrm{Fe}_{20}$ wires with diameter $\SI{70}{\nano\meter}$. The distribution spans in the range $\SI{0-12}{\milli\tesla}$, with an average value $\SI{6}{\milli\tesla}$. No clear correlation could be made between the exact location of pinning sites, and possible structural defects, as investigated with electron holography combined with structural transmission electron microscopy\citep{bib-STA2016b}. Domain walls could be moved quasistatically in $\SI{150}{\nano\meter}$-diameter $\mathrm{Ni}_{60}\mathrm{Co}_{40}$ under similar fields\citep{bib-FRU2018}. \citet{bib-SIN2010} reported anisotropic magneto-resistance measurements of Ni wires with $\SI{200}{\nano\meter}$ diameter, suggesting the observation of stochastic quantized levels in resistance, which they interpret as the spontaneous motion of walls. However, the weak dependance with external field, and absence of imaging, makes this interpretation dubious. \citet{bib-WON2016} report the effect of electric current on a contacted FeNi wire. However the diameter was $\SI{350}{\nano\meter}$, for which the DW length is expected to largely exceed one micrometer\citep{bib-FRU2015b}, and thus be loosely defined. In conclusion, there is still room for a general picture of pinning, as well as the development of materials with very soft properties.

\begin{figure}[htbp]
\centering
    \includegraphics[width=128mm]{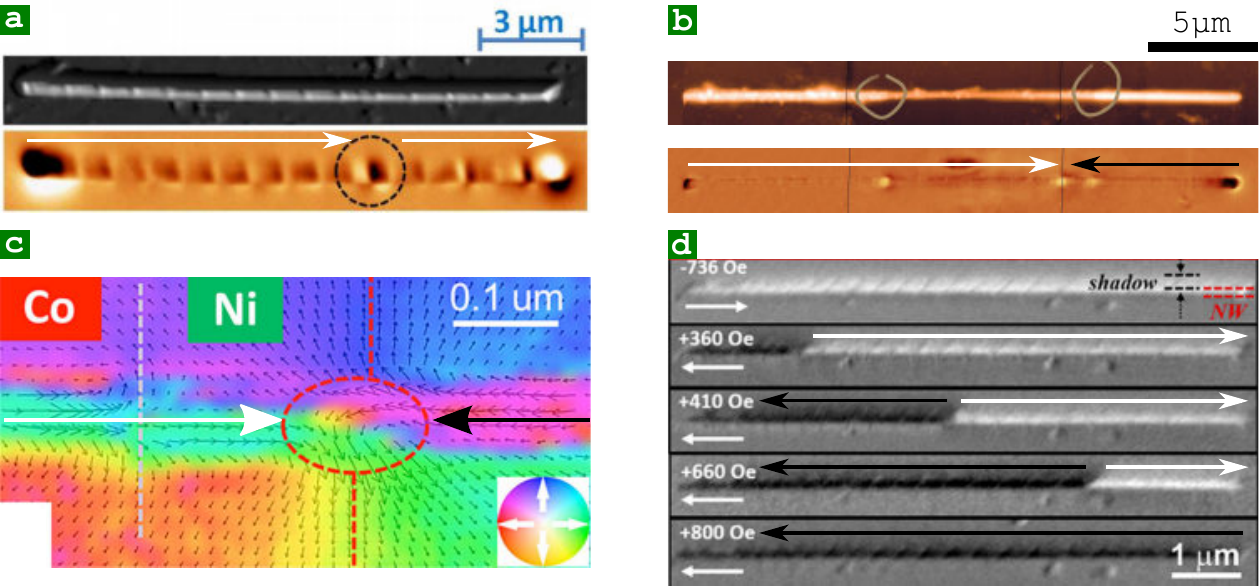}
\caption{\textbf{Interplay of domain walls with modulations}. The black arrows depict the local direction of magnetization. (a)~AFM~(top) and MFM~(bottom) images of periodic protrusions of diameter of $\thicknm{170}$ in an otherwise $\thicknm{150}$-diameter $\mathrm{Fe}_{28}\mathrm{Co}_{67}\mathrm{Cu}_5$ wire, so-called bambou-like. A claim is made for a DW pinned between two protrusions, in the course of magnetization switching, however this is not compatible with the opposite end contrasts. This may be a $\angledeg{360}$ wall, however the contrast seems weak for this\citep{bib-BER2016} (b)~AFM~(top) and MFM~(bottom) images of a $\mathrm{Co}_{40}\mathrm{Ni}_{60}$ wire $\approx\SI{10}{\micro\meter}$-long segments of diameters $\thicknm{200/150/200}$. The modulations are highlighted with circles in the AFM image. A DW is stopped ahead of a modulation of diameter in the course of magnetization  reversal\dataref{ERL CoNi G2 2015.04.10}. Sample courtesy S.~Bochmann, J.~Bachmann. (c)~DW stopped ahead of a composition-modulation in a Co/Ni wire with diameter $\thicknm{80}$, imaged with TEM in the DPC Foucault mode. Reprinted with permission from \citet{bib-IVA2016b}. Copyright 2016 American Chemical Society. (d)~Shadow XMCD-PEEM of segmented $\mathrm{Fe}_{35}\mathrm{Co}_{65}/\mathrm{Cu}$ wire with diameter $\thicknm{120}$, increasing magnetic segment length from left to right, Cu segment length $\thicknm{30}$. Each imaged is taken at remanence, following application of a quasistatic magnetic with magnitude and direction as indicated. Adapted with permission from \citet{bib-BRA2018}. Copyright 2018 American Chemical Society.}%
\label{img_wall-and-modulations}
\end{figure}

The strength of artificial pinning sites has also been investigated~(see \secref{sec-fabricationEngineered} for synthesis). A first motivation is the study of a model object in place of a defect. A second motivation is controlling pinning sites, that would be necessary in a DW-based device such as a race-track memory\citep{bib-PAR2008}. Here we focus on single-object measurements, letting aside assembly measurements mentioned in \secref{sec-MagnetismDipolarSwitching}. A first way to modulate the properties of DWs, is varying the diameter along the axis. One expects that DWs like to remain in the narrow section, for the sake minimizing both their dipolar and exchange energy. Experiments have been done for wires, however not yet for tubes. Experiments were pioneered by \citet{bib-PIT2011}, monitoring with focused MOKE Ni wires with segments of $\SI{80}{\nano\meter}$ and $\SI{160}{\nano\meter}$ diameter. However, the switching field is around $\SI{20}{\milli\tesla}$, so that several effects may compete such as nucleation, propagation and pinning at modulation, so that it was not conclusive. Separation of nucleation and propagation is also not fully comprehensive by \cite{bib-PAL2015,bib-BER2016}, considering CoFeCu with weak modulations\bracketsubfigref{img_wall-and-modulations}a. \subfigref{img_wall-and-modulations}b shows a DW stopped in front a larger modulation in $\mathrm{Co}_{40}\mathrm{Ni}_{60}$ wire. There is a number of micromagnetic simulations\citep{bib-ALL2009, bib-FRA2011,bib-ARZ2017,bib-FER2018}, quite detailed and informative, however often on specific cases. A nice overview is achieved through analytical modeling for nanowires and nanotubes\citep{bib-ALL2011}, clearly highlighting the physics of pinning, depinning and wall precession~(TWs are considered). However, this is done at the expense of complexity for formulas. \citet{bib-FRU2018c} developed a less accurate however simpler model, which delivers an interestingly universal scaling law for the pinning of a wall: $\Hp\approx\Ms\times\mathrm{Slope}$, where \textsl{Slope} is the slope of the cross-section of the wire. Micromagnetic simulations proved the good accuracy of this scaling law fro smooth and rather extended modulations. The second family of modulations is segments of different composition, either two magnetic materials, or one magnetic and not the other. In the former case a DW may by pinned at the interface. This pinning has been tracked under field through AMR\citep{bib-MOH2016} and direct imaging\citep{bib-IVA2016b}\bracketsubfigref{img_wall-and-modulations}c, for Co and Ni segments in a wire with diameter $\SI{80}{\nano\meter}$. The (weak) assistance of field-propagation through spin-tranfer torque has been reported\citep{bib-IVA2017}. In magnetic/non-magnetic segments, each magnetic segment is disconnected from the next by exchange, however may be coupled through dipolar fields\citep{bib-FRU2017}. This has been used to propose a ratchet-type switching, \ie, unidirectional\citep{bib-BRA2018}\bracketsubfigref{img_wall-and-modulations}d. Modulations of diameter and composition have also been combined\citep{bib-SAL2018}. As regards tubes, except for magnetometry on large arrays\citep{bib-PIT2009}, there exists only simulations. \citet{bib-SAL2013b,bib-NEU2013} considered wire-tube elements. Besides pinning from the thinnest to the thicker cross-section, an interesting feature is the transformation of wall type from one medium to the next. This could be used as a DW selector. More generally, pinning and impact on DW speed has been considered for non-straight wires\citep{bib-MOR2017}.

The previous paragraphs concern quasistatic investigations of wall motion in wires and tubes, which are emerging. To the contrary, the gap is striking with the fascinating and detailed predictions made by theory and simulation of their dynamics. No convincing report has been made on features such as mobility of walls as a function of type, selection of circulation for BPWs and vortex walls in tubes, plateau of speed and emission of spin waves. \citet{bib-FRU2018} reported the first hint of dynamics, by monitoring the inner structure of DWs before and after application of a pulse of magnetic field. They showed experimentally that, although initially not predicted, the type of wall can change during motion. This raises new frontiers in their understanding. This said, there is an intriguing similarity with glass-coated amorphous microwires\citep{bib-VAZ2015}. This is a large and long-standing family of wires, with various types of anisotropy. Their diameter is rather in the range of micrometers, so, not expected to be a textbook case for the micromagnetic and one-dimensional predictions. Still, wall speed in the range $\SI{1-10}{\kilo\meter\per\second}$ was measured is such wires with the so-called inductive  Sixtus-Tonk method, \eg, for FeSi and FeNiSi compositions\citep{bib-VAR2005,bib-VAR2006,bib-VAR2008b}. Interestingly, the topics and communities of microwires on one side, and nanowires and nanotubes on the other side, are closing the gap, starting to shed light on the two branches. For instance, \citet{bib-STU2014} used vectorial Kerr microscopy to elucidate the domain configuration under the influence of either axial or azimuthal field, in $\SI{100}{\micro\meter}$-diameter wires, highlighting well-defined helical domains, very similar to those in magnetostrictive\citep{bib-FRU2017e} or angle-deposited-wire \citep{bib-ZIM2018} tubes. Visualization of the DWs is key in understanding their dynamics. For instance, \citet{bib-CHI2016} evidenced the impact of the tilt of $\SI{180}{\deg}$ walls with respect to the normal to the wire axis: mobility is enhanced for tilted walls, and explained simply by the geometry of motion. The recent availability of glass-coated wires with sub-micrometer diameters, and their structural and magnetic comparison, provides an experimental playground to bring both topics ever closer together\citep{bib-OVI2014}. This leaves exciting challenges for the future.

\subsection{Ferromagnetic resonance and spin waves}
\label{sec-MagnetismFMRAndMagnonics}

\subsubsection{Ferromagnetic resonance and Giant Magneto-Impedance}
\label{sec-MagnetismFMR}

We mentioned in \secref{sec-MagnetismDipolar} the application of ferromagnetic resonance~(FMR) to arrays of magnetic nanowires. Here, FMR has been used as a characterization technique, providing information on magnetization, dipolar interactions, damping parameter. FMR has been applied to microwires\citep[chap. ~15]{bib-VAZ2015} however not really to single nanowires or nanotubes. Nevertheless, resonance effects play a key role in giant magneto-impedance~(GMI)\citep[chap.~7, 8]{bib-VAZ2015}. GMI results from the losses of ac axial charge currents. The usual physical effect is interaction of the resulting \OE rsted field with magnetization. However, the emergence of the spin-transfer torque phenomenon requires that direct effect of the current is considered\citep{bib-JAN2018}. It may play a key role in thin-walled nanotubes of small diameter, in which the \OE rsted field is weak for a given current density in the material.

GMI relies on the competition of an external axial field, with an anisotropy or \OE rsted field favoring azimuthal magnetization. Spontaneous azimuthal magnetization is ideal, as found in glass-coated strained microwires, either with an axial core and azimuthal shell, or a non-magnetic core and simply the azimuthal shell\citep[chap.~7, 8]{bib-VAZ2015}. The later is analogous to rolled-up\citep{bib-STR2014} or electroless-deposited magnetostrictive\citep{bib-FRU2017e} or angle-coated\citep{bib-ZIM2018} nanotubes. The asset of GMI is the sensing of low magnetic fields. The underlying physics may be rich and complex, and is only outlined here.

The low-frequency regime, below typically the GHz, involves DW motion. We described shortly this situation in \secref{sec-MagnetismDWsPropagation} \citet{bib-JAN2018} described this situation with analytics and simulations in nanotubes, highlighting GMI exceeding $\SI{100}{\%}$. However, the system is on the verge of the Walker regime, implying possible non-linearities depending on extrinsic effects and thus distributions from tube to tube. Also, there may be an asymmetry whether field is parallel or antiparallel to the initial N\'{e}el wall core. So, this may be problematic for sensors.

The high-frequency regime involves uniform FMR of the shell of the wire or tube. At remanence or under a dc bias of \OE  rsted field, the ground state consists of domains with azimuthal magnetization. Upon increasing the axial field the ground state becomes axial uniform magnetization. At the transition the system becomes magnetically soft, giving rise to a soft FMR mode. It is this soft mode and the associated losses and their sharp dependence with applied field, which provides the high sensitivity of GMI for sensors~(see also \secref{sec-Applications}). Other effects in GMI may include the interaction between an axially-magnetized core and azimuthally-magnetized shell, of the finite skin depth at high frequency.

\subsubsection{Spin waves}
\label{sec-MagnetismSpinWaves}

We consider here non-uniform precessional magnetization dynamics, \ie, associated with localized or propagative modes. One may be interested in spin waves for several reasons, among others: fundamental knowledge; using wires and tubes as wave guides or filters; understand their interaction with DWs, either the spin waves emitted by DWs, or the possibility to move DWs using spin waves.

We first consider axially-magnetized structures, starting with wires. As demagnetizing fields are homogeneous in uniformly-magnetized cylinders, the uniform mode is perfectly described by Kittel's formula\citep{bib-KIT1948} for demagnetizing factors $N_x=N_y=1/2$:

\begin{equation}
  \label{eqn-kittelWire}
  \omega_0=\gamma_0(H_0+\Ms/2).
\end{equation}
$H_0$ is the magnetic field applied along the wire axis. Letting aside dipolar interactions, this formula is suitable for analyzing FMR experiments on arrays of wires\bracketsecref{sec-MagnetismDipolarAnisotropy}.

\begin{figure}[htbp]
\centering
    \includegraphics[width=94.877mm]{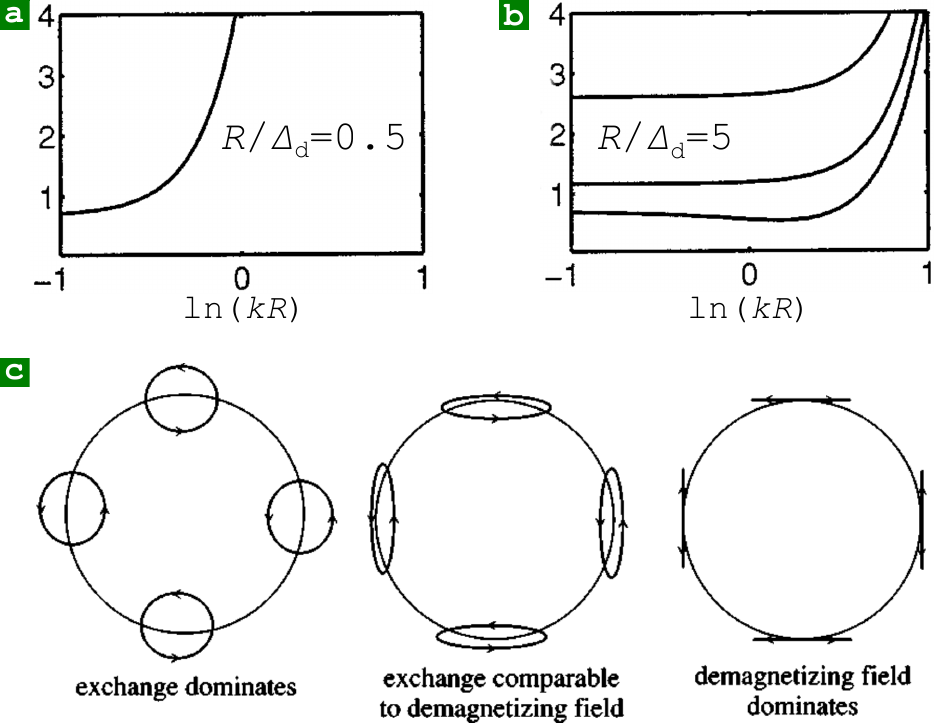}
\caption{\textbf{Ferromagnetic resonance in wires and tubes}. (a)~Zero$^\mathrm{th}$-azimuthal-order spin wave dispersion in wires with radius as indicated, and applied field $H=0.19\Ms$ parallel to magnetization. Reprinted from \citet{bib-ARI2001}. Copyright 2001 by the American Physical Society. (b)~Sketch for the phase of oscillation around the azimuth for a thin nanotube. Reprinted with permission from \citet{bib-LEB2004}. Copyright 2004 by the American Physical Society.}%
\label{img_fmr}
\end{figure}

Non-uniform modes induce both dipolar and exchange energy, responsible for the dispersion curve~$\omega(k)$, $k$ being the axial wave vector. Besides, the existence of radial and azimuthal modes gives rise to several branches of dispersion. Dipolar interactions are independent of the size $L$ of a system, while exchange energy scales with~$1/L^2$. Consequently, for large system size, one can neglect exchange. This was done by \citet{bib-JOS1961}, yielding so-called magnetostatic modes. A key feature is that all modes have a negative group velocity~$v_\mathrm{g}=\linediff{\omega}{k}$. The reason is that modes with finite $k$ give rise to an alternation of magnetic charges of opposite sign, always decreasing energy compared to the uniform mode. This is analogous to the case of stripe domains in thin films\citep{bib-MUR1966,bib-MUR1967}. The complete theory of spin waves taking into account dipolar and exchange energy, was developed by \citet{bib-ARI2001} for circular cross-section, later generalized to an arbitrary cross-section\citep{bib-ARI2004}. Exchange energy always provides a positive curvature to the dispersion curve, as it implies a cost in energy. Besides, one can show that the gain in dipolar energy is dominant for long wave length~(small $k$), while exchange becomes dominant for short wave length~(large $k$). So, the group velocity is always first negative and then positive again, for increasing~$k$. In practice however, the negative group velocity is noticeable only when $R$ is significantly larger than~$\DipolarExchangeLength$. Similarly, only in the latter case the splitting between the various branches associated with higher-order modes is sufficiently small, that several branches can be observed\bracketsubfigref{img_fmr}a. The physics can be quite complex, with crossings and hybridizations between different modes for large radius. A recent report of \citet{bib-RYC2018} revisits this theory, in particular with mode hybridization and control through external field, in mind. Spin wave modes for multiple wires coupled through dipolar field have also been addressed\citep{bib-ARI2003}. Such collectives modes have been seen in experiments on arrays of wires by Brillouin Light Scattering~(BLS)\citep{bib-WAN2002}. We are not aware of experimental measurements of spin waves in single wires.

The case of nanotubes is more complex. \citet{bib-LEB2004} proposed a theory for thin-walled tubes, taking into both dipolar and magnetostatic energy. They derive the following analytical formula:

\begin{equation}
  \label{eqn-kittelTube}
  \omega_0=\sqrt{\left[{H_0+\frac{2A}{\muZero}\left({k^2+\frac1{R^2}}\right)}\right] \cdot \left[{\Ms+H_0+\frac{2A}{\muZero}\left({k^2+\frac1{R^2}}\right)}\right]}.
\end{equation}
This equation highlights all the physics at play. For $R\rightarrow\infty$ and $k=0$, \eqnref{eqn-kittelTube} boils down to Kittel's formula for a thin film of soft magnetic material, with in-plane polarizing field. This means, the curvature is so low that locally the material feels like a thin film with infinite lateral dimensions. The term with $k^2$ is the usual contribution of exchange for spin waves. The term is $1/R^2$ arises from the fact that precession would tend to be shifted linearly with the azimuth, around the tube\bracketsubfigref{img_fmr}b, under the influence of the dipolar field. This implies a cost in energy scaling like this, as seen in \secref{sec-MagnetismCylindricalSpecific} In practice, the phase of oscillation depends on the balance of dipolar energy and exchange, so again, on the ratio~$R/\DipolarExchangeLength$. More complete theories exist to account for the different radial and azimuthal modes in thick-walled tubes\citep{bib-DAS2011b}, delivering qualitatively similar messages as for wires. Other contributions include the case of transverse applied field based on an atomic lattice case\citep{bib-NGU2006}. Few experiments exist, only on short tubes investigated by BLS. However only uniform modes are reporting, used as the signature of the magnetization state\citep{bib-WAN2005c,bib-STA2009b,bib-CHE2011c}.

Yet another case is tubes with azimuthal magnetization. A key finding is the non-reciprocity of spin wave propagation along opposite directions of the tube axis\citep{bib-OTA2016}. The reason for this is the following. This geometry is the so-called Damon-Eshbach one\citep{bib-ESH1960,bib-DAM1961}, of magnetostatic surface modes. The outward normal to the material surface and the direction of magnetization determine the propagation direction $\unitvect k$, so that waves on opposite surfaces propagate along opposite directions. As we consider rather thin-walled tubes the wave extends throughout the entire material. Nevertheless, it decays exponentially from the tube surface, inner or outer depending on the propagation direction. As these two surfaces are not equivalent due to the curvature and dipolar field, the two dispersion curves are different. The linewidth and thus decay length of the spin waves are also non-reciprocal\citep{bib-OTA2018}. Note that, due to the rolled geometry, the first-order azimuthal mode may be of lower energy that the uniform mode\citep{bib-OTA2017}. While azimuthal magnetization may be obtained from a soft magnetic material through an \OE rsted field, or in short tubes\citep{bib-WYS2017}, or in arbitrarily-long tubes with an azimuthal magnetic anisotropy\citep{bib-FRU2017e,bib-ZIM2018}, so far these predictions have not been confirmed experimentally. Care should be taken when interpreting future experiments, as it was shown on arrays of wires that the asymmetry of the magneto-optical efficiency alone may yield an asymmetry of the Stokes-Antistokes peaks\citep{bib-STA2009b}.

The interaction between spin waves and DWs is also being considered. \citet{bib-GON2010} derived the spin wave spectrum in tubes with axial magnetization, in the presence of a vortex wall. Compared with a uniformly-magnetized tube, information such as the phase shift upon reflection or transmission, or modes localized inside the DW, are extracted. Conversely, \citet{bib-YAN2011c} argued that a magnon may be transmitted through a head-to-head domain DW in wires. From theoretical arguments, as it changes the sign of its quantum of angular momentum upon the process, the difference must be provided by the DW, which is expected to move along the direction opposite the the spin wave. This is the phenomenon of transfer of angular momentum. This is different from the suggestion made in flat strips, for which reflection is expected, contributing to a dragging of the wall. This process is sometimes called linear momentum transfer\citep{bib-WAN2015}, which is most efficient when the frequency of the SW matches one of those of the DW\citep{bib-HAN2009b}. \citet{bib-YAN2015b} considered on their side tubes with a vortex wall and showed by micromagnetic simulations that SWs are largely reflected, leading to the forward motion of the DW. The frequency-dependence of the pressure is maximum in a window, limited in the low-frequency regime by propagation of SWs, and in the high-frequency regime by damping and the emergence of higher-order radial modes.  The reflection of the spin wave should be particularly effective in tubes with azimuthal magnetization, due to the non-reciprocity of spin waves on either side of the wall\citep{bib-WAN2015b}.

The spin wave can be viewed as a spin magnonic current, which can also be generated by a source of heat\citep{bib-HIN2011}. \citet{bib-ZIM2018} provided a preliminary report of the expulsion of DWs from a nanotube submitted to a long radio-frequency excitation, however with no quantitative analysis at this stage.

The topic of modulated media and spin waves is active in general, for the sake of building magnonic crystals, \ie, metamaterial displaying band gap or other filtering features. This case has also considered for modulated nanowires, producing band gaps\cite{Tkachenko2010,bib-LI2016}.

\subsection{Techniques}
\label{sec-MagnetismTechniques}

This section reviews some techniques for measuring magnetic properties, plus one on micromagnetic modelling. We focus on those techniques of specific use for magnetic nanowires and nanotubes, and highlight their specific assets and perspectives in this context. We provide references as illustrations, however, we do not detail again the physics of the various systems mentioned. This is to be found in the previous sections.


\subsubsection{Magnetometry}
\label{sec-MagnetismTechniquesMagnetometry}

\begin{figure}[thbp]
\centering\includegraphics[width=127.856mm]{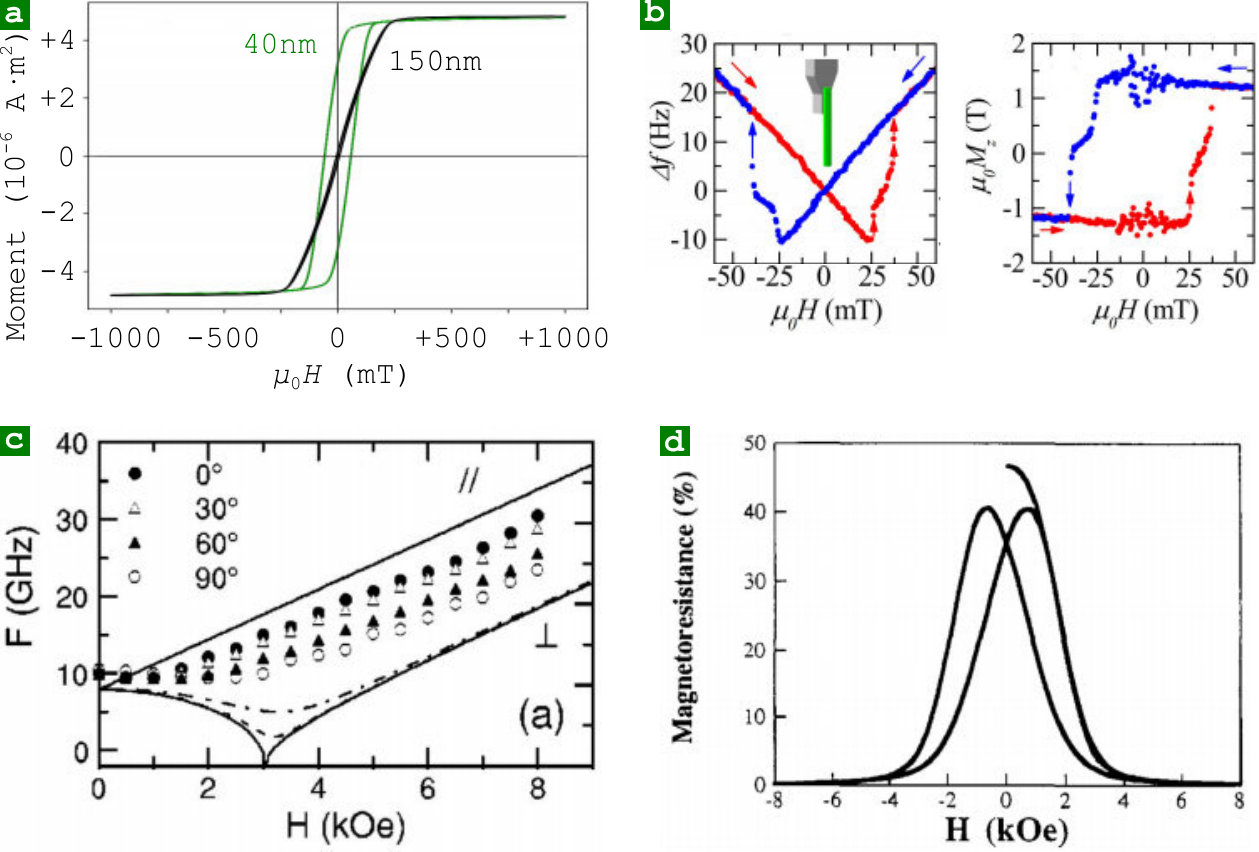}
\caption{\textbf{Illustration of some non-microscopy techniques} (a)~VSM-Squid hysteresis loop of a millimeter-sized array of $\mathrm{Co}_{40}\mathrm{Ni}_{60}$ with diameter $\thicknm{40}$ and $\thicknm{150}$. Sample courtesy S.~Bochmann, J.~Bachmann. Measurement courtesy M.~Sch\"{o}bitz. (b)~Cantilever magnetometry of a single CoFeB nanotubes with diameter $\thicknm{250}$, wall thickness $\thicknm{30}$ and magnetic field applied along the axis: frequency shift and reconstructed loop. Adapted with permission from \citet{bib-GRO2016}. Copyright 2016 by the American Physical Society. (c)~Frequency of FMR peak versus magnitude of applied field, for various direction of the latter, from parallel to the axis~(oop, \angledeg{0}) to perpendicular to the axis~(ip, \angledeg{90}). Full lines stand for models for these two extreme cases. Adapted from \citet{bib-ENC2001}. Copyright 2001 by the American Physical Society. (d)~$\tempK{77}$ Giant magneto-resistance measurements in multilayered Co[\thicknm{4}]/Cu[\thicknm{10}] wires electroplated in polycarbonate pores with diameter $\thicknm{90}$, under ip magnetic field. Reprinted from \citet{bib-PIR1996}, Copyright 1996, with permission from Elsevier.}\label{img_techniques-loops}
\end{figure}

Magnetometry refers to the measurement of the global magnetic moment of a sample as a function of an external parameter. The hysteresis loop is the most common case, consisting of moment versus external magnetic field\bracketsubfigref{img_techniques-loops}a. Measurements versus temperature, pressure, stress, electric field \etc, also exist. Magnetometry is a basic characterization technique of magnetic materials, as it provides clues about general features of a system: is it magnetic or not, magnetically soft or hard, anisotropic or not. Magnetization processes may be guessed through modelling quantities such as remanence, coercivity, susceptibility or more complex measurements such as FORC~[first-order reversal curves, see \citet{bib-DOB2013}]. Common techniques for the measurement of the flux are inductive or SQUID~(Superconducting quantum interference device, with higher sensitivity), implemented in an extraction or vibrating sample magnetometer~(VSM) scheme. Probing techniques other than through flux can be used, such as MOKE\bracketsecref{sec-MagnetismTechniquesMagnetoOptics}, magnetic dichroism\bracketsecref{sec-MagnetismTechniquesMagnetoOptics} \etc.

 Magnetometry has been largely employed by many groups, for the characterization of arrays of nanowires and nanotubes. Coercivity of a large ensemble is similar to the single-object coercivity, as expected from mean-field modelling\citep{bib-FRU2011g} and measured experimentally\citep{bib-IVA2013b}. The angular variation of coercive or switching field and its comparison with models, can be employed to distinguish switching modes of the wires and tubes, \ie, curling or transverse \citep{Bachmann2009,Albrecht2011}. The initial susceptibility or saturation field along the easy axis depend on the switching field distribution, the inter-wire/tube dipolar interactions, and single-object shape and magnetocrystalline anisotropy. All this can be modelled, however, information on the sample may be required~(\eg, geometry) to disentangle one from another effect, based on this single measurement\bracketsecref{sec-MagnetismDipolar}. Minor loops and FORC help in this regard, allowing to separate hysteretic~(switching) from anhysteretic effects~(\eg, anisotropy), as well as interactions and distributions. Single objects can be measured with local probes such as microSQUID devices prepared by lithography\citep{bib-WER1996,Buchter2013}\bracketsubfigref{img_techniques-loops}b.

 To conclude, magnetometry is a key tool for magnetic nanowires and nanotubes, providing information about the material and switching processes. Recent trends consist in refining models to extract information, and measuring single objects.

\subsubsection{Torquemetry / Cantilever magnetometry}
\label{sec-MagnetismTechniquesTorque}

Torque measurements consist in measuring the mechanical torque exerted on a magnetic sample, when subjected to an external magnetic field of varying magnitude and/or direction. Magnetization is at rest under the balanced Zeeman and internal energy torques, so that measuring the mechanical torque gives access to magnetic moment and internal energy, upon modelling. It is more reliable than hysteresis loops to extract magnetic anisotropy, as measurements are often done under magnetic field large enough to saturate the sample, which largely avoids hysteretic contributions, thus directly reflecting intrinsic properties.

Global torque measurements can be applied to large arrays of nanowires, \eg, as done for Co\citep{Ounadjela1997,Rivas2002} and Ni\citep{bib-VEG2011}. Besides anisotropy, fine information may be extracted through micromagnetic modelling, \eg, details about end domains\citep{bib-ROT2017}. Thanks to progress in nanofabrication techniques, torque measurements are increasingly being applied to single objects attached to a micro-cantilever\citep{Stipe2001,Shamsudhin2016}. In particular, the dynamic mode is very sensitive, tracking changes in oscillation frequency of the cantilever\citep{Weber2012,Buchter2013,Gross2016}. Modelling remains necessary to extract relevant information, such as magnetization~$\Ms$.

To conclude, torque measurements provide information on material and possibly switching processes. Besides global measurements in standard torquemetry setups, single-object measurements are one of the key techniques for the fine characterization of magnetic wires and tubes.


\subsubsection{Ferromagnetic resonance}
\label{sec-MagnetismTechniquesFMR}

Ferromagnetic resonance (FMR) consists in probing the precession of magnetization, which occurs typically at microwave frequencies~(GHz). FMR gives access to spontaneous magnetization, magnetic anisotropy and Land\'{e} factor~$g$ through analysis of the resonance frequencies, and magnetic damping through analysis of the line width. It is therefore another technique for extracting material properties. Standard FMR is implemented in a resonant cavity. Its sensitivity is sufficient for the investigation of large arrays of nanostructures. Since roughly two decades, it is also used based on planar striplines and microresonators prepared by lithography, that are suitable also for single nanostructures\citep{Banholzer2011}. Electrical detection via a vector network analyzer (VNA-FMR) is possible as well, with an enhanced VNA employing an interferometer being able to probe single nanostructures\citep{Tamaru2014}. References to other variants (optical heterodyne FMR, spintorque FMR, magnetic force microscopy-based FMR, and anomalous Hall effect-based FMR) can be found in work by~\citet{Tamaru2014}.

Investigations were reported on arrays of nanowires from the early days of the topic\citep{bib-ENC2001,Demand2002,Darques2009} \bracketsubfigref{img_techniques-loops}c, as well as on arrays of tubes \citep{Ahmad2017}. Information about $\Ms$ and anisotropy are extracted, while information on damping may be problematic due to the inhomogeneity of demagnetizing fields. Single-wire/tube measurements have been restricted so far to microwires, as summarized by \citet[ch.~15]{bib-VAZ2015}, or to YIG nanofibers\citep{Jalalian2011}. Wires of diameter below typically a few hundreds of nanometers present the advantage that higher-order exchange modes are significantly split\bracketsecref{sec-MagnetismFMR}.

To conclude, FMR is an important technique for characterizing the material in wires and tubes, especially for extracting anisotropy and magnetization. It remains to be applied to single objects, especially to extract the damping parameter, of key importance for domain-wall motion and other spintronic effects.



\subsubsection{Magnetoresistance}
\label{sec-MagnetismTechniquesMagnetoresistance}
\label{sec_magnetoresistance}

Magnetoresistance refers to the change of resistance of a system, when subjected to an external magnetic field. Besides direct effects of the field such as Lorentz magneto-resistance and the Hall effect, of particular interest for us result are effects resulting from the interaction of conduction electrons with magnetization: anisotropic magnetoresistance~(AMR), giant magnetoresistance~(GMR), tunnel magnetoresistance (TMR), and giant magneto-impedance~(GMI). AMR arises in materials, GMR and TMR require the nanosized juxtaposition of at least two magnetic systems, and GMI is specific to some wires and tubes. Magnetoresistance may be used as a probe of material parameters, however, it may also reveal the physics of interaction of conduction electrons with magnetization. This is therefore a key in spintronic investigations, to use magnetoresistive effects for themselves, or pave the way to spin-torque experiments using the reverse effect, that of current on magnetization.

Magnetoresistance measurements are more delicate than other techniques, since they require electrical contacting of possibly a single wire or tube. Despite this, it investigations of magnetoresistance have been reported by many groups.

\paragraph{GMR} The first measurements date back to the 90's, at the time of intense activity on GMR\bracketsubfigref{img_techniques-loops}d. Indeed, in planar stacks the natural and easiest contacting geometry is current-in-plane~(cip), while patterns with a vertical geometry suitable for the current-perpendicular-to-plane are very challenging to achieve. They are nevertheless interesting, as they rely on a different physics~(spin accumulation) and yield higher GMR ratios. Segmented wires such as Co/Cu\citep{Piraux1994,Blondel1994,Heydon1997,bib-PIR1996} and FeNi/Cu\citep{bib-DUB1999} are ideally suited for this. These provided GMR ratio at room temperature in the range $\SI{14-20}{\%}$. From varying the length of the magnetic and non-magnetic segmented it was possible to extract the spin diffusion length, of the order of $\SI{3}{\nano\meter}$ in permalloy\citep{bib-DUB1999}, $\SI{50}{\nano\meter}$ in Co\citep{bib-PIR1996,Piraux1998}, and $\SI{150}{\nano\meter}$ in Cu~(see \citet{bib-BAS2007} for a review of cpp GMR). These figures are at liquid nitrogen temperature, however, are not decreased much at room temperature. The cpp geometry has been used later in tubes: \citet{Davis2006,Davis2010} reported $\SI{4.5}{\%}$ GMR at room temperature. cip has been demonstrated as well for core-shell nanowire spin valves: CoO[(10\,nm)/Co (5\,nm)/Cu (5\,nm)/Co (5\,nm), deposited through sputter deposition around chemical-vapour-deposited Ni NWs\citep{Chan2010}. These core-shell wires exhibited a GMR ratio of $\SI{9}{\%}$, comparable to similar multilayers in the form of a planar film. cip is suitable for monitoring the location of a DW along a one-dimensional structure. It is a standard technique for planar systems\citep{bib-GRO2003b}, however it has not been applied to wires or tubes so far.

\paragraph{AMR} The motivation for measuring AMR in NWs and NTs has been largely for the sake of material characterization.  This was done for various wires, \eg for Co\citep{bib-EBE2000,bib-VIL2002}, Co and Ni\citep{Ohgai2003}, single-crystal Ni\citep{Kan2018}. This has been done also for various tubes, such as Ni\citep{Ruffer2012}, CoFeB and permalloy\citep{Ruffer2014,Baumgaertl2016}. AMR has also been applied to segmented wires alternating cobalt and nickel segments\citep{bib-MOH2016,Mohammed2017}. AMR can be used to track the presence of one or more domain walls between the voltage leads\citep{bib-EBE2000}. It has been claimed to be used to monitor DW stochastic motion\citep{bib-SIN2010,bib-MUK2011}, however the results are not consistent with those expected for a DW under field of different magnitudes. One may think of using magnon magnetoresistance\citep{bib-NGU2011}, already demonstrated in wires by \citet{bib-SER2015}, to monitor the location of a wall.

\paragraph{TMR} TMR requires the use of an insulating spacer layer. This is \apriori challenging for electroplated wires and compatible only with very thin insulating layers, for the sake of the electroplating of the second electrode. Nevertheless, this kind of structure was achieved by \citet{bib-DOU1997}, in the form Ni/NiO/Co~(the NiO was natural oxidation). However, although changes of resistance were measured under magnetic field, their sign and amplitude was seen to change over time. This was interpreted as resulting from trapped charges and therefore a barrier of insufficient structural stability, creating resonant states for tunneling\citep{bib-SOK2003}. Metal/oxide/metal structures are \apriori easier to fabricate based on tubes and has already been demonstrated\citep{MS_thesis}, however the core-shell contacting strategy may be more challenging.

\paragraph{GMI} Giant magneto-impedance is an effect specific to high-frequency excitation of microwires, with a complex physics related to skin depth, magnetocrystalline anisotropy and precessional dynamics\bracketsecref{sec-MagnetismFMR}. It has been mentioned theoretically to be applicable to tubes with orthoradial magnetization\citep{bib-JAN2018}, however there exists no experimental realization yet.

\medskip

To conclude, giant magnetoresistance has been measured in a number of various wire and tube systems, from AMR to GMR, and possibly TMR. Challenges remains in the ability to track the position of DWs, or demonstrate the synthesis of integrated magnetic field sensors directly in chips, not repositioned like for microwires.

\subsubsection{Magneto-optics}
\label{sec-MagnetismTechniquesMagnetoOptics}

Magneto-optics relies on change of light polarization or intensity upon interaction with a magnetic sample\citep{Visnovsky2006}. It is called the magneto-optical Kerr effect~(MOKE) in the reflection geometry, and Faraday effect in transmission. Depending on the geometry it may be sensitive to magnetization planar and/or perpendicular, with respect to the reflecting/transmitting surface. It is commonly used as a global or local magnetometry method, through a large beam or a beam focused down to the wavelength of light, typically a few hundreds of nanometers, or as an imaging technique in an optical microscope. The probing depth in metals is around~$\SI{10}{\nano\meter}$. A key feature of MOKE is the compatibility with time resolution, through single-shot or pump-probe experiments, with time resolution going down to the femtosecond range.

Due to its limited spatial resolution, magneto-optics has been used only in the global or local magnetometry modes on wires and tubes. MOKE with a large beam size has been used to measure arrays of wires and tubes embedded in templates\citep{bib-ZEN2002,bib-WAN2008a,Pathak2015}. An asset of MOKE is the surface sensitivity, able to highlight curling end domains, compared with bulk magnetometry\citep{bib-WAN2008a}. The Faraday geometry was demonstrated by \citet{Peng2003}. Single nanowires have been investigated as well\citep{bib-PIT2011,bib-VEG2012,Fernandez-Pacheco2013,Vidal2015,bib-PAL2015,bib-FRU2017e,bib-BRA2018}. However, one should be careful about two aspects for wires and tubes: curved surfaces may lead to partial beam depolarization, not speaking of surface roughness. Also, these and tubes in particular, can be affected or even completely destroyed by heating induced by the focused beam, due to the poor heat transfer with the supporting surface. Microwires have been investigated as well, imaged, \eg, by \citet{Chizhik2009,bib-STU2014,Chizhik2018}.

To conclude, MOKE is a simple yet powerful laboratory technique. It has been significantly used for wires and tubes, however the signal modelling may be more challenging than for thin films, due to the effect or curvature and three-dimensional magnetization. The time resolution, a key asset of magneto-optics, still remains to be demonstrated. The reproducibility of magnetization processes, required for pump-probe experiments, will require a good control over the material properties.

\subsubsection{Magnetic force microscopy}
\label{sec-MagnetismTechniquesMFM}


Magnetic force microscopy~(MFM) is a member of the family of scanning probe techniques, commonly available in a laboratory~[\citet[chap. 11 and 12]{Hopster2005}, \citet[section 2.6.1]{bib-HUB1998b}]. MFM measures point by point the interaction between a magnetic sample and a sharp magnetic tip, mounted on a flexible cantilever operated at mechanical resonance, to optimize the sensitivity of detection. While quantitative analysis is a difficult task, images may be understood qualitatively as reflecting the vertical component of the sample stray field, arising from its magnetic charges\citep{bib-HUB1997}. Samples can be measured without special preparation, and MFM is compatible with external magnetic field, injected current or varying temperature. Assets of MFM are a reasonable spatial resolution~($\SI{25-50}{\nano\meter}$) with time per image between one to a few tens of minutes. Delicate aspects include image analysis, mutual magnetic perturbation of sample and tip, and imaging of three-dimensional samples for issues of both scanning and disentangling topographic from magnetic contrast.

\begin{figure}[thbp]
\centering\includegraphics[width=128mm]{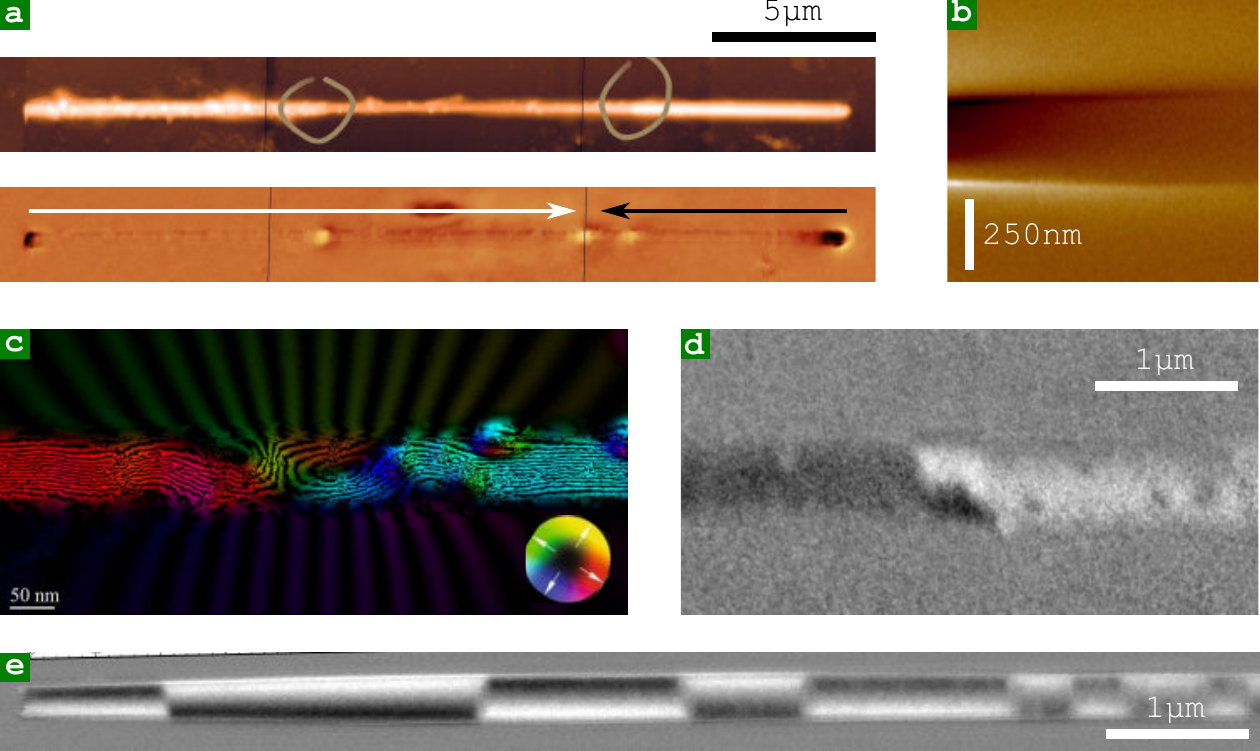}
\caption{\textbf{Illustration of some microscopy techniques} (a)~AFM~(top) and MFM~(bottom) images of a uniformly-magnetized $\mathrm{Co}_{40}\mathrm{Ni}_{60}$ wire, consisting of three segments of diameters $\SI{200/150/200}{\nano\meter}$. Contrast due to the stray field arises both at the wire ends, and at the modulations of diameter. (b)~Contrast versus lift height above the end domain of a wire such as in (a), from $\thicknm{20}$ to $\thicknm{200}$ from the left to the right of the image. The peak-to-peak oscillation amplitude is $\thicknm{50}$, and the peak-to-peak contrast amplitude is $\angledeg{1.3}$. (c)~Electron holography of a TVW wall in a $\thicknm{140}$-diameter $\mathrm{Co}_{40}\mathrm{Ni}_{60}$ wire. This is a phase map contour, colored with the local direction of induction~(see color wheel). Image courtesy A.~Masseboeuf. (d)~$\SI{4\times2}{\micro\meter}$ shadow-PEEM image of a BPW in a $\thicknm{150}$-diameter $\mathrm{Co}_{40}\mathrm{Ni}_{60}$ wire. The beam is tilted $\angledeg{60}$ away from the wire axis to image primarily the wall, however still the domains, with a lighter contrast. Imaging courtesy CIRCE beamline, ALBA synchrotron. Sample courtesy S.~Bochmann, J.~Bachmann in all the above. (e)~STXM XMCD image of a CoNiB nanotube with diameter $\thicknm{350}$ displaying azimuthal domains at remanence, and a gradient of properties along the tube axis. Sample courtesy S.~Schaefer, W.~Ensinger. Image courtesy HERMES beamline, synchrotron SOLEIL.}\label{img_techniques-microscopy}
\end{figure}

MFM has been performed both at the surface of arrays of nanowires still in templates\citep{Nielsch2002-Ni_array,bib-WAN2008a,Tabasum2014} as well as on isolated wires\citep{Heydon1997,bib-HEN2001,bib-VOC2014,bib-IGL2015,Vidal2015,bib-FRU2016c,Bran2016} including multisegmented ones \citep{Mohammed2017,Bochmann2017,bib-BER2017}. Imaging of tubes is less common, but feasible on arrays\citep{Tabasum2014} as well as isolated tubes\citep{MS_thesis, Li2008}. Some aspects of MFM specific to wires and tubes are the following.

\paragraph{Contrast} arises from magnetic charges, present at structures' ends, domain walls, modulations of any kind: magnetization, diameter, wire/tube. Thus, one should be careful not to confuse the signal of diameter\citep{bib-PIT2011} or magnetization\citep{bib-BER2017} modulations with a domain wall\bracketsubfigref{img_techniques-microscopy}a. Nevertheless, a wall may be trapped at a modulation, in which case the contrast gets higher\citep{bib-FRU2016c}, allowing one to discriminate the two situations. Also, related to the geometry of the MFM setup, in most cases the tip moment is not exactly perpendicular to the surface imaged, neither is the direction of oscillation\citep{bib-FRU2016b}. As a result, a misleading asymmetry of contrast usually arises at the front versus back side of a wire, when aligned perpendicular to the cantilever\bracketsubfigref{img_techniques-microscopy}b. For this reason, it is preferable to have wires and tubes aligned parallel to the cantilever.

\paragraph{Scanning conditions} are important to optimize MFM images. First, the large height and steep slopes on the sides of wires and tubes, require a rather slow scanning~(typically a few micrometers per second at most), preferentially performed perpendicular to the wire axis for a better removal of drift along the slow direction. Second, the lift height does not require to equal the wire diameter. A few tens of nanometers at most is generally sufficient. The value of lift height influences the contrast and the afore-mentioned asymmetry related to the geometry of the instrument\citep{bib-FRU2018b}. Third, if in need of contrast, the optimum amplitude of oscillation of the cantilever can be set equal to the wire diameter. This is roughly the extent of the stray field.

\paragraph{Sensitivity} is sufficient to measure magnetic domain walls in isolated nanowires, but it could be challenging in case of wires with very small diameter or isolated nanotubes. The constraint is even higher when low-moment tips need to be used to reduce the sample-tip interaction, such as when tracking DWs. Depending on the wire diameter, magnetization and tip, the contrast may range from a few degrees to a few millidegrees. The sensitivity can be improved by working under vacuum with a phase-lock loop\citep{bib-IGL2015} and/or lower temperatures.

Aside from MFM, there exist other more sophisticated scanning probe microscopy techniques, which can be used for nanowire or nanotube imaging. \citet{Vasyukov2018} used scanning SQUID magnetometry (SQUID on a tip), measuring the vertical component of the stray field from magnetic nanotubes. \citet{Lee2018} used nitrogen-vacancies~(NV) centers in diamond to quantify the magnetic stray field of an electrodeposited Co nanowire. Note that no scanning was performed on this very experiment, although it is often implemented for imaging. \citet{Maertz2010} demonstrated a field resolution of 20\,$\micro$T with NV-center microscopy. The limits of the technique are much lower, ultra-pure diamonds with the vacancy have sensitivity down to 3\,nT \citep{Maze2008}. So, these two techniques have a higher sensitivity than MFM and do not influence the sample. However, they do not necessarily have a better spatial resolution.

To conclude, MFM is an easy technique to implement, however even more than for other samples, imaging conditions and contrast analysis may be delicate. Other scanning probe techniques may be required on the basis of sample perturbation or sensitivity.

\subsubsection{Magnetic imaging with electron microscopy}
\label{sec-MagnetismTechniquesElectrons}


The most common technique for magnetic imaging with electrons is the transmission electron microscopy. This uses the charge of the probing electrons, namely the Lorenz force, or the change of phase of a coherent electron wave. These give rise to the so-called Lorentz and Foucault modes for the former, and electron holography for the latter, both with spatial resolution down to a few nanometers\citep{McCartney2007,Lichte2008,Kasama2011}. In both cases, the contrast arises from components of induction~(magnetization plus magnetic field) transverse to the beam of electrons. Imaging is performed in full-field~(no scanning), which may provide a close-to-video rate~(note that, technically, Lorentz and Foucault can be performed in scanning mode, however this is quite rare).

Most investigations have been performed using electron holography on wires\citep{Beeli1996,bib-BIZ2013,bib-REY2016,bib-IVA2016,Rodriguez2016,MS_thesis}\bracketsubfigref{img_techniques-microscopy}c and more recently also on nanotubes~\citep{Diehle2015,MS_thesis}. Lorenz microscopy, including the improved differential phase contrast method, has been used for imaging domain walls in nanowires by, \eg, \citet{bib-IVA2016b,bib-IVA2017}. These techniques are well suited for wires and tubes, as the ideal sample thickness lies in the range $\SI{10-100}{\nano\meter}$, as a compromise between sensitivity and transmission. An extra asset of these techniques for wires and tubes is the quantitative probing of magnetization in the bulk of a sample, which is required to solve the three-dimensional magnetization textures expected. Note however, that this benefits from comparison with numerical modelling\citep{bib-BIZ2013,bib-STA2016b} and/or image acquisition at various angles (tomography) to obtain a true 3D vectorial map of magnetization\citep{bib-TAN2015}. As a quantitative technique, electron holography is also suitable for the determination of magnetization\citep{bib-BEL1997}, the accuracy being mainly limited by the precise determination of the sample cross-section.

Other electron-based microscopy techniques exploit the spin of the electrons. This is the case of scanning electron microscopy with polarization analysis\citep[SEMPA, or spin-SEM]{Allenspach2000,Unguris2001}, analyzing the spin-polarization of secondary electrons emitted from the magnetic sample, while being scanned with a non-polarized electron beam of an SEM. The technique is very surface sensitive~(probing depth around 1\,nm, so it requires ultra-high vacuum and a clean sample surface). It can provide a vectorial magnetization map at the sample surface, with spatial resolution down to 10\,nm. While the technique has been used mostly for thin films, it can map also the surface of magnetic (nano)wires and 3D curved structures. Its specific asset for 3D structures is the large depth of focus, as demonstrated by~\citet{bib-WIL2017b}. No imaging of wires or tubes has been reported with spin-polarized low-energy electron microscopy\citep{bib-BAU2014}, another electron-based techniques exploiting the spin of the electron.

To conclude, electron microscopies are very powerful for imaging magnetic nan\-owires and nanotubes, especially for electron holography. The further development of vectorial tomography to fully resolve the magnetization distribution in three-dimensional objects, has a very high potential for wires and tubes.

\subsubsection{Magnetic imaging with soft polarized X-rays}
\label{sec-MagnetismTechniquesXRays}

Magnetic imaging with synchrotron soft X-rays relies on X-ray magnetic circular dichroism for ferromagnets~(XMCD, the difference in resonant absorption of left and right circularly-polarized light), and linear dichroism (XMLD) for antiferromagnets. For former gives access to the vectorial projection of magnetization along the beam, while the latter gives access to the projection of the domain direction. The X-ray photon energy is tuned to the absorbtion edge of a given element, which enables element-specific measurements. This is a key for complex systems, such as here segmented and core-shell tubes/wires. Spectroscopic measurements give also information about composition and oxidation states, making it a powerful material characterization technique. Even though antiferromagnetic films have been (extensively) investigated by XMLD, no investigation of antiferro magnetic nanostructures in cylindrical geometry has been reported so far. Therefore, we will focus below exclusively on XMCD.

Imaging can be performed in two main arrangements: detecting transmitted X-ray photons in (X-ray transmission microscopy -- XTM) or collecting photoelectrons emitted from the sample upon absorption of X-rays (photoemission electron microscopy -- PEEM). In both cases, the ability to have access to the distribution of magnetization inside wires and tubes is a key asset. Spatial resolution, discussed in subsections below, is typically 20-40\,nm. As in case of other synchrotron techniques, time-resolved experiments can be performed. More information on both X-PEEM and TXM can be found in a review by \citet{Fisher2015} or book by \citet{Stohr2006magnetism}.

\paragraph{Transmission X-ray microscopy}
In TXM the beam is focused on a sample, and the transmitted light is collected to form an image. There are setups for wide-field imaging (a first zone-plate lens focusing the beam to the field-of-view, a second lens for forming the image), as well as scanning (a single lens focusing the beam to typically 20-30\,nm). As (S)TXM relies on transmission, the sample thickness is limited to around 100-200\,nm, and an X-ray transparent substrate is required, such as a SiN membrane or a TEM grid. (S)TXM gives information about the volume-integrated projection of sample magnetization along the beam direction. As only photons are employed, and also the device is grounded and maintained in primary or secondary vacuum, imaging under (high) magnetic field, and application of current (pulses), are easier to implement compared with X-PEEM.

So far, (S)TXM has been used mainly for imaging magnetic tubes\citep{bib-STR2015,bib-FRU2017e,bib-ZIM2018}\bracketsubfigref{img_techniques-microscopy}e. Indeed, even for large diameter such as  (300-400\,nm), the sample thickness around the axis is not too high, thanks to the hollow core. The easier implementation of magnetic field, surrounding devices and RF feedthroughs provides it with a great potential for forthcoming experiments related to wall motion and magnonics in wires and tubes, in the time or frequency domains.

\paragraph{(Shadow) X-ray photoemission electron microscopy}
In X-ray photoemission electron microscopy\citep[PEEM]{Locatelli2008,Cheng2012PRP}, a full-field image is formed in a low-energy electron microscope\citep{bib-BAU2014}, using the secondary electrons arising from the absorbtion of X-rays illuminating the entire field of view. In most setups the X-ray beam arrives typically around 16$^{\circ}$ above the substrate plane, so that one obtains information on mostly planar magnetization, with a spatial resolution of 25-40\,nm. As the microscope needs to handle electrons, and as the sample is maintained at high voltage in most setups and is kept under ultra-high vacuum, the technique is less flexible than (S)TXM for implementation of magnetic field and RF setups, although key progress has been made recently\citep{bib-FOE2016}.

\begin{figure}[thbp]
\centering\includegraphics[width=88.354mm]{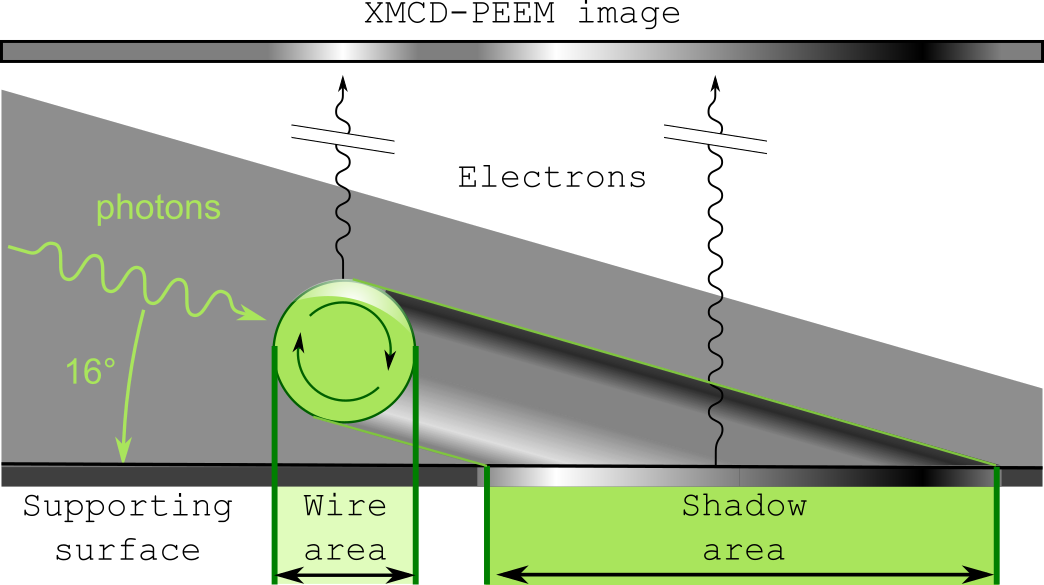}
\caption{\textbf{Shadow XMCD-PEEM}. Illustration in the case of azimuthal magnetization in a wire, leading to a bipolar contrast in the shadow.}\label{img_techniques-shadow-peem}
\end{figure}

A specific and interesting aspect when imaging wires and tubes is the following. While historically PEEM was developed for imaging thin films by collecting excited photoelectrons by PEEM from the sample surface, as the beam arrives at a small angle with respect to the substrate, 3D objects cast a shadow on the supporting surface, besides the direct image at their top surface. If the object is not too thick ($\lesssim200$\,nm), the transmitted photons provide information about sample volume, more precisely about the projection of magnetization integrated along the beam\bracketfigref{img_techniques-shadow-peem}. This makes it complementary to magnetic imaging in TEM, for which the transverse component only is imaged, instead. Note that, thanks to the grazing incidence of the beam, the resolution in the shadow is increased roughly by a factor of 3.6 (1/$\sin 16^{\circ}$) along the beam direction.

The shadow XMCD-PEEM technique was pioneered by \citet{bib-KIM2011b}, and further developed quantitatively in our group\citep{bib-FRU2014,bib-FRU2015c} when associated with micromagnetic modelling of the contrast. It has been used by us and other groups as well, for wires\citep{bib-BRA2016,bib-BRA2017,bib-KAN2018} and tubes\citep{bib-STR2014b,bib-STR2015,bib-FRU2017e,bib-WYS2017}, making it an established technique. The bulk imaging capacity of the shadow mode allowed in the above cited works, to prove the existence of structures such as the Bloch-point wall in wires\bracketsubfigref{img_techniques-microscopy}d and curling of magnetization in tubes.

Fine points of the technique are the following. First, curved surfaces themselves act as a lens for the collected electrons. This implies that the focusing settings of the objective lens is different at the wire/tube surface, compared to the supporting surface, in a way that depends on the radius of curvature, and so-called start voltage. Also, the work function is different on the object and on the supporting surface, so that the start voltage for highest intensity is different. These may be a handicap, or on the reverse be used advantageously. Second, part of the shadow is obscured by the structure itself. A work-around to get access to the full shadow can be by suspending the structure, investigating thin sections of diameter-modulated structures\citep{bib-FRU2014}, or investigate vertical objects\citep{bib-STR2015,bib-FRU2018}.

To conclude, shadow X-PEEM has quickly become a key technique for the investigation of three-dimensional magnetization textures in wires and tubes, which shows an excellent complementarity with Lorentz microscopy and holography.

\subsubsection{X-ray ptychography}
\label{sec-MagnetismTechniquesPtychography}

X-ray ptychography\citep{Pfeiffer2018} has features of both scanning transmission X-ray microscopy, and coherent diffraction imaging. A sample is illuminated with a coherent X-ray beam and diffraction patterns are recorded in far-field over multiple overlapping areas, from which a reconstruction algorithm provides a real-space map of the sample, including the magnetization map\citep{bib-DON2016}. 2D projections can be acquired for various sample orientations to reconstruct information about 3D structures in tomography\citep{bib-DON2015}. So far, it was based on hard X-rays, \ie, in the range of the K edges of the 3d elements~ (\eg, 7.71\,keV for Co). A disadvantage is the low dichroic contrast at the K~edge; an advantage is that samples with thickness up to several micrometers can be probed \citep{bib-DON2017}. Currently, the best spatial resolution in magnetic imaging is few tens of nanometers, but with prospects for even lower values.

So far the technique has not been applied to nanowires and nanotubes, although the magnetization texture observed around Bloch-points and possibly anti-Bloch-points in the bulk of a micrometer-sized $\mathrm{GdCo}_2$ pillar\citep{bib-DON2017}, is very similar to the ones of concern in nanowires. Potentially, ptychography could provide a better spatial resolution than both (S)TXM and X-PEEM.

\subsubsection{Other experimental techniques}
\label{sec-MagnetismTechniquesOthers}

Arrays of magnetic nanowires have been investigated by polarized small angle neutron scattering (PSANS)\citep{Maurer2014,Gunther2014,bib-GRU2017}. The power of this method is that fine information may be extracted from diffraction peaks of the lattice in case an ordered array is considered, the same way structural crystallography provides fine information about lattice symmetry and unit cell. There are probably niche applications for nanowires and nanotubes with SANS, which remain to be exploited.

Ensembles of both wires\citep{Chen2002} and tubes\citep{Kozlovskiy2015,Kozlovskiy2016} can be probed by M\"{o}ssbauer spectroscopy to yield the magnetic hyperfine splitting of nuclear energy levels in the surrounding magnetic field. This can be used for the determination of the internal magnetic field, of particular interest for estimating the strength of exchange. This can be useful in case the material is not known perfectly, \eg, in the case of inclusion of boron or phosphorous through electroless plating.

Brillouin light scattering~(BLS) spectroscopy has been utilized to investigate spin wave dynamics in both nanowire\citep{Wang2002,bib-CHE2011c} and nanotube\citep{Wang2005} arrays. In the content of pending experimental issues on magnonics in wires and tubes, BLS is a technique with a high potential, although BLS on single wire would be challenging as regards heating effects due to the high power used in BLS.

\subsubsection{Micromagnetic modelling}
\label{sec-MagnetismTechniquesMicromagnetism}

Numerical micromagnetics, \ie, micromagnetic simulations, is usually based on the hypothesis of micromagnetic theory, which maps magnetization on a vector field of uniform and constant magnitude\citep{bib-BRO1963b}. It has become an essential tool for analyzing experiments, and even more, predict new static and dynamic features of magnetic systems. The predictive aspect is striking for magnetic nanowires and nanotubes, for which theory is largely preceding experiments. We review below a few aspects specific to nanowires and nanotubes.

Discretization is a major issue in numerical micromagnetism. There are two main schemes: finite differences, dividing the system in regular prisms, and finite elements, dividing the system in tetrahedrons. The former is more powerful, relying on more rigorous interpolation schemes and error estimators, and is suitable for the use of a fast Fourier transform for the computation of dipolar fields. A drawback is the artificial roughness introduced when describing curved surfaces\citep{bib-KRI2014}, although smoothing schemes have been proposed to reduce this issue\citep{bib-RIA2013}. Finite elements has a less firm mathematical basis although in practice it provides good results, and has a larger computation time for dipolar fields, making use of fast multiple or similar techniques, instead of the fast fourier transform. A key advantage is the better description of curved surfaces, which may be interesting to reduce discretization artifacts. We mentioned in \secref{sec-MagnetismSingleSoft} that numerical roughness is probably the reason for the common misconception that curling states with opposite circulations, are of lower energy.

The Bloch point is another serious issue. Indeed, it violates the hypothesis of uniform magnitude of magnetization, as intrinsically it requires its local cancellation. In standard micromagnetic codes, the magnetization texture is spontaneously such that the extrapolated locus of a Bloch point lies in-between nodes where the constant modulus is imposed. For instance, in square nanowires an even number of nodes is required for the BPW to be symmetric\citep{bib-THI2006}. It is not clear presently, whether the spiralling of the Bloch-point sometimes witnessed during wall motion\citep{bib-HER2016}, may not be related. More generally, the finite size of the mesh induces an artificial pinning upon motion of BPWs\citep{bib-THI2003}. Progress has been made recently with multiscale codes\citep{bib-LEB2012,bib-HAN2014,bib-LEB2014,bib-EVA2014}. However, these still consider Heisenberg spins, so that the constraint of uniform magnetization is not fully lifted, or consider an effective penalty for exchange through LLG-Bloch equations\citep{bib-GAR1997b,bib-GAL1993}. Therefore, care should still be taken about simulations involving Bloch points.

We finish by statements about data analysis and display. First, it is sometimes not easy to display on a printed support, a three-dimensional vectorial configuration.
Special graphical views and the extraction of numerical markers can be very useful, for instance: single\citep{bib-HER2004a} or multiple\citep{bib-BIZ2011} cross-sections, open views with iso-value surfaces of a magnetization-component and their intercepts\citep{bib-HER2006,bib-HER2015,bib-FER2018}, open views of lines of magnetization in the volume\citep{bib-FRU2014}, unrolled surfaces of wires and tubes\citep{bib-YAN2011b,bib-FRU2015b}, surface maps of $\vect m\dotproduct \vect n$\citep{bib-FRU2015b}, winding numbers\citep{bib-WIE2004a,bib-FRU2018}, $m_x$, $m_y$ and $m_z$, $m_\varphi$ \etc. versus time\citep{bib-HER2004a} or position\citep{bib-USO2006}, curling or rotational\citep{bib-USO2006,bib-FRU2015b}, position versus time of special features such Bloch point, surface vortices etc.\citep{bib-FRU2018}. Second, a DW in a nanowire or nanotube is closely described by the one-dimensional Becker-Kondorski model for coercivity\citep{bib-BEC1932,bib-KON1937}. Therefore, it make sense to compute the energy of a domain wall as a function of position. However, only the energy of local minima can be computed by simple methods, \ie, when the micromagnetic state is an equilibrium one. One solution is to use a high damping, typically $\alpha\simeq1$, to mimic a quasistatic equilibrium. This fairly well reproduces the true energy landscape\citep{bib-FRU2018c}.

\section{Applications}
\label{sec-Applications}

Here we review existing and potential applications for magnetic nanowires and nanotubes. For each of them we highlight the specific assets ofr magnetic wires and tubes for the function considered, and state the level of technological readiness.

\subsection{Bio-medical applications and catalysis}
\label{sec-Applications-Particles}

Magnetic nanoparticles are being increasingly considered for various applications, when in suspension in liquid. For some of these applications, magnetic nanowires and nanotubes could be an interesting substitute to magnetic nanoparticles. First, their magnetic anisotropy and switching field can be tailored over typically two orders of magnitude, through the proper choice of diameter and length. Second, tubes, core-shell and segmented structures can provide more active surface, flexibility or multifunctional properties.

The principle of these applications is to exploit the ability to address those particles at a distance with a magnetic field, to create forces, torques, heating, a local magnetic field. Applications involve (bio)chemical separation\citep{Lee2007}, catalysis\citep{Meffre2015,bib-BOR2016}, magnetic resonance imaging\citep{Liao2011}, drug delivery \citep{Son2005,Ozkale2015}, magnetic hyperthermia\citep[and included references]{bib-CAR2011}.


\paragraph{Cancer cell destruction}
Magnetic nanoparticles are at an advanced stage of clinical tests for magnetic hyperthermia, with remote control of local heating is induced by an ac magnetic field to trigger the death of the neighboring cells\citep{bib-CAR2011}. The mechanical rotational motion of anisotropic particle through an applied torque, is also an option to trigger apoptosis, \ie, the self-destruction of the cell. This has been first demonstrated for microdisks\citep{Kim2010,bib-LEU2015}, while it is now extended to other shapes. In the latter case, fluid friction can also lead to a local temperature increase\citep{Egolf2016}. Wires and tubes could be interesting for the flexibility to tune first the coercive and anisotropy fields, two key parameters to adjust for ac heating and torque. Second, the steric effect of the particle and the force created on a cell, could be largely tuned independently from the magnetic torque, through the use of core-shell structures. Iron-based materials such as ALD $\mathrm{Fe}_3\mathrm{0}_4$\citep{Bachmann2009} are interesting for biocompatibility, to the contrary of Ni- and Co-containing materials, even if capped. Aside from experiments, simulations are being conducted in order to find optimal material and geometrical parameters as well as frequency of external magnetic field, both for wires\citep{Fernandez-Roldan2018} and tubes\citep{Gutierrez-Guzman2017}.


\paragraph{Catalysis and sensors}
Local heating induced by mechanical friction or magnetic losses during hysteresis are considered to facilitate the catalysis of chemical reactions. Local heating could considerably reduce the energy footprint of a reaction, as well as preserve from heat other sensitive constituents of a fluid. This has been demonstrated on core-shell particles, for the efficient conversion of carbon dioxide into methane\citep{Meffre2015,bib-BOR2016}. Again, wires and tubes could enhance the flexibility of magnetic and mechanical parameters. These could combine magnetic layers with non-magnetic catalysts. The magnetic part could be exploited not only for heating, but also to re-collect the expensive catalyst, as shown by \citet{Schaefer2016} in the case of CoNiB tubes covered with Pd seeds. Note that catalytic properties of tubes are potentially more interesting than those of solid wires, thanks to the larger surface/volume ratio. These can be exploited also for dye degradation and decomposition of dangerous molecules\citep{Li2014,Muench2013}, \eg, in the context of water purification. Magnetic tubes can be also employed as gas/liquid sensors and electrodes for lithium-ion batteries\citep{Chen2005}.

\subsection{Hard magnetic materials}
\label{sec-Applications-PermanentMagnets}

Permanent magnets are indispensable for a rising number of applications, in the context of intelligent systems and energy conversion. These include remote sensing, energy production and electric motors. High-performance permanent magnets use rare-earth elements to achieve the required high coercivity, some of those elements needed such as dysprosium being rare enough to raise criticality concerns. Therefore, aside from recycling rare-earths, there is a growing interest in alternatives materials.

This impetus revived investigations of various kinds of materials, including 3d ferromagnets and their alloys. While Fe and Co are desired to achieve a high energy product thanks to their high magnetization, the coercivity of bulk compounds is relatively low. This is where nanowires come into play, owing to the sharply increasing nucleation field for decreasing diameters\citep{bib-ZHE2000,bib-ZEN2002}. Co-based systems\citep{Li2017} are quite extensively investigated in this regard. \citet{Gandha2014} reported a coercivity at room temperature of about 820\,kA/m ($\SI{10.3}{\tesla}$ in terms of induction) for single crystal Co wires with diameter around 15\,nm and 200 to 300\,nm in length. Ensembles of nanowires in a matrix of polymer or epoxy need to be used to reach a high packing density of such wires, and thus a large energy product. Reported room-temperature values are in the range $\SI{100-160}{\kilo\joule\per\cubic\meter}$. Improving the uniformity of shape and length, and minimize surface oxidation, could allow to reach $\SI{200}{\kilo\joule\per\cubic\meter}$\citep[ch. 21, Soft chemistry nanowires for permanent magnet fabrication]{bib-VAZ2015}.

While it is clear that the performance of sintered NdFeB magnets cannot be beaten by nanowires, applications may exist for moderate performance and volumes. For instance, similar to AlNiCo and ferrites, the energy product in the case of wires does not decrease as sharply with temperature compared to NdFeB\citep{Maurer2007}. So, nanowires could compete with SmCo or bonded NdFeB.

\subsection{Magnetic force microscopy tips}
\label{sec-Applications-MFM}

Magnetic nanowires and potentially also nanotubes could be used as probes for magnetic force microscopy. Their assets could be high coercivity, high aspect ratio, high spatial resolution, quantitative analysis. Magnetic nanowires can be grown by electrodeposition~(or other methods) in a template, later dissolved and attached to the apex of a Si tip on a standard cantilever\citep{Yang2005}. Alternatively, one can perform the electroplating of wires directly on a Si tip covered with a conductive layer such as platinum, as demonstrated,\eg,  by~\citet{Alotaibi2018}, however, the probes might not be so sharp. Another possibility is the direct synthesis of high-aspect ratio magnetic rods on tips by focused electron beam induced deposition \citep[see also \secref{sec-fabricationDepositionFEBID}]{Lau2002,Belova2012,Gavagnin2014}. The advantage of FEBID is the precise control of positioning as well as the tilt of the wires. This way, a perfectly vertical magnetic moment of the probe can be achieved, which is essensial for quantitative analysis of MFM contrast\citep{bib-FRU2016b}.  Imaging employing magnetic tubes is not so common aside from carbon nanotubes filled/coated with magnetic particles \citep{bib-VOC2010}. Quantitative analysis was put forward in the latter case, making use of the well-defined monopole-like feature of a wire, giving reliable access to the first spatial derivative of the stray field arising from the sample\citep{bib-KRA1995b}. Some companies are currently adding such tips to their catalog. These tips display interesting features. However, unless they can be batch-fabricated, they will remain a niche product.

\subsection{Sensors of magnetic field based on magneto-impedance}
\label{sec-Applications-GMI}

Giant magnetoimpedance effect (GMI) in amorphous single microwires \citep{Peng2016-GMI_microwires} is used in magnetic field sensors\citep[see also section~\ref{sec-MagnetismFMR}]{Chen2018}, with products already on the market (companies such as Microfir Tehnologii Industriale, Tamag Ib\'{e}rica S.L., Aichi Steel). GMI sensors have a low power consumption and very good sensitivity, better than the nT in product and ultimately down to the pT\citep{Uchiyama2012}. Magnetic microwires can be also used as wire-less stress sensors \citep{Marin2017}. Recently, \citet{bib-JAN2018} calculated that the GMI can be exploited also in magnetic (nano)tubes with azimuthal magnetization, displaying multiple domain walls. Suitable samples with long length have been already fabricated by \citet{bib-FRU2017e}. In principle, segmented and core-shell magnetic nanowires, as well as multilayered tubes, could be used as magnetoresistance sensors, possibly directly integrated on chips.


\subsection{Microwave applications and magnonics}
\label{sec-Applications-Microwave}

First, properties of large assemblies or wires can be used in a device for processing large RF signals. The material and physics are ready for applications.  Magnetic microwires can be exploited as microwave absorbers and electromagnetic interference shielding, when embedded in polymeric matrices\citep{Peng2016-GMI_microwires}. As mentioned in \secref{sec-MagnetismFMRAndMagnonics}, the precession of magnetic moments in ferromagnets can be excited with microwaves. If the microwave frequency matches the resonant frequency, significant microwave absorption occurs. These can be exploited in filters and other devices.

The resonant frequency highly depends on the material, but also on the shape of the magnetic body. Unlike in films (well-decribed by Kittel formula), resonance in rod-like objects (and thus microwave absorption) occurs also without application of external magnetic field (\eg, around $\SI{30}{\giga\hertz}$ for $\mathrm{Co}_{50}\mathrm{Fe}_{50}$ nanowires). The range of working frequency can be selected by a choice of material and packing density of the wires. The frequency can be further tuned over an even wider range by application of magnetic field, \eg, $\SI{25-40}{\giga\hertz}$ for arrays of $\SI{50}{\nano\meter}$ Co NWs\citep{Darques2009}. These arrays in polymeric\citep{Darques2009} and alumina\citep{Darques2010} matrices can be used in tunable microwave devices such as circulators, filters, and phase shifters\citep{Sharma2014}. The asset of nanowires here is the easy and compact integration on a chip.

Second, single wires and tubes could be used as a medium for magnonics, for transferring information, and possibly computing. These possibilities lie more at the stage of a concept. The asset of magnonics against charge current is the reduction of power consumption. While some magnonic physical components have been demonstrated using the low-damping material YIG~(interferometers, transistors, circuits, splitters,\,\ldots\citep{Chumak2014,Vogt2014}, the size of these prototypes is in hundreds of micrometers, if not millimetres. Downscaling of these components is needed in order to have viable technology to compete with current silicon (and charge-based) devices. Magnetic nanostructures such as magnetic wires and tubes~(see section~\ref{sec-MagnetismSpinWaves}) could be good candidates for components in miniaturized devices. In particular, nanotubes with azimuthal magnetization are of interest as these could serve as a non-reciprocal magnonic waveguide\citep{bib-OTA2017}. There has been also a theoretical modelling of bandgap in magnonic crystals based on segmented cylindrical nanowires\citep{Tkachenko2010,bib-LI2016}, however, there have been no corresponding experiments so far.

Third, segmented wires such as Co/Cu/Co, have been used to implement a spin-torque nano-oscillator~(STNO)\citep{bib-MOU2008,bib-ARA2013,bib-ARA2017}. The asset of nanowires is the prospect to obtain the synchronization of a large number of oscillators, either along the same wire or in neighboring wires, which would help increase the power generated by STNO in the $\micro\watt$ range required for applications. Nevertheless, only single oscillators have been measured so far.


\subsection{Data storage}
\label{sec-Applications-Storage}

\citet{bib-PAR2004} proposed the so-called race-track concept of a three-dimensional solid-state storage medium, based on a dense array of vertical magnetic nanowires. Information would be encoded with domain walls, which would be moved with the then recently-proposed\citep{bib-SLO1996,bib-BER1996} and demonstrated\citep{bib-MYE1999,bib-GRO2003b} spin-transfer-torque phenomenon. This concept has caught the attention even though it was associated with severe technological bottlenecks, because it would provide a disruptive way to greatly enhance the capacity of mass-storage media. This further motivated the scientific community to explore the underlying physics, which has been done in flat strips, both with planar\citep{{bib-YAM2004b,bib-THO2006}} and perpendicular \citep{bib-BUR2009,bib-MIR2010,bib-MIR2011} magnetization, and lead also to full demonstrators\citep{bib-ANN2011}.

Such strips cannot be realized at sufficiently small size to be competitive with solid-state mass storage devices such as flash. Flash gained its competitiveness by pilling up many layers of storage, however, this remains an incremental process. A true 3D implementation of a storage device may reach sufficiently high areal density to compete on the market. Porous template layers and their filling, leading to wires and tubes, are the only viable route for the deep fabrication of such a dense array. While there remain acute challenges on the way, this is an incentive to demonstrate viable spintronic building blocks based on magnetic wires and tubes.

\section{Conclusion and perspective}
\label{sec-Perspectives}

While we can for sure expect further predictions of novel physical effects and functionalities as regards both wires and tubes, one thing is clear: as far as single-objects are concerned, theory has always been and remains largely ahead of experiments. This points at the interdisciplinary and sometimes acute experimental challenges existing in the field of magnetic nanowires and nanotubes. These pertain to material science, synthesis engineering, device integration, instrumentation and simulation, and are discussed below.

While magnetic NWs and NTs have been synthesized and investigated for three decades, it is only recently that their properties are being monitored at the scale of single objects, such as for instance DW motion. Thus, one now realizes that materials need to be optimized in the light of the investigation of single-object processes, not like previously on the basis of global processes such as hysteresis loops. From that respect, the field lies two or three decades behind that of flat nanomagnetism. For instance, it would be highly desirable to simply understand the source of pinning of DWs in wires and tubes, and design new strategies to reduce it. Currently pinning fields are in the range $\SI{1-10}{\milli\tesla}$, which is about one order or magnitude larger than in flat strips. Besides, our capacities for producing heterostructures must be expanded to match those under control in spintronics thin-film technology: so far we are lacking convincing ferromagnetic/antiferromagnetic exchange bias, spin-Hall bilayers, RKKY interlayer exchange coupling. For all three above-mentioned effects the control of the interface from the point of view of magnetic exchange or spin transparency is crucial, and sets new challenges in material synthesis techniques.

Integration of wires in fundamental devices is emerging, as regards manipulation with local and pulsed field, injection of spin waves or spin-polarized current. This will for sure expand, and unlock the demonstration of new physics pertaining to the dynamics of DW motion, and magnonics. The gap is indeed currently very large between theoretical predictions and experimental demonstrations: high DW mobility for BPWs, magnonic regime, non-reciprocity of spin waves, spin-wave-driven DW motion. Nevertheless, the refinement of our understanding of the atomic nature and behavior of the Bloch point remains a frontier with great challenges.

To do so, however, there is the need for pushing the power of imaging techniques further, combining high spatial resolution with time or frequency resolution. This requires also operando setups, in which wires can be excited and read electrically, at the same time as imaging. However, as combining high spatial and time resolution relies on stroboscopic pump probe experiments in general, we come back to the request of better control of the materials, to be able to investigate reproducible events. Finally, ptychography and vectorial tomography, which are still emerging both for electron and X-ray microscopy, are certainly called for a bright future. While single wires and tubes are intrinsically three-dimensional on their own, and already benefit from these techniques, the emergence of three-dimensional magnetic scaffolds calls even further for these.

As regards applications, we have reviewed a number of growing interests for sensing and bio-applications, for instance. In parallel and on the long run, although the proposal for a 3D race-track memory is extremely ambitious, and full of challenges of many types,  academic researchers should nevertheless be ambitious in expending the functions of spintronics to wires and tubes. These are indeed the natural objects to build three-dimensional devices, related to the existing synthesis techniques for pores in the academics as well as in the semiconductor industry~(vertical interconnects). We may argue that too many challenges are ahead, however technology will not dwell on those materials not capable of switching to~3D. 2017 was a key year in that respect, as for the first time the areal density of NAND flash memory exceeded that of hard disk drives. This was made possible by the move of the flash industry to stack up to tens of actives layers for storage nowadays, which is the reason why we all have flash sticks and disks in our devices. It is interesting to see that the MRAM industry is sneaking into 3D to push forward the retention of cells at dimensions below $\SI{20}{\nano\meter}$\citep{bib-WAT2018,bib-PER2018}. Thus, we should aim at demonstrating functions of wires and tubes for reading, writing, sensing, computing, to remain competitive on the long run. Going to 3D is the fate of high-performance technologies. However, functional 3D devices would probably always require planar processing and capacities, so that interconnecting planar and vertical structures, involving bends and branching, is probably another frontier to look at, from the design to the realization.

\section{Acknowledgements}
\label{sec-Acknowledgements}

M.S. acknowledges support by Laboratoire d'excellence LANEF in Grenoble (ANR-10-LABX-51-01) during his previous stay at Institut NEEL, Grenoble, France. We acknowledge support from the European Union Seventh Framework Programme (FP7/2007-2013) under grant agreement n\textsuperscript{$\circ$} 309589 (M3d). We thank R.~Hertel, M.~Vazquez for sharing data; A.~Masseboeuf for sharing unpublished data; S.~Bochmann and J.~Bachmann for sharing samples. We acknowledge helpful comments from J.~Bachmann.

\section{Appendices}
\label{sec-appendices}

\subsection{Symbols}
\label{sec-appendicesSymbols}

\noindent%
\begin{tabularx}{\linewidth}{lp{10.5  cm}}

$\varphi$     & Azimuthal angle in the cylindrical coordinates, referencing points in real space.\\
$\theta$        &  Polar angle in spherical coordinates, referencing magnetization direction. $\theta$ is defined with respect to the axis of the wire/tube.\\
$\phi$     & Azimuthal angle in local spherical coordinates, referencing magnetization direction.\\
$\beta$  &  Ratio of inner over outer radius of a tube. $\beta=0$ stands for a wire, $\beta\lesssim1$ stands for a thin-walled tube.\\
$\rho$  & Radial coordinate in the cylindrical coordinates, referencing points in real space.\\
$\gamma$              & Gyromagnetic ratio, equals $ge/2m$.\\
$\DipolarExchangeLength$     & Dipolar exchange length, equals $\sqrt{A/\Kd}=\sqrt{2A/\muZero\Ms^2}$. Also sometimes called: exchange length. \\
$\delta$  &  Wall width\\
$\Delta_\mathrm{W}$     & Wall parameter, so that $\delta=\pi\Delta_\mathrm{W}$ is the wall width defined by the intercept of asymptotes for the magnetization angle versus position, for a Bloch wall.\\
$\muZero$        & Magnetic permeability in vacuum: $\muZero=\unit[\scientific{4\pi}{-7}]{\mathrm{H}/\mathrm{m}}$.\\
$\omegaZero$  & Ferromagnetic resonance angular frequency.\\
$A$     &  Exchange stiffness, with unit $\joule\per\meter$. Defines the micromagnetic volume density of exchange energy $\Eech=A\left({\vectNabla{\vect m}}\right)^2$. \\
$\vect B$     &  Magnetic induction, with unit Tesla.\\
$d$     & Wire or tube external diameter.\\
$D$     & Axis-to-axis distance betwen wires or tubes in a template.\\
$D$   & Micromagnetic Dzyaloshinskii-Moriya parameter.\\
$E_\mathrm{d}$     &   Volume density of dipolar energy, with unit $\mathrm{J}/\mathrm{m}^3$.\\
$\Eex$     &   Volume density of exchange energy, with unit $\mathrm{J}/\mathrm{m}^3$.\\
$E_\mathrm{Z}$     &   Volume density of Zeeman energy, with unit $\mathrm{J}/\mathrm{m}^3$.\\
$g$                  & Land\'{e} factor. $g=1$ for magnetic moments of purely orbital origin, and $g\approx2$ for magnetic moments of purely spin origin.\\
$\vect H$     &  Magnetic field, with unit $\mathrm{A}/\mathrm{m}$.  \\
$\vectHd$     &  Dipolar field, or demagnetizing field.\\
$\Ha$     &  Anisotropy field.\\
$\vect j$           &  Volume density of electric current, with unit $\ampere\per\meter\squared$.\\
$k$  &  Wave number of a spin wave.\\
$\Kd$                 & Dipolar constant or coefficient, with unit $\joule\per\cubic\meter$. $\Kd=\frac12\muZero\Ms^2$. \\
$\Ku$                 & Uniaxial anisotropy coefficient, with unit $\joule\per\cubic\meter$. For the case of magnetic anisotropy: $\Ea=\Ku\sin^2\theta$, with unit $\joule\per\cubic\meter$. \\
$L$   &  Tube or wire length.\\
$\vectM$, $M$         & Magnetization (vector and magnitude). \\
$\vect m$  & Local unit vector parallel to magnetization. \\
$\Ms$      & Spontaneous magnetization, with unit $\ampere\per\meter$. \\
$N$  &   Demagnetizing coefficient. \\
$p$    & Porosity of a template, \ie, the ratio of the volume of pores versus total volume. When considering the resulting magnetic material we switch to calling this the \textsl{packing factor}. For wires in an alumina matrix with pitch $D$, $p=[\pi/(2\sqrt{3})]/(d/D)^2$.\\
$Q$  &  Quality factor. $Q=\Ku/\Kd$.\\
$r$  & Inner radius of a tube.\\
$r$   & Radial coordinate in the spherical system, referencing magnetization direction.\\
$R$  & Radius of a wire; outer radius of a tube.\\
$t$ & Tube wall thickness. $t=R-r=(1-\beta)R$\\
$u$  & Equivalent velocity of the spin polarized current, used in the generalized LLG equation.\\

\end{tabularx}

\subsection{Acronyms}
\label{sec-appendicesAcronyms}

\noindent%
\begin{tabularx}{\linewidth}{lX}
    AFM  & Atomic Force Microscopy \\
    ALD  & Atomic Layer Deposition\\
    AMR     & Anisotropic Magnetoresistance\\
    BPW(s)  & Bloch-point wall(s)\\
    cip  & Current in-plane, with reference to the direction of electric current with respect to material interfaces\\
    cpp  & Current perpendicular to plane, with reference to the direction of electric current with respect to material interfaces\\
    DPC & Differential phase contrast, an imaging technique used in TEM magnetic microscopy\\
    DW(s)  & Domain wall(s)\\
    FMR  & Ferromagnetic resonance \\
    FORC  &  First-Order Reversal Curves\\
    GMR     & Giant Magnetoresistance\\
    ip     & In-Plane. In the case of templates, this means, a direction transverse to the axis of wires and tubes\\
    LLG       &   Landau-Lifshitz-Gilbert (for the equation describing the time evolution of magnetization) \\
    oop     & Out-Of-Plane. In the case of templates, this means, a direction parallel to the axis of wires and tubes\\
    MFM       &   Magnetic Force Microscopy  \\
    MOKE  &  Magneto-Optical Kerr Effect\\
    NT(s)  & Nanotube(s) \\
    NW(s)   &  Nanowire(s) \\
    PEEM & Photo-emission electron microscopy \\
    SEMPA     &   Scanning Electron Microscopy with Polarization Analysis. Also called spin-SEM by some\\
    SHE  & Spin-Hall effect\\
    SQUID     &   Superconducting Quantum Interference Device\\
    STNO & Spin-Torque Nano-Oscillator\\
    VLS  &  Vapour-Liquid-Solid \\
    VSM   &   Vibrating Sample Magnetometer\\
    TW(s)  & Transverse wall(s)\\
    TVW(s)  & Transverse-vortex wall(s)\\
    TXM       &   Transmission X-ray Microscopy\\
    XMCD & X-ray magnetic circular dichroism\\
    XMLD & X-ray magnetic linear dichroism\\
\end{tabularx}



%

\bibliographystyle{elsarticle-harv}



\end{document}